\pdfoutput=1

\documentclass[11pt,twoside,a4paper,cmspaper,final,collab]{cms-tdr}
\def\svnVersion{266697}\def\cmsCernNo{2013-037}\def\cmsCernDate{\today}\def\cmsMessage{Published in the Journal of High Energy Physics as \href{http://dx.doi.org/10.1007/JHEP10(2014)160}{\doi{10.1007/JHEP10(2014)160.}}}

\begin{document}\cmsNoteHeader{HIG-13-021}

\hyphenation{had-ron-i-za-tion}
\hyphenation{cal-or-i-me-ter}
\hyphenation{de-vices}

\RCS$Revision: 263588 $
\RCS$HeadURL: svn+ssh://svn.cern.ch/reps/tdr2/papers/HIG-13-021/trunk/HIG-13-021.tex $
\RCS$Id: HIG-13-021.tex 263588 2014-10-10 16:37:27Z alverson $
\newlength\cmsFigWidth
\ifthenelse{\boolean{cms@external}}{\setlength\cmsFigWidth{0.85\columnwidth}}{\setlength\cmsFigWidth{0.4\textwidth}}
\ifthenelse{\boolean{cms@external}}{\providecommand{\cmsLeft}{top}}{\providecommand{\cmsLeft}{left}}
\ifthenelse{\boolean{cms@external}}{\providecommand{\cmsRight}{bottom}}{\providecommand{\cmsRight}{right}}
\newcommand{\vecMET}{\ensuremath{\vec{E}_\mathrm{T}^\mathrm{miss}}\xspace}
\providecommand{\tauh}{\ensuremath{\Pgt_\mathrm{h}}\xspace}
\providecommand{\CLs}{\ensuremath{\mathrm{CL}_\mathrm{s}}\xspace}

\newcommand{\PA}{\ensuremath{\cmsSymbolFace{A}}\xspace}
\providecommand{\Ph}{\ensuremath{\cmsSymbolFace{h}}\xspace}

\cmsNoteHeader{HIG-13-021}
\title{Search for neutral MSSM Higgs bosons decaying to a pair of tau leptons in pp collisions}

\date{\today}

\abstract{
A search for neutral Higgs bosons in the minimal supersymmetric extension of the standard model (MSSM) decaying to tau-lepton pairs in pp collisions is performed, using events recorded by the CMS experiment at the LHC.
The dataset corresponds to an integrated luminosity of 24.6\fbinv, with 4.9\fbinv at 7\TeV and 19.7\fbinv at 8\TeV. To enhance the sensitivity to neutral MSSM Higgs bosons, the search
 includes the case where the Higgs boson is produced in association with a b-quark jet. No excess is observed in the tau-lepton-pair invariant mass spectrum. Exclusion limits are presented in the MSSM parameter space for different benchmark
scenarios, $m_\mathrm{h}^\text{max}$, $m_\mathrm{h}^\text{mod$+$}$, $m_\mathrm{h}^\text{mod$-$}$, light-stop, light-stau, $\tau$-phobic, and low-$m_\PH$. Upper limits on the cross section times branching fraction  for gluon fusion and b-quark associated Higgs boson production are also given.
}

\hypersetup{%
pdfauthor={CMS Collaboration},%
pdftitle={Search for neutral MSSM Higgs bosons decaying to a pair of tau leptons in pp collisions},%
pdfsubject={CMS},%
pdfkeywords={Higgs, SUSY, MSSM, Taus}}

\maketitle

\section{Introduction}
A broad variety of precision measurements have shown the overwhelming success of the standard model (SM)~\cite{Glashow:1961tr,Weinberg:1967tq,SM3} of fundamental interactions,
which includes an explanation for the origin of the mass of the weak force carriers,  as well as for the quark and lepton masses.
In the SM, this is achieved via the Brout--Englert--Higgs mechanism~\cite{Englert:1964et,Higgs:1964ia,Higgs:1964pj,Guralnik:1964eu,Higgs:1966ev,Kibble:1967sv}, which predicts the existence of a scalar boson, the Higgs boson.
However, the Higgs boson mass in the SM is not protected against quadratically divergent quantum-loop corrections at high energy, known as the hierarchy problem.
In the model of supersymmetry (SUSY)~\cite{Golfand:1971iw,Wess:1974tw},
which postulates a symmetry between the fundamental bosons and fermions,
a cancellation of these divergences occurs naturally. The Higgs sector of the minimal supersymmetric extension of the standard model (MSSM)~\cite{Fayet:1974pd,Fayet:1977yc} contains two scalar doublets that result in five physical Higgs bosons: a light and a heavy CP-even Higgs boson $\Ph$ and $\PH$, a CP-odd Higgs boson $\PA$, and two charged Higgs bosons $\PHpm$. At tree level the Higgs sector can be expressed in terms of two parameters which are usually chosen as the mass of the CP-odd Higgs boson $m_\PA$ and $\tan\beta$, the ratio of the two vacuum expectation values of the two Higgs doublets.

\begin{figure*}[htbp]\begin{center}
  \includegraphics[width=0.32\textwidth]{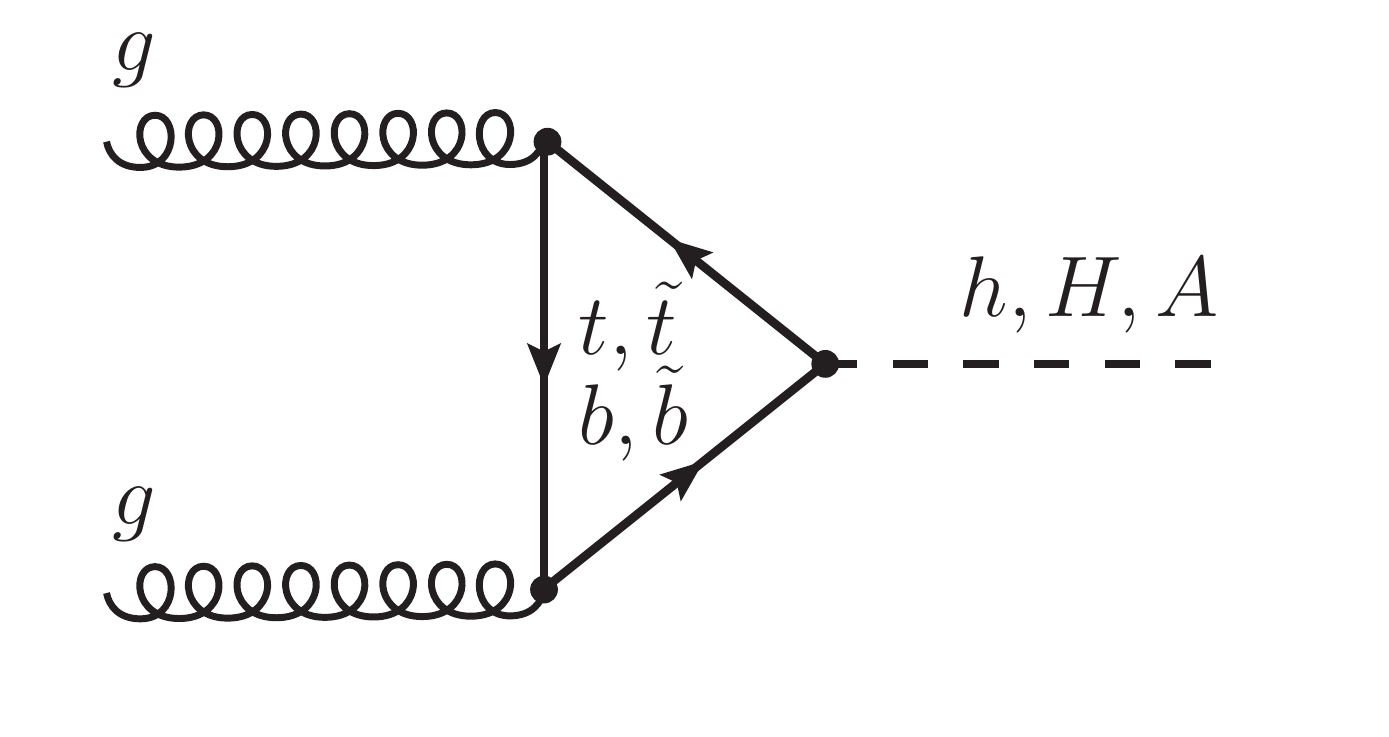}
 \includegraphics[width=0.32\textwidth]{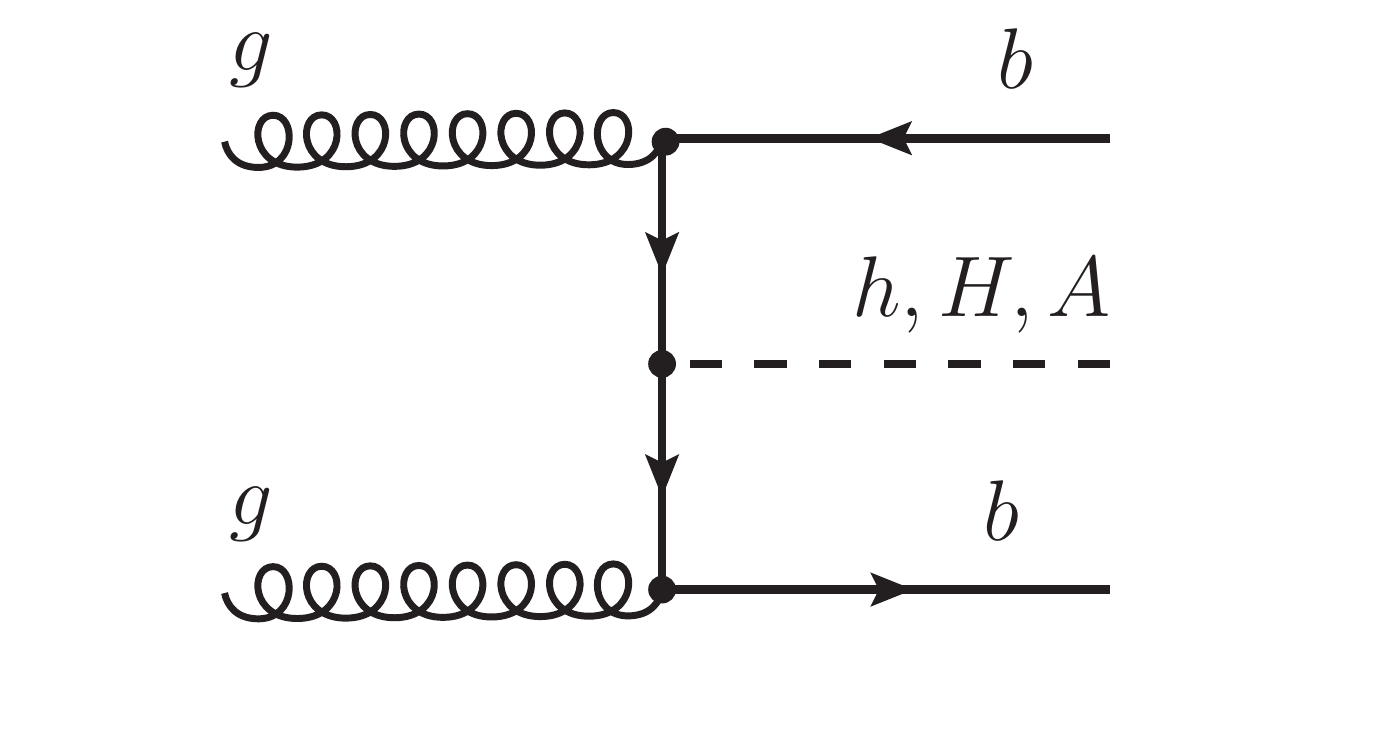}
 \includegraphics[width=0.32\textwidth]{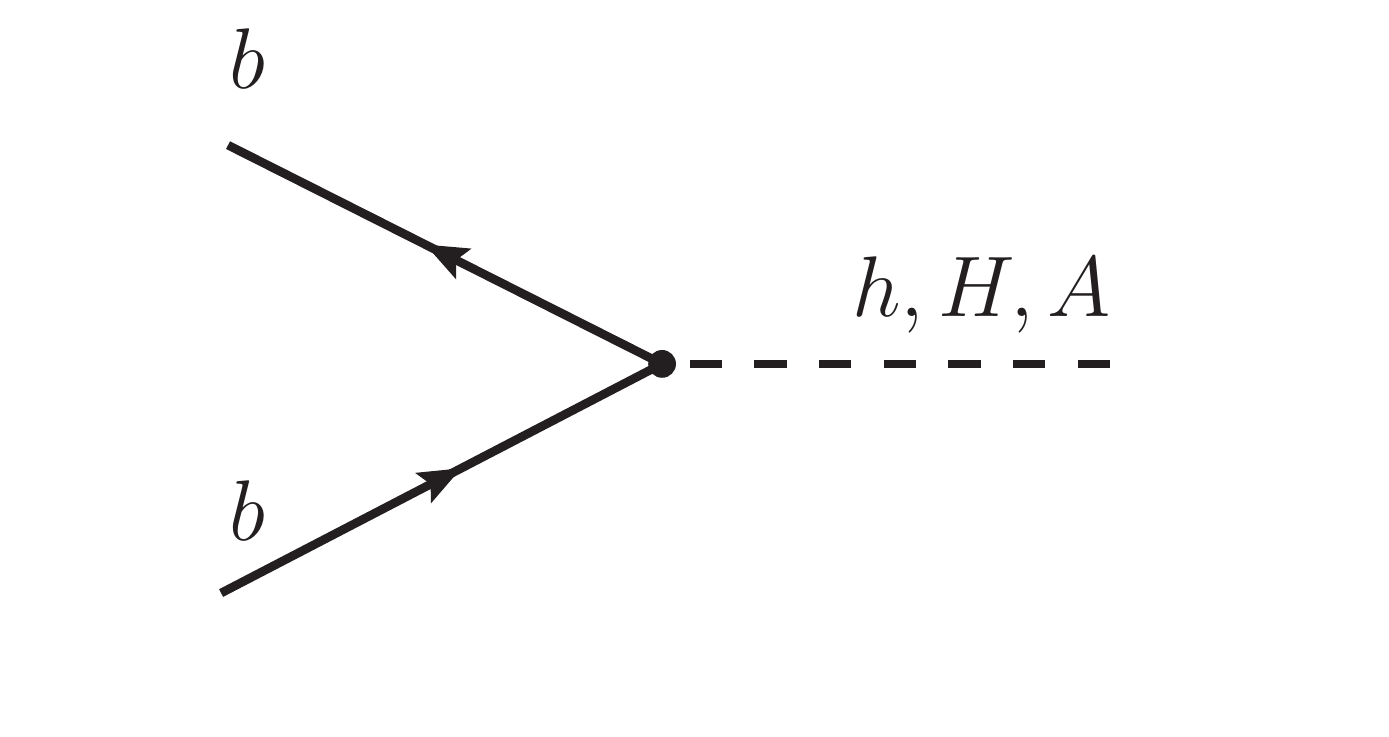}
  \caption{Leading-order diagrams of the gluon fusion (left) and  b-quark associated Higgs boson production, in the four-flavor (center) and the five-flavor (right) scheme.}
   \label{fig:FeynDiagrams}\end{center}\end{figure*}

The dominant neutral MSSM Higgs boson production mechanism is the gluon fusion process for small and moderate values of $\tan\beta$. At large values of $\tan\beta$ b-quark associated production is the dominant contribution, due to the enhanced Higgs boson Yukawa coupling to b quarks. Figure~\ref{fig:FeynDiagrams} shows the leading-order diagrams for the gluon fusion and  b-quark associated Higgs boson production, in the four-flavor and in the five-flavor scheme.
In the region of large $\tan\beta$ the branching fraction to tau leptons is also enhanced, making the search for neutral MSSM Higgs bosons in the $\Pgt\Pgt$ final state particularly interesting.

This paper reports a search for neutral MSSM Higgs bosons in $\Pp\Pp$ collisions at $\sqrt{s}=7\TeV$ and 8\TeV in the $\Pgt\Pgt$ decay channel.
The data were recorded with the CMS detector~\cite{Chatrchyan:2008aa} at the CERN LHC
and correspond to an integrated luminosity of 24.6\fbinv, with 4.9\fbinv at 7\TeV and 19.7\fbinv at 8\TeV.
Five different $\Pgt\Pgt$ signatures are studied,
$\Pe\tauh, \Pgm\tauh, \Pe\Pgm$, $\Pgm\Pgm$, and $\tauh\tauh$, where $\tauh$ denotes a hadronically decaying $\Pgt$.
These results are an extension of previous searches by the CMS and ATLAS experiments~\cite{Chatrchyan:2011nx,Chatrchyan:2013qga,Aad:2011rv} at 7\TeV,
and are complementary to the searches in $\Pp\Pap$ and $\Pep\Pem$ collisions at the Tevatron~\cite{Aaltonen:2009vf,Abazov:2010ci,Abazov:2011jh,Aaltonen:2011nh} and LEP~\cite{Schael:2006cr}, respectively.

The results are interpreted in the context of the MSSM with different benchmark scenarios described in Section~\ref{sec:mssm} and also in a model independent way,
in terms of upper limits on the cross section times branching fraction $\sigma\cdot\mathcal{B}(\phi\to\Pgt\Pgt)$ for gluon fusion ($\cPg\cPg\phi$) and b-quark associated ($\cPqb\cPqb\phi$) neutral Higgs boson production,
where $\phi$ denotes a single resonance with a narrow width compared to the experimental resolution.

\subsection{MSSM Higgs boson benchmark scenarios}\label{sec:mssm}
Traditionally, searches for MSSM Higgs bosons are expressed in terms of benchmark scenarios where the parameters $\tan\beta$ and $m_\PA$ are varied, while the other parameters that enter through radiative corrections are fixed to certain benchmark values. At tree level the masses of the
neutral MSSM scalar Higgs bosons $\Ph$ and $\PH$ can be expressed in terms of $\tan\beta$ and $m_{\PA}$ as follows
\begin{equation}
m_{\PH,\Ph}^2=\frac{1}{2} \left [ m_{\PA}^2 + m_\cPZ^2 \pm \sqrt{(m_{\PA}^2+m_\cPZ^2)^2 - 4 m_\cPZ^2m_{\PA}^2(\cos^{2}2\beta)} \right ],
\label{eq:mass}
\end{equation}
which gives an upper bound on the light scalar Higgs boson mass, $m_\Ph$, in terms of the $\cPZ$-boson mass of $m_{\Ph} \leq m_\cPZ \cos 2\beta$, which is below the excluded value of the LEP experiments~\cite{Schael:2006cr}. After radiative corrections, values of the mass larger than the LEP limits are obtained with a maximum value of $m_{\Ph}\sim$ 135\GeV~\cite{Degrassi:2002fi}.

Taking into account higher-order corrections, the following extended set of parameters defines the MSSM Higgs sector:
$M_\mathrm{SUSY}$
denotes the common soft-SUSY-breaking third-generation squark masses; $\mu$ is the higgsino
mass parameter; $M_{1}$ ($M_{2}$) is the U(1) (SU(2)) gaugino mass parameter; $X_t$ is the stop mixing parameter; $A_{t}$,
 $A_{b}$ and $A_{\tau}$ are the trilinear Higgs--stop, Higgs--sbottom and Higgs--stau-lepton couplings, respectively; $m_{\PSg}$ ($m_{\tilde{{l}_{3}}}$) is the gluino (stau) mass.
$A_{t}$ is obtained by the relation $A_{t} = X_t + \mu / \tan \beta$ and
the value of the U(1)-gaugino mass parameter $M_{1}$ is generally fixed via the unification relation $M_{1} = (5/3) M_{2} \tan^{2}
\theta_{w}$, where $\cos\theta_{w}=m_{\PW}/m_{\cPZ}$.

Previous MSSM Higgs searches~\cite{Chatrchyan:2011nx,Chatrchyan:2013qga,Aad:2011rv,Aaltonen:2009vf,Abazov:2010ci,Abazov:2011jh,Aaltonen:2011nh,Schael:2006cr}
were interpreted in the $m_{\Ph}^\text{max}$ benchmark scenario~\cite{Carena:2002qg,Carena:2005ek}, which allows
the mass of the light scalar Higgs boson $\Ph$ to reach its maximum value of $\sim$135\GeV.
The ATLAS and CMS experiments have reported the observation of a new boson with
mass around 125\GeV~\cite{Aad:2012tfa,Chatrchyan:2012ufa,Chatrchyan:2013lba}. Evidence that this new boson also decays into tau lepton pairs has recently been
reported by  CMS~\cite{Chatrchyan:2014nva}. If the new boson is interpreted as the light scalar MSSM Higgs boson $\Ph$, a large part of the
$\tan\beta$ and $m_{\PA}$ parameter space in the $m_{\Ph}^\text{max}$ scenario is excluded. However, changes in some of the parameters open up a
large region of the allowed parameter space again~\cite{Heinemeyer:2011aa}. New benchmark scenarios~\cite{Carena:2013qia} have thus recently
been proposed where the mass of one of the scalar Higgs bosons, $\Ph$ or $\PH$, is compatible with the mass of the recently discovered Higgs
boson of 125\GeV within a range of $\pm$3\GeV. This uncertainty is a conservative estimate of the theoretical uncertainty of the MSSM Higgs boson mass calculations~\cite{Degrassi:2002fi}.
Table~\ref{table:defBenchmarkScenarios} summarizes the main parameters of the benchmark scenarios considered in this study.

The traditional $m_{\Ph}^\text{max}$ scenario has been slightly modified to the $m_{\Ph}^\text{mod$+$}$ and $m_{\Ph}^\text{mod$-$}$ scenarios, where the different values of the stop mixing parameter yield a smaller light scalar Higgs boson mass than the maximal value of $\sim$135\GeV. Other scenarios which have recently been proposed due to their interesting Higgs sector phenomenology compared to the SM are the light-stop scenario, which allows for a modified gluon fusion rate; the light-stau, which gives a modified $\PH\to\Pgg\Pgg$ rate; and the $\Pgt$-phobic scenario, which gives a reduced Higgs decay rate to down-type fermions of up to 30\% at large values of $\tan\beta$ and $m_{\PA}$. The value of $m_{\PA}$ is generally varied between 90 and 1000\GeV. In the light-stop scenario the scan is only performed up to 600\GeV, because the calculation of the SUSY next-to-leading order (NLO) QCD corrections loses validity at larger masses. The range of $\tan\beta$ values studied for each scenario is chosen such that the calculation of the light scalar Higgs boson mass is well defined.
In contrast to the other scenarios, that interpret the light scalar Higgs $\Ph$ as the recently discovered Higgs boson, the low-$m_{\PH}$ scenario assumes the heavy scalar MSSM Higgs $\PH$ as the new discovered state. In this scenario, the parameters have been chosen such that the mass of the light scalar Higgs $\Ph$ is not excluded by the LEP results~\cite{Schael:2006cr}. The mass of the pseudoscalar Higgs boson is set to $m_{\PA} = 110\GeV$ and the higgsino mass parameter $\mu$ and $\tan\beta$ are varied as shown in Table~\ref{table:defBenchmarkScenarios}.

\begin{table*}[tbh]
\begin{center}
\topcaption{
  MSSM benchmark scenarios.}
\begin{tabular}{l|ccc}
Parameter           & $m_{\Ph}^\text{max}$         & $m_{\Ph}^\text{mod$+$}$        & $m_{\Ph}^\text{mod$-$}$   \\
\hline
$m_{\PA}$               & 90--1000\GeV          & 90--1000\GeV          & 90--1000\GeV     \\
$\tan\beta$          & 0.5--60                & 0.5--60                & 0.5--60           \\
$M_\mathrm{SUSY}$          & 1000\GeV             & 1000\GeV             & 1000\GeV        \\
$\mu$               & 200\GeV              & 200\GeV              & 200\GeV          \\
$M_{1}$             & (5/3) $M_2$ $\tan^2\theta_W$ & (5/3) $M_2$ $\tan^2\theta_W$ & (5/3) $M_2$ $\tan^2\theta_W$ \\
$M_{2}$             & 200\GeV              & 200\GeV              & 200\GeV         \\
$X_{t}$             & 2 $M_\mathrm{SUSY}$          & 1.5 $M_\mathrm{SUSY}$        & -1.9 $M_\mathrm{SUSY}$  \\
$A_b, A_t, A_\tau$  & $A_b=A_t=A_\tau$      & $A_b=A_t=A_\tau$      & $A_b=A_t=A_\tau$ \\
$m_{\PSg}$     & 1500\GeV             & 1500\GeV             & 1500\GeV        \\
$m_{\tilde{{\rm l}_{3}}}$ & 1000\GeV             & 1000\GeV             & 1000\GeV        \\
\hline
\end{tabular}

\vspace*{.4cm}

\begin{tabular}{l|cccc}
Parameter           & light-stop & light-stau  & $\Pgt$-phobic      & low-$m_{\PH}$ \\
\hline
$m_{\PA}$               & 90--600\GeV         & 90--1000\GeV       & 90--1000\GeV      & 110\GeV \\
$\tan\beta$          & 0.7--60               & 0.5--60             & 0.9--50            & 1.5--9.5\\
$M_\mathrm{SUSY}$          & 500\GeV             & 1000\GeV          & 1500\GeV         & 1500\GeV\\
$\mu$               & 400\GeV             & 500\GeV           & 2000\GeV         & 300-3100\GeV \\
$M_{1}$             & 340\GeV             & (5/3) $M_2 \tan^2\theta_W$ & (5/3) $M_2 \tan^2\theta_W$ & (5/3) $M_2 \tan^2\theta_W$ \\
$M_{2}$             & 400\GeV             & 200\GeV           & 200\GeV          & 200\GeV\\
$X_{t}$             & 2 $M_\mathrm{SUSY}$      & 1.6 $M_\mathrm{SUSY}$     & 2.45 $M_\mathrm{SUSY}$   & 2.45 $M_\mathrm{SUSY}$\\
$A_b, A_t, A_\tau$  & $A_b=A_t=A_\tau$  & $A_b=A_t, A_\tau=0$   & $A_b=A_t=A_\tau$  & $A_b=A_t=A_\tau$ \\
$m_{\PSg}$     & 1500\GeV            & 1500\GeV          & 1500\GeV         & 1500\GeV \\
$m_{\tilde{{l}_{3}}}$ & 1000\GeV            & 245\GeV           &  500\GeV         & 1000\GeV \\
\hline
\end{tabular}
\label{table:defBenchmarkScenarios}
\end{center}
\end{table*}

The neutral MSSM Higgs boson production cross sections and the corresponding uncertainties are provided by the LHC Higgs Cross Section Group~\cite{Dittmaier:2011ti}.
The cross sections for the gluon fusion process in the $m_{\Ph}^\text{max}$ scenario have been obtained with the NLO QCD program \textsc{HiGlu}~\cite{Spira:1995rr,Spira:1995mt}, for the contribution of the top loop, the bottom loop, and the interference. The top loop contribution has been further corrected using the next-to-next-to-leading order (NNLO) program \textsc{ ggh@nnlo}~\cite{Harlander:2002wh,Anastasiou:2002yz,Ravindran:2003um,Harlander:2002vv,Anastasiou:2002wq}. In the case of the other benchmark scenarios, the program \textsc{ SusHi}~\cite{Harlander:2012pb} has been used as it includes the SUSY NLO QCD corrections~\cite{Harlander:2004tp,Harlander:2005rq,Degrassi:2010eu,Degrassi:2011vq,Degrassi:2012vt} that are of importance in these alternative scenarios. In the \textsc{SusHi} calculations, the electroweak corrections due to light-fermion loop effects~\cite{Aglietti:2004nj,Bonciani:2010ms} have also been included. For the $\cPqb\cPqb\phi$ process, the four-flavor NLO QCD calculation~\cite{Dittmaier:2003ej,Dawson:2003kb} and the five-flavor NNLO QCD calculation, as implemented in \textsc{bbh@nnlo}~\cite{Harlander:2003ai} have been combined using the Santander matching scheme~\cite{Harlander:2011aa}.
In all cross section programs used, the Higgs boson Yukawa couplings have been calculated with \textsc{FeynHiggs}~\cite{Heinemeyer:1998yj,Heinemeyer:1998np,Degrassi:2002fi,Frank:2006yh}. The Higgs boson branching fraction to tau leptons in the different benchmark scenarios has been obtained with \textsc{FeynHiggs} and \textsc{hdecay}~\cite{Spira:1997dg,Djouadi:1997yw, Djouadi:2006bz}, as described in Ref.~\cite{Heinemeyer:2013tqa}.

\section{Experimental setup, event reconstruction, and simulation}
 The central feature of the CMS apparatus is a superconducting solenoid of 6\unit{m} internal diameter, providing a magnetic field of 3.8\unit{T}. Within the superconducting solenoid volume are a silicon pixel and strip tracker, a lead tungstate crystal electromagnetic calorimeter, and a brass/scintillator hadron calorimeter, each composed of a barrel and two endcap sections. Muons are measured in gas-ionization detectors embedded in the steel flux-return yoke outside the solenoid. Extensive forward calorimetry complements the coverage provided by the barrel and endcap detectors. The first level of the CMS trigger system, composed of custom hardware processors, uses information from the calorimeters and muon detectors to select the most interesting events in a fixed time interval of less than 4\mus. The High Level Trigger processor farm further decreases the event rate from around 100\unit{kHz} to less than 1\unit{kHz}, before data storage. A more detailed description of the CMS detector, together with a definition of the coordinate system,  
can be found in Ref.~\cite{Chatrchyan:2008aa}.

An average of 9\,(21) $\Pp\Pp$ interactions occurred per LHC bunch crossing in 2011 (2012).
For each reconstructed collision vertex the sum of the  $\pt^2$ of all tracks associated to the vertex is computed and the one with the largest value is taken as the primary collision vertex, where $\pt$ is the transverse momentum. The additional $\Pp\Pp$ collisions are referred to as pileup. 

A particle-flow algorithm~\cite{CMS-PAS-PFT-09-001,CMS-PAS-PFT-10-001} is used to combine information from all CMS subdetectors to identify and reconstruct individual particles in the event, namely muons, electrons, photons, charged hadrons, and neutral hadrons. The resulting particles are used to reconstruct jets, hadronically decaying tau leptons, and the missing transverse energy vector \vecMET, defined as the negative of the vector sum of the transverse momenta of all reconstructed particles, and its magnitude \MET.

Jets are reconstructed using the anti-\kt jet algorithm~\cite{Cacciari:2011ma,Cacciari:2005hq} with a distance parameter of 0.5. To correct for the contribution to the jet energy due to pileup, a median transverse momentum density ($\rho$) is determined event by event. The pileup contribution to the jet energy is estimated as the product of $\rho$ and the area of the jet and subsequently subtracted from the jet transverse momentum~\cite{Cacciari:2007fd}. 
Jet energy corrections~\cite{Chatrchyan:2011ds} are also applied as a function of the jet \pt and pseudorapidity $\eta = - \ln[\tan(\theta/2)]$, where $\theta$ is the polar angle. To tag jets coming from b-quark decays the combined secondary vertex algorithm is used, that is based on the reconstruction of secondary vertices, together with track-based lifetime information~\cite{Chatrchyan:2012jua}. Jets with $\abs{\eta} < 4.7$ and b-tagged jets with $\abs{\eta} < 2.4$ are used. 

Hadronically-decaying tau leptons are reconstructed using the hadron-plus-strips algorithm~\cite{Chatrchyan:2012zz}. The constituents of the reconstructed jets are used to identify individual $\Pgt$ decay modes with one charged hadron and up to two neutral pions, or three charged hadrons. The presence of extra particles within the jet, not compatible with the reconstructed decay mode of the $\Pgt$, is used as a criterion to discriminate $\tauh$ decays from jets. Additional discriminators are used to separate $\tauh$ decays from electrons and muons.

Tau leptons from Higgs boson decays are expected to be isolated in the detector, while leptons from heavy-flavor (c and b) decays and decays in flight are expected to be found inside jets. A measure of isolation is used to discriminate the signal from the QCD multijet background, based on the charged hadrons, photons, and neutral hadrons falling within a cone around the lepton momentum direction.
Electron, muon, and tau lepton isolation are estimated as
\begin{equation}\begin{aligned}
I_{\Pe,\Pgm} &=  \sum_{\rm charged}  \pt + \text{max}\left( 0, \sum_{\rm neutral}  \pt
                                        +  \sum_{\gamma} {\pt} - 0.5 \sum_{\rm charged, pileup} \pt  \right ), \\
I_{\tauh} &=  \sum_{\rm charged}  \pt + \text{max}\left( 0, \sum_{\gamma} {\pt} - 0.46 \sum_{\rm charged, pileup} \pt  \right ),
\label{eq:reconstruction_isolation}
\end{aligned}\end{equation}
where $\sum_\text{charged}\pt$ is the scalar sum of the transverse momenta of the charged hadrons, electrons, and muons from the primary vertex located in a cone centered around the lepton direction of size $\Delta R = \sqrt{(\Delta\eta)^2+(\Delta\phi)^2}$ of 0.4 for electrons and muons and 0.5 for tau leptons.
The sums $\sum_\text{neutral}\pt$ and $\sum_{\gamma} \pt$ represent the same quantities for neutral hadrons and photons, respectively. In the case of electrons and muons the innermost region is excluded
to avoid the footprint in the calorimeter of the lepton itself from entering the sum.
Charged particles close to the direction of the electrons are excluded as well, to prevent tracks originating from the conversion of photons emitted by the bremsstrahlung process from spoiling the isolation. In the case of $\tauh$, the particles used in the reconstruction of the lepton are excluded. The contribution of pileup photons and neutral hadrons
is estimated from the scalar sum of the transverse momenta of charged hadrons from pileup vertices in the isolation cone $\sum_\text{charged, pileup}$. This sum is multiplied by a factor of 0.5 that approximately corresponds to the ratio of neutral-to-charged hadron production in the hadronization process of inelastic $\Pp\Pp$ collisions. In the case of $\tauh$, a value of 0.46 is used, as the neutral hadron contribution is not used in the computation of $I_{\tauh}$. An $\eta$, \pt, and lepton-flavor dependent threshold on the isolation variable is applied.

In order to mitigate the effects of pileup on the reconstruction of \MET, a multivariate regression correction is used where the inputs are separated in those components coming from the primary vertex and those which are not~\cite{CMS-JME-12-002}.
The correction improves the \MET resolution in $\cPZ\to\Pgm\Pgm$ events by roughly a factor of two in the case where 25 additional pileup events are present.

The MSSM neutral Higgs boson signals are modelled with the event generator \PYTHIA 6.4~\cite{Sjostrand:2006za}.
For the background processes, the \MADGRAPH 5.1~\cite{Alwall:2011uj} generator is used for $\cPZ$+jets, $\PW$+jets, $\cPqt\cPaqt$ and di-boson production, and {\POWHEG} 1.0~\cite{Nason:2004rx,Frixione:2007vw,Alioli:2009je,Alioli:2010xd} for single-top-quark production.
The \POWHEG and \MADGRAPH generators are interfaced with \PYTHIA for parton shower and fragmentation. All generators are interfaced with  \TAUOLA~\cite{Davidson:2010rw} for the simulation of the $\Pgt$ decays. Additional interactions are simulated with \PYTHIA and reweighted to the observed pileup distribution in data. All generated events are processed through a detailed simulation of the CMS detector based on {\GEANTfour}~\cite{Agostinelli:2002hh} and are reconstructed with the same algorithms as the data.
The missing transverse energy in Monte Carlo (MC) simulated events is corrected for the difference between data and simulation measured using a sample of $\cPZ\to\Pgm\Pgm$ events~\cite{Khachatryan:2010xn}.

\section{Event selection}

The events in this analysis have been selected with dedicated triggers that use a combination of electron, muon and tau lepton trigger objects~\cite{CMS-PAS-EGM-10-004,Chatrchyan:2012xi,Chatrchyan:2011nv}. The identification criteria and transverse momentum thresholds of these objects were progressively tightened as the LHC instantaneous luminosity increased over the data-taking period.

In the $\Pe\tauh$ and $\Pgm\tauh$ final states, events are selected in the 2011 (2012) dataset with an electron of $\pt >20$\,(24)\GeV  or a muon of $\pt >17$\,(20)\GeV and $\abs{\eta} < 2.1$, and an oppositely charged $\tauh$ of $\pt > 20\GeV$ and $\abs{\eta} < 2.3$. The tau lepton is required to have $I_{\tauh}$ of less than 1.5\GeV.
To reduce the $\cPZ\to\Pe\Pe, \Pgm\Pgm$ contamination, events with two electrons or muons of $\pt >15\GeV$, of opposite charge, and passing loose isolation criteria are rejected.

In the $\Pe\Pgm$ and $\Pgm\Pgm$ final states, events with two oppositely charged leptons are selected, where the highest (second-highest) \pt lepton is required to have $\pt>20$\,(10)\GeV. Electrons  with  $\abs{\eta}<2.3$ and muons with $\abs{\eta}<2.1$ are used.
The large background arising from $\cPZ\to \Pgm\Pgm$ events in the $\Pgm\Pgm$ channel
is reduced by a multivariate  boosted decision tree discriminator~\cite{Hocker:2007ht} using different muon kinematic variables, including the distance of closest approach of the muon pair.

In the $\tauh\tauh$ final state, events with two oppositely charged hadronically decaying tau leptons with $\pt > 45\GeV$ and $\abs{\eta} < 2.1$ are selected, where the isolation $I_{\tauh}$ of both tau leptons is required to be less than 1\GeV.

In order to reject events coming from the $\PW$+jets background,
a dedicated selection is applied.
In the $\Pe\tauh$ and $\Pgm\tauh$ final states, the transverse mass of the electron or muon and the \MET
\begin{equation}
M_{T} = \sqrt{2 \pt \MET (1-\cos\Delta\phi)},
\label{eq:mt}
\end{equation}
is required to be less than 30\GeV, where \pt is the lepton transverse momentum and $\Delta\phi$ is the difference in the azimuthal angle between  the lepton momentum and the \vecMET. In the $\Pe\Pgm$  final state, a discriminator to reject $\PW$+jets events is formed by considering the bisector of the directions of the visible $\Pgt$ decay products transverse to the beam direction, and is denoted as the $\zeta$ axis.  From the projections of the visible decay product momenta and the $\vecMET$ onto the $\zeta$ axis,
two values are calculated,
\begin{equation}
P_{\zeta} = \left( \vec{p}_\mathrm{T,1} + \vec{p}_\mathrm{T,2} + \vecMET \right) \cdot \frac{\vec{\zeta}}{|\zeta|}  \qquad \text{and} \qquad
P_{\zeta}^{\text{vis}} = \left( \vec{p}_\mathrm{T,1} + \vec{p}_\mathrm{T,2} \right) \cdot \frac{\vec{\zeta}}{\abs{\zeta}},
\end{equation}
where $p_\mathrm{T, 1}$ and $p_\mathrm{T, 2}$ indicate the transverse momentum
of the two reconstructed leptons.  Events are selected with $P_\zeta - 1.85 P_\zeta^{\text{vis}} > -20$\GeV.

To further enhance the sensitivity of the search to Higgs bosons, the sample of selected events is split into
two mutually exclusive categories:
\begin{itemize}
\item {b-tag:} At least one b-tagged jet with $\pt>20$\GeV is required  and not more than one jet with $\pt>30$\GeV, in order to reduce the contribution from the $\cPqt\cPaqt$ background. This event category is intended to exploit the production of Higgs bosons in association with b quarks which is enhanced in the MSSM.
\item {no b-tag:} Events are required to have no b-tagged jets with $\pt>20$\GeV. This event category is mainly sensitive to the gluon fusion Higgs boson production mechanism.
\end{itemize}
This analysis uses a simpler event categorization than the dedicated SM Higgs boson search in the $\Pgt\Pgt$ decay mode~\cite{Chatrchyan:2014nva}, to reduce possible model dependencies in the result interpretation. The sensitivity to the SM Higgs boson in this analysis is thus reduced, as the contributions from vector boson fusion and boosted gluon fusion Higgs boson production are not enhanced.

\section{Background estimation}
\label{sec:background}

The estimation of the shapes and yields of the major backgrounds in each of the channels is obtained from the observed data.

The $\cPZ\to\Pgt\Pgt$ process is the largest source of background events in the $\Pe\tauh$, $\Pgm\tauh$, and $\Pe\Pgm$ channels. This background is estimated using a sample of $\cPZ\to\Pgm\Pgm$ events from data where the reconstructed muons are replaced by the reconstructed particles from simulated $\Pgt$ decays. The normalization for this process is determined from the measurement of the $\cPZ\to\Pgm\Pgm$ yield in data. This technique substantially reduces the systematic uncertainties due to the jet energy scale and the missing transverse energy, as these quantities are modelled with collision data.

Another significant source of background is QCD multijet events, which can mimic the signal in various ways. For example, two jets may be misidentified as $\tauh$ decays in which case the event will contribute to the $\tauh\tauh$ channel. Or, in the $\Pe\tauh$ and $\Pgm\tauh$ channels, one jet is misidentified as an isolated electron or muon and a second jet as $\tauh$.
In the $\Pe\tauh$ and $\Pgm\tauh$ channels, the shape of the QCD background is estimated using a sample of same-sign (SS)
$\Pgt\Pgt$ events in data. The yield is obtained by scaling the observed number of SS events by the ratio of the opposite-sign (OS) to SS event yields obtained in a QCD-enriched region with loose lepton isolation. In the $\tauh\tauh$ channel, the shape is obtained from OS events with loose $\Pgt$ isolation. The yield is obtained by scaling these events  by the ratio of SS events with tight and loose $\Pgt$ isolation.

$\PW$+jets events in which there is a jet misidentified as a $\tauh$ are another sizable source of background in the $\Pe\tauh$ and $\Pgm\tauh$ channels. The background shape is modelled using a MC simulation and the rate is estimated using a control region of events with large transverse mass.

The Drell--Yan production of muon pairs is the largest background in the $\Pgm\Pgm$ channel. The  $\cPZ\to \Pgm\Pgm$ event yield is obtained from a fit to the distance of closest approach of the muon pairs observed in data, after subtracting all backgrounds.
In the $\Pe\tauh$ and $\Pgm\tauh$ channels, the  contribution of Drell--Yan production of electron and muon pairs is estimated from the simulation, after rescaling the simulated yield to the one derived from $\cPZ\to\Pgm\Pgm$ data. In the $\Pe\tauh$ channel, the $\cPZ\to\Pe\Pe$ simulation is further corrected using the $\Pe\to\tauh$ fake-rate measured in data using a ``tag-and-probe'' technique~\cite{Khachatryan:2010xn} on $\cPZ\to \Pe\Pe$ events.

In the $\Pe\Pgm$ final state, the $\PW +$jets and multijet background rate is obtained by measuring the number of events with one good lepton and a second one that passes relaxed selection criteria, but fails the nominal lepton selection.
This rate is extrapolated to the signal region using the efficiencies
for such loose lepton candidates to pass the nominal lepton selection. These efficiencies
are measured in data using multijet events.

The $\cPqt\cPaqt$, di-boson and single-top-quark background contributions are estimated from simulation.
The event yield of the $\cPqt\cPaqt$ background is checked in a sample of $\Pe\Pgm$ events with two b-tagged jets.

The observed number of events for each category, the expected number of events from various background processes, and the expected signal yields and efficiencies, are shown in Tables~\ref{table:events:etau}--\ref{table:events:tautau}.
The uncertainties are obtained after the likelihood fit described in Section~\ref{sec:stat}.

\begin{table*}[!b]
\begin{center}
    \topcaption{Observed and expected number of events in the two event categories in the $\Pe\tauh$ channel, where the combined statistical and systematic uncertainty is shown.
      The expected signal yields for $\Ph,\PH,\PA\to\Pgt\Pgt$
in the $m_{\Ph}^\text{max}$ scenario for $m_{\PA}$ = 160\GeV and $\tan\beta = 8$ and the signal efficiency times acceptance for a MSSM Higgs boson of 160\GeV mass are also given.
}
\begin{tabular}{lr@{$ \,\,\pm\,\, $}lr@{$ \,\,\pm\,\, $}lr@{$ \,\,\pm\,\, $}lr@{$ \,\,\pm\,\, $}l}
\hline
\multicolumn{9}{c}{$\Pe\tauh$ channel} \\
\hline
& \multicolumn{4}{c}{$\sqrt{s} = 7\TeV$} & \multicolumn{4}{c}{$\sqrt{s} = 8\TeV$} \\
Process & \multicolumn{2}{c}{no b-tag} & \multicolumn{2}{c}{b-tag} & \multicolumn{2}{c}{no b-tag} & \multicolumn{2}{c}{b-tag}\\
\hline
$\cPZ\to \Pgt\Pgt$            & 11819      & 197       & 135        & 4         & 30190      & 345       & 453        & 14        \\
QCD                                   & 4163       & 212       & 78         & 11        & 11894      & 544       & 194        & 27        \\
$\PW$+jets                            & 1344       & 112       & 29         & 7         & 5646       & 385       & 113        & 25        \\
$\cPZ$+jets ($\Pe$, $\Pgm$ or jet faking $\Pgt$)     & 1334       & 130       & 9          & 1         & 6221       & 360       & 83         & 7         \\
$\cPqt\cPaqt$                         & 43         & 3         & 19         & 3         & 290        & 22        & 102        & 12        \\
Di-bosons + single top                & 46         & 5         & 7          & 0.9       & 224        & 23        & 30         & 4         \\
\hline
Total background                      & 18750      & 144       & 278        & 11        & 54464      & 259       & 975        & 29        \\
$\Ph,\PH,\PA\to\Pgt\Pgt$          & 128        & 13        & 10         & 1         & 466        & 43        & 37         & 5         \\
Observed data                                  & \multicolumn{2}{c}{18785     }& \multicolumn{2}{c}{274       }& \multicolumn{2}{c}{54547     }& \multicolumn{2}{c}{975       }\\
\hline
\multicolumn{9}{l}{Efficiency $\times$ acceptance}\\
\hline
gluon fusion Higgs         & \multicolumn{2}{c}{ $1.39\times 10^{-2}$}& \multicolumn{2}{c}{ $1.24\times 10^{-4}$}& \multicolumn{2}{c}{ $9.48\times 10^{-3}$}& \multicolumn{2}{c}{ $1.11\times 10^{-4}$}\\
b-quark associated Higgs & \multicolumn{2}{c}{ $1.12\times 10^{-2}$}& \multicolumn{2}{c}{ $2.10\times 10^{-3}$}& \multicolumn{2}{c}{ $7.78\times 10^{-3}$}& \multicolumn{2}{c}{ $1.49\times 10^{-3}$}\\
\hline
\end{tabular}
\label{table:events:etau}
\end{center}
\end{table*}

\begin{table*}[!h]
\begin{center}
    \topcaption{Observed and expected number of events in the two event categories in the $\Pgm\tauh$, $\Pe\Pgm$, and $\Pgm\Pgm$ channels. Other details are as in Table~\ref{table:events:etau}.
}
\begin{tabular}{lr@{$ \,\,\pm\,\, $}lr@{$ \,\,\pm\,\, $}lr@{$ \,\,\pm\,\, $}lr@{$ \,\,\pm\,\, $}l}
\hline
\multicolumn{9}{c}{$\Pgm\tauh$ channel} \\
\hline
& \multicolumn{4}{c}{$\sqrt{s} = 7\TeV$} & \multicolumn{4}{c}{$\sqrt{s} = 8\TeV$} \\
Process & \multicolumn{2}{c}{no b-tag} & \multicolumn{2}{c}{b-tag} & \multicolumn{2}{c}{no b-tag} & \multicolumn{2}{c}{b-tag}\\
\hline
$\cPZ\to \Pgt\Pgt$            & 26838      & 244       & 284        & 8         & 87399      & 497       & 1118       & 31        \\
QCD                                   & 5495       & 258       & 131        & 18        & 18056      & 811       & 552        & 62        \\
$\PW$+jets                            & 2779       & 201       & 55         & 14        & 12845      & 793       & 237        & 52        \\
$\cPZ$+jets ($\Pe$, $\Pgm$ or jet faking $\Pgt$)     & 716        & 109       & 11         & 2         & 3704       & 454       & 54         & 9         \\
$\cPqt\cPaqt$                         & 82         & 6         & 36         & 5         & 564        & 41        & 194        & 22        \\
Di-bosons + single top                & 94         & 11        & 12         & 2         & 506        & 51        & 60         & 7         \\
\hline
Total background                      & 36004      & 205       & 530        & 18        & 123075     & 407       & 2214       & 44        \\
$\Ph,\PH,\PA\to\Pgt\Pgt$          & 226        & 23        & 17         & 2         & 929        & 85        & 67         & 9         \\
Observed data                                  & \multicolumn{2}{c}{36055     }& \multicolumn{2}{c}{542       }& \multicolumn{2}{c}{123239    }& \multicolumn{2}{c}{2219      }\\
\hline
\multicolumn{9}{l}{Efficiency $\times$ acceptance}\\
\hline
gluon fusion Higgs          & \multicolumn{2}{c}{ $2.34\times 10^{-2}$}& \multicolumn{2}{c}{ $2.49\times 10^{-4}$}& \multicolumn{2}{c}{ $1.78\times 10^{-2}$}& \multicolumn{2}{c}{ $2.32\times 10^{-4}$}\\
b-quark associated Higgs         & \multicolumn{2}{c}{ $1.96\times 10^{-2}$}& \multicolumn{2}{c}{ $3.54\times 10^{-3}$}& \multicolumn{2}{c}{ $1.53\times 10^{-2}$}& \multicolumn{2}{c}{ $2.66\times 10^{-3}$}\\
\hline
\end{tabular}

\vspace*{0.4 cm}

\begin{tabular}{lr@{$ \,\,\pm\,\, $}lr@{$ \,\,\pm\,\, $}lr@{$ \,\,\pm\,\, $}lr@{$ \,\,\pm\,\, $}l}
\hline
\multicolumn{9}{c}{$\Pe\Pgm$ channel} \\
\hline
& \multicolumn{4}{c}{$\sqrt{s} = 7\TeV$} & \multicolumn{4}{c}{$\sqrt{s} = 8\TeV$} \\
Process & \multicolumn{2}{c}{no b-tag} & \multicolumn{2}{c}{b-tag} & \multicolumn{2}{c}{no b-tag} & \multicolumn{2}{c}{b-tag}\\
\hline
$\cPZ\to \Pgt\Pgt$            & 13783      & 134       & 165        & 5         & 48218      & 300       & 679        & 8         \\
QCD                                   & 804        & 114       & 14         & 3         & 4302       & 356       & 148        & 17        \\
$\cPqt\cPaqt$                         & 467        & 29        & 309        & 18        & 2215       & 158       & 1183       & 48        \\
Di-bosons + single top                & 501        & 55        & 63         & 8         & 2367       & 248       & 308        & 38        \\
\hline
Total background                      & 15556      & 128       & 551        & 21        & 57102      & 257       & 2318       & 37        \\
$\Ph,\PH,\PA\to\Pgt\Pgt$          & 114        & 11        & 9          & 1         & 455        & 43        & 34         & 4         \\
Observed data                                 & \multicolumn{2}{c}{15436     }& \multicolumn{2}{c}{558       }& \multicolumn{2}{c}{57285     }& \multicolumn{2}{c}{2353      }\\
\hline
\multicolumn{9}{l}{Efficiency $\times$ acceptance}\\
\hline
gluon fusion Higgs         & \multicolumn{2}{c}{ $1.14\times 10^{-2}$}& \multicolumn{2}{c}{ $1.11\times 10^{-4}$}& \multicolumn{2}{c}{ $8.83\times 10^{-3}$}& \multicolumn{2}{c}{ $8.85\times 10^{-5}$}\\
b-quark associated Higgs           & \multicolumn{2}{c}{ $9.70\times 10^{-3}$}& \multicolumn{2}{c}{ $1.80\times 10^{-3}$}& \multicolumn{2}{c}{ $7.59\times 10^{-3}$}& \multicolumn{2}{c}{ $1.33\times 10^{-3}$}\\
\hline
\end{tabular}

\vspace*{0.4 cm}

\begin{tabular}{lr@{$ \,\,\pm\,\, $}lr@{$ \,\,\pm\,\, $}lr@{$ \,\,\pm\,\, $}lr@{$ \,\,\pm\,\, $}l}
\hline
\multicolumn{9}{c}{$\Pgm\Pgm$ channel} \\
\hline
& \multicolumn{4}{c}{$\sqrt{s} = 7\TeV$} & \multicolumn{4}{c}{$\sqrt{s} = 8\TeV$} \\
Process & \multicolumn{2}{c}{no b-tag} & \multicolumn{2}{c}{b-tag} & \multicolumn{2}{c}{no b-tag} & \multicolumn{2}{c}{b-tag}\\
\hline
$\cPZ\to \Pgt\Pgt$            & 6838       & 118       & 34         & 1         & 20931      & 376       & 101        & 5         \\
$\cPZ\to \mu\mu$              & 562008     & 764       & 1435       & 34        & 1894509    & 1566      & 5125       & 69        \\
QCD                                   & 380        & 51        & 4          & 2         & 1131       & 108       & 31         & 7         \\
$\cPqt\cPaqt$                         & 183        & 15        & 83         & 7         & 809        & 62        & 324        & 15        \\
Di-bosons + single top                & 1114       & 221       & 10         & 2         & 5543       & 625       & 48         & 8         \\
\hline
Total background                      & 570523     & 763       & 1566       & 35        & 1922923    & 1439      & 5629       & 72        \\
$\Ph,\PH,\PA\to\Pgt\Pgt$          & 57         & 5         & 3          & 0.4       & 195        & 17        & 8          & 1         \\
Observed data                                  & \multicolumn{2}{c}{570616    }& \multicolumn{2}{c}{1559      }& \multicolumn{2}{c}{1922924   }& \multicolumn{2}{c}{5608      }\\
\hline
\multicolumn{9}{l}{Efficiency $\times$ acceptance}\\
\hline
gluon fusion  Higgs          & \multicolumn{2}{c}{ $5.66\times 10^{-3}$}& \multicolumn{2}{c}{ $2.55\times 10^{-5}$}& \multicolumn{2}{c}{ $3.73\times 10^{-3}$}& \multicolumn{2}{c}{ $2.29\times 10^{-5}$}\\
b-quark associated  Higgs         & \multicolumn{2}{c}{ $5.33\times 10^{-3}$}& \multicolumn{2}{c}{ $6.65\times 10^{-4}$}& \multicolumn{2}{c}{ $3.58\times 10^{-3}$}& \multicolumn{2}{c}{ $3.42\times 10^{-4}$}\\
\hline
\end{tabular}
\label{table:events:mutau_emu_mumu}
\end{center}
\end{table*}

\begin{table*}[!h]
\begin{center}
    \topcaption{Observed and expected number of events in the two event categories in the $\tauh\tauh$ channel. Other details are as in Table~\ref{table:events:etau}.
}
\begin{tabular}{lr@{$ \,\,\pm\,\, $}lr@{$ \,\,\pm\,\, $}l}
\hline
\multicolumn{5}{c}{$\tauh\tauh$ channel} \\
\hline
& \multicolumn{4}{c}{$\sqrt{s} = 8$\TeV} \\
Process & \multicolumn{2}{c}{no b-tag} & \multicolumn{2}{c}{b-tag}\\
\hline
$\cPZ\to \Pgt\Pgt$            & 2511       & 97        & 60         & 3         \\
QCD                                   & 20192      & 236       & 273        & 21        \\
$\PW$+jets                            & 630        & 165       & 17         & 5         \\
$\cPZ$+jets ($\Pe$, $\Pgm$ or jet faking $\Pgt$)     & 115        & 19        & 2          & 0.4       \\
$\cPqt\cPaqt$                         & 38         & 4         & 16         & 2         \\
Di-bosons + single top                & 63         & 13        & 5          & 1         \\
\hline
Total background                      & 23548      & 174       & 374        & 19        \\
$\Ph,\PH,\PA\to\Pgt\Pgt$          & 325        & 34        & 30         & 5         \\
Observed data                                  & \multicolumn{2}{c}{23606     }& \multicolumn{2}{c}{381       }\\
\hline
\multicolumn{5}{l}{Efficiency $\times$ acceptance}\\
\hline
gluon fusion Higgs          & \multicolumn{2}{c}{ $9.11\times 10^{-3}$}& \multicolumn{2}{c}{ $1.32\times 10^{-4}$}\\
b-quark associated Higgs         & \multicolumn{2}{c}{ $7.26\times 10^{-3}$}& \multicolumn{2}{c}{ $1.37\times 10^{-3}$}\\
\hline
\end{tabular}
\label{table:events:tautau}
\end{center}
\end{table*}

\section{Tau lepton-pair invariant mass}
\begin{figure*}[!h]\begin{center}
  \includegraphics[width=0.39\textwidth]{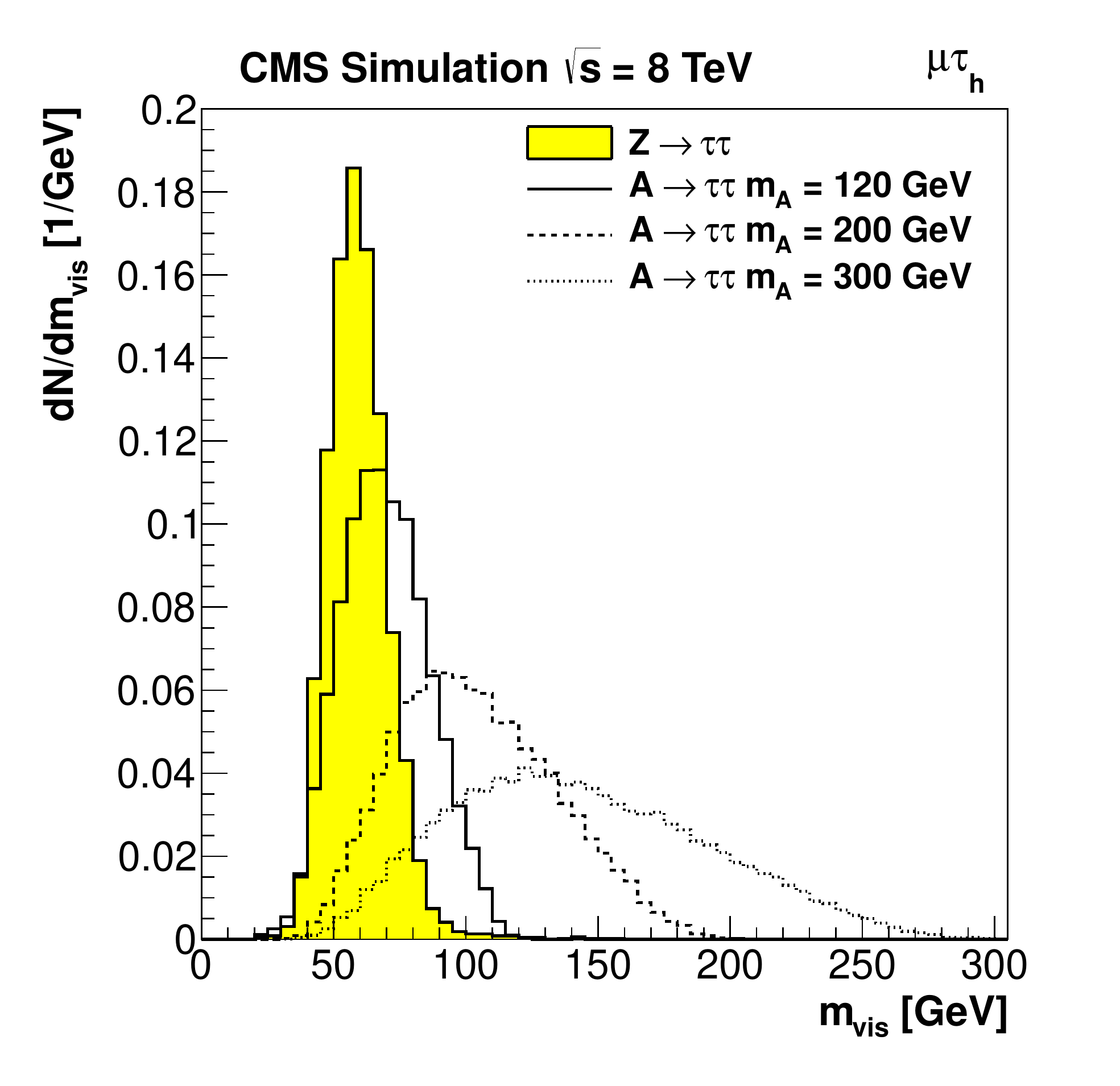}
 \includegraphics[width=0.39\textwidth]{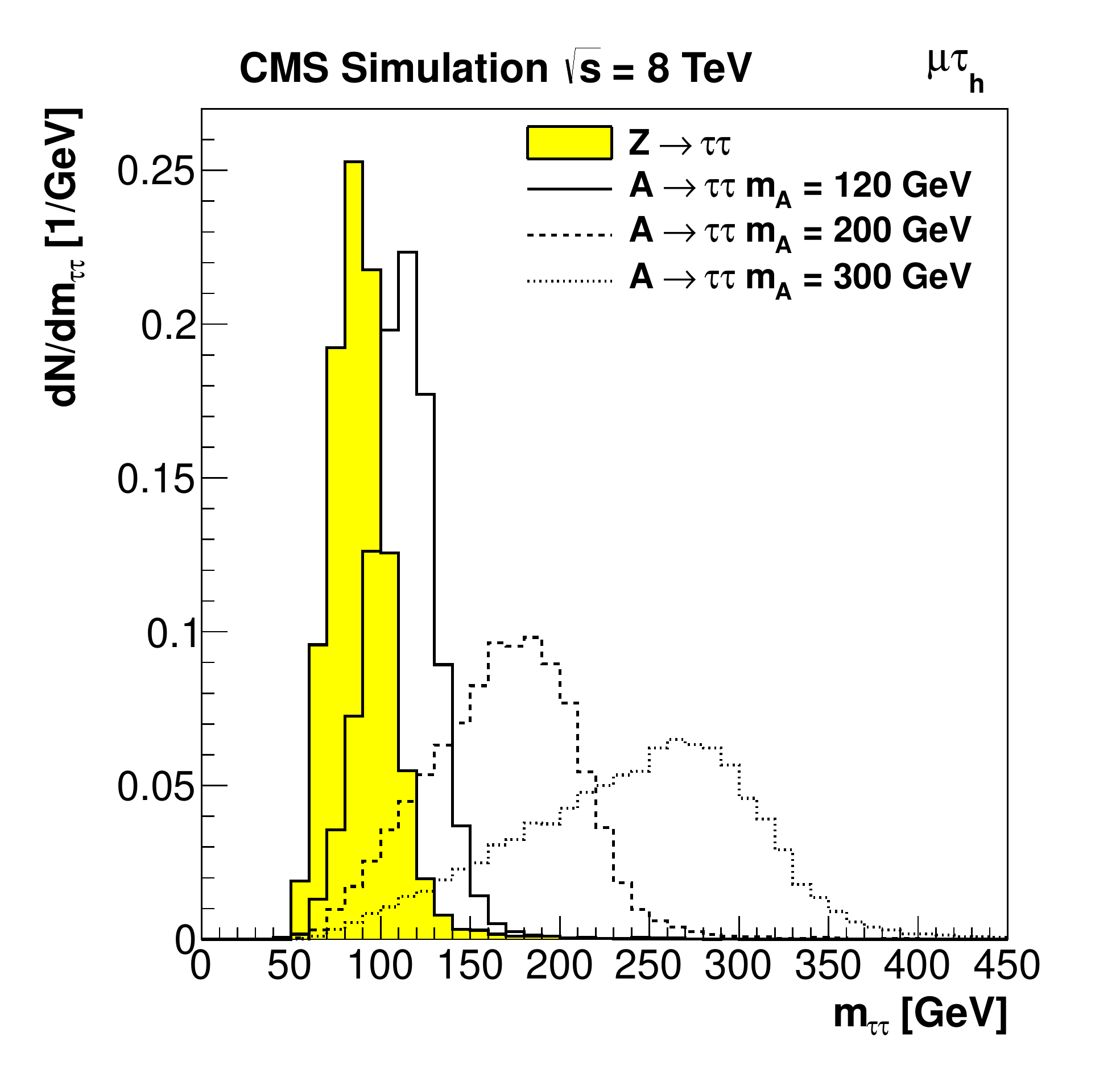}
  \caption{Visible mass $m_\text{vis}$ (left) and $m_{\Pgt\Pgt}$ reconstructed mass (right)
         for simulated events: $\cPZ\to\Pgt\Pgt$ and MSSM $\PA \to \Pgt\Pgt$ signal with $m_{\PA} = 120$, 200 and 300\GeV, in the $\Pgm\tauh$ final state. Distributions are normalized to unity.}
   \label{fig:expectedSVfitMassDistributions}\end{center}\end{figure*}

To distinguish Higgs boson signals from the background, the tau-lepton pair invariant mass, $m_{\Pgt\Pgt}$, is reconstructed using a maximum likelihood technique~\cite{Chatrchyan:2014nva}.
The $m_{\Pgt\Pgt}$ resolution
 for $\cPZ\to\Pgt\Pgt$ events depends on the final state considered but typically amounts to 20\%,
relative to the true mass value.
Distributions of the mass of the visible decay products $m_\text{vis}$ and $m_{\Pgt\Pgt}$ for simulated events
are shown in Fig.~\ref{fig:expectedSVfitMassDistributions}.
The reconstruction of  $m_{\Pgt\Pgt}$ improves the separation power between the main $\cPZ\to\Pgt\Pgt$ background
and the hypothetical MSSM Higgs boson $\PA$ signals.

\begin{figure*}[!h]\begin{center}
 \includegraphics[width=0.45\textwidth]{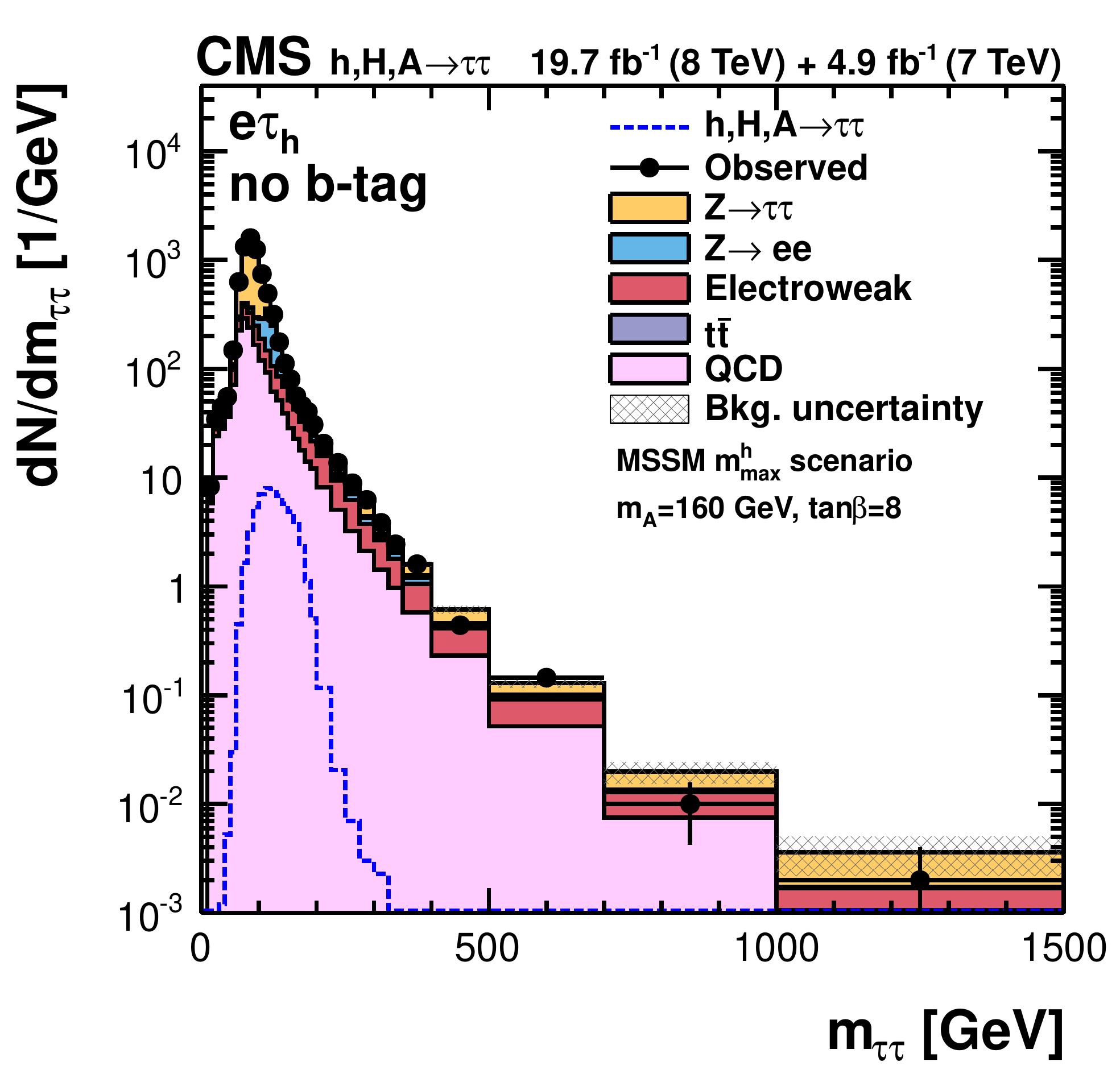}
  \includegraphics[width=0.45\textwidth]{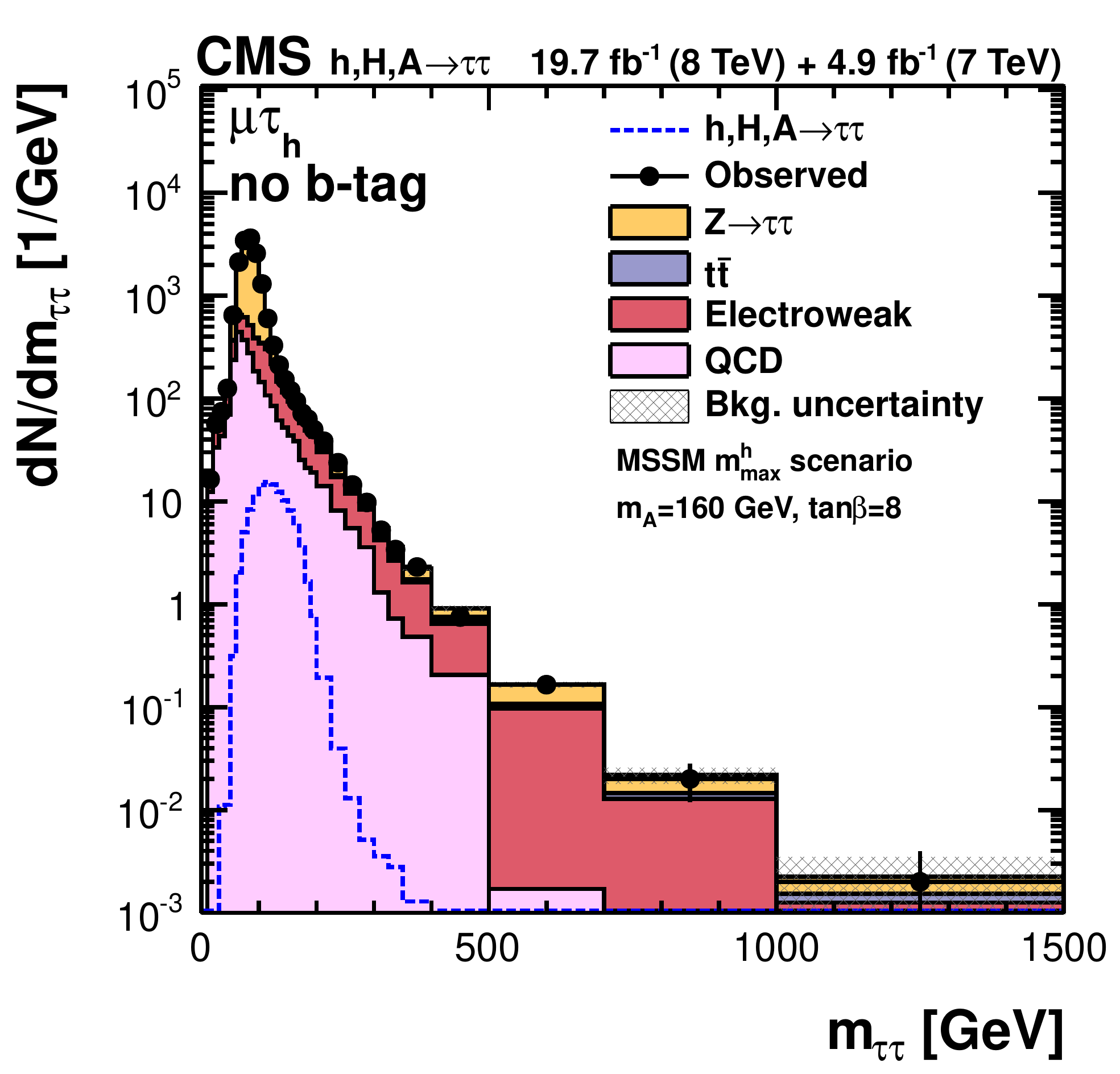}
 \includegraphics[width=0.45\textwidth]{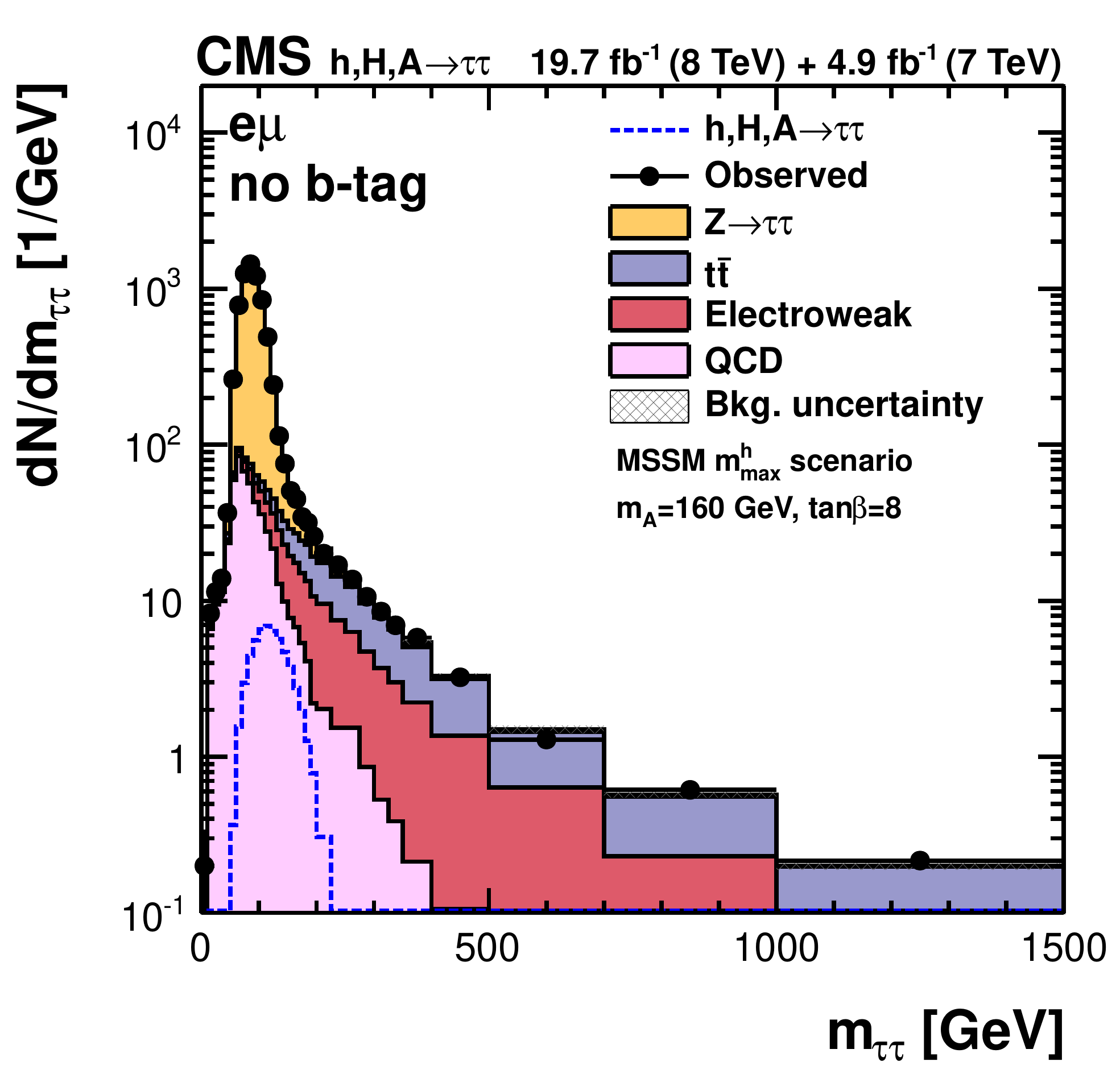}
 \includegraphics[width=0.45\textwidth]{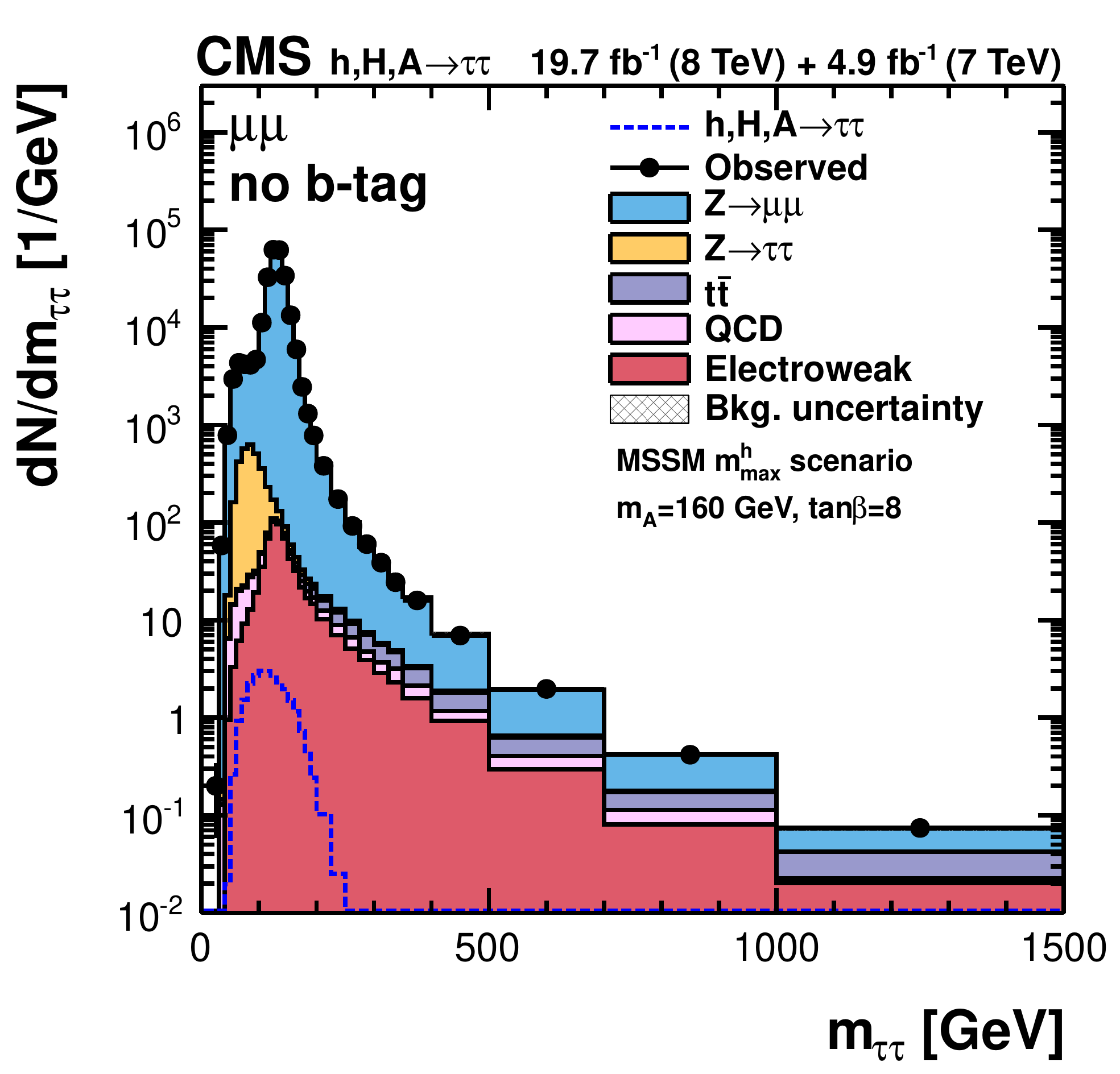}
 \includegraphics[width=0.45\textwidth]{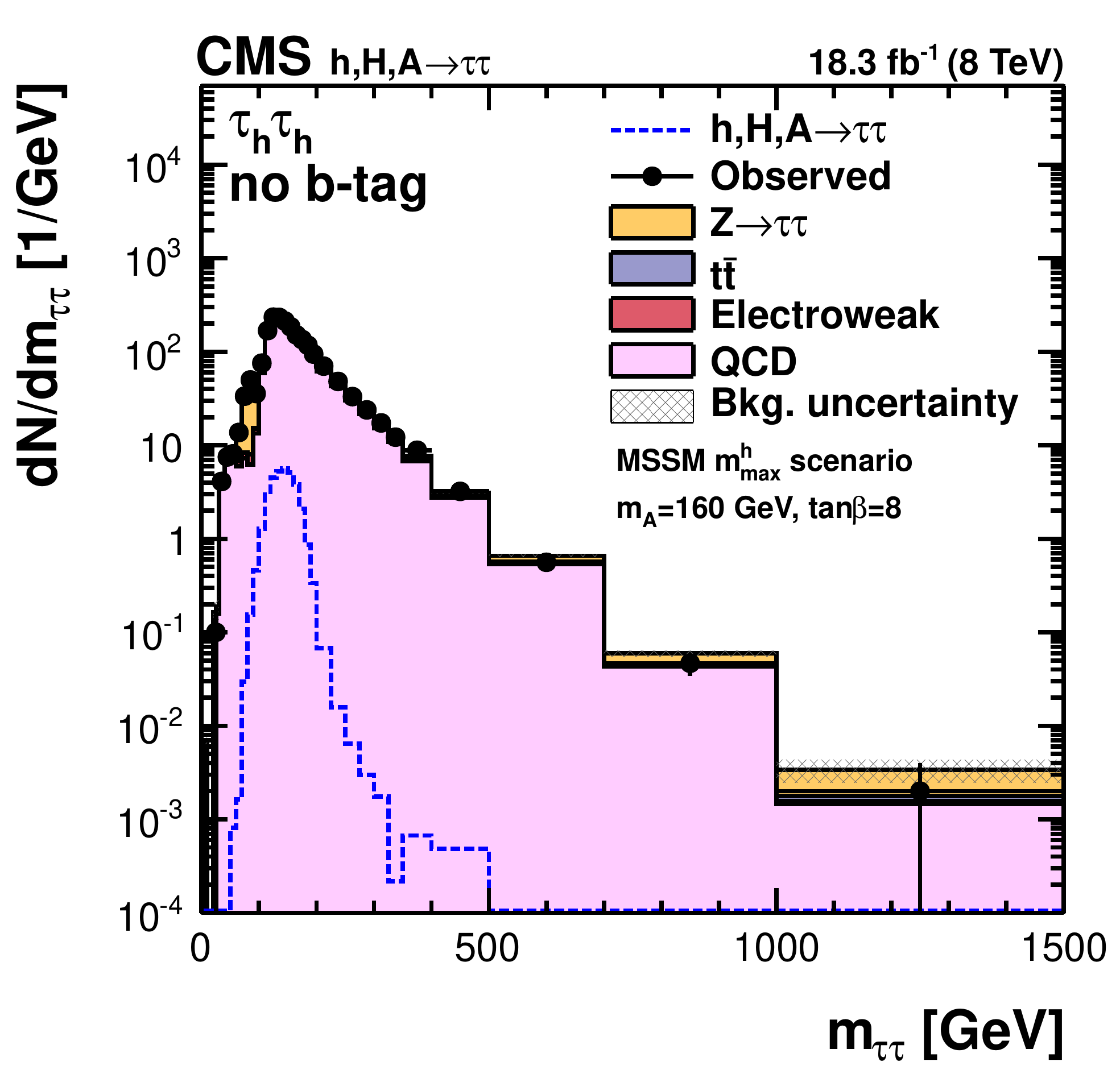}
  \caption{Reconstructed $\Pgt\Pgt$ invariant-mass in the no b-tag category for the $\Pe\tauh$, $\Pgm\tauh$, $\Pe\Pgm$, $\Pgm\Pgm$ and $\tauh\tauh$ channels. The electroweak background includes the contributions from $\cPZ\to\Pe\Pe$, $\cPZ\to\Pgm\Pgm$, $\PW$, di-bosons, and single top. In the $\Pgm\Pgm$ channel, the $\cPZ\to\Pgm\Pgm$ contribution is shown separately. The expected signal yield of the MSSM Higgs bosons $\Ph$, $\PH$, and $\PA$ for $m_{\PA}$ = 160\GeV and $\tan\beta=8$ in the $m_{\Ph}^\text{max}$ scenario is also shown.}
   \label{fig:mass_no_b_tag}\end{center}\end{figure*}

\begin{figure*}[!h]\begin{center}
 \includegraphics[width=0.45\textwidth]{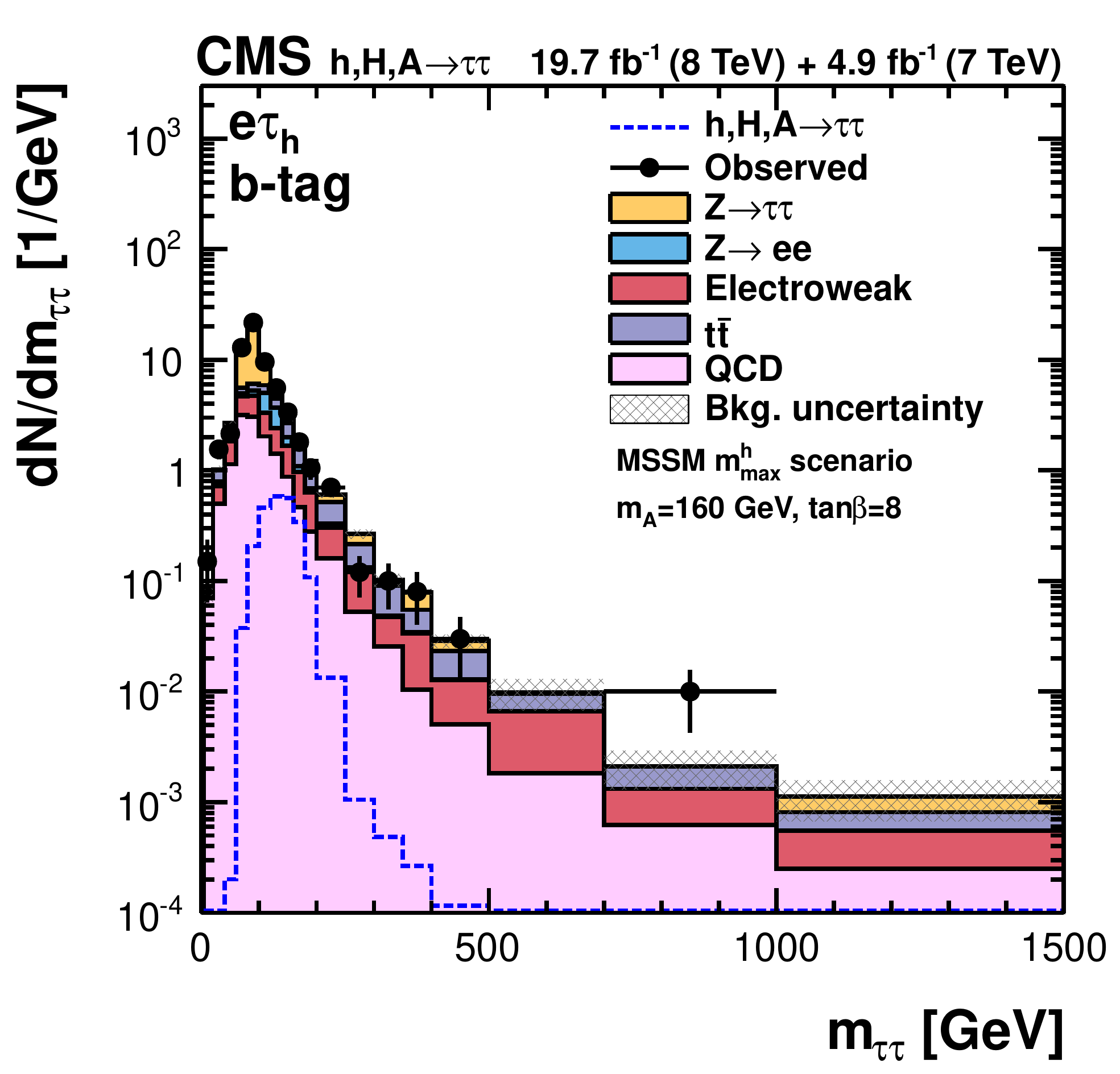}
 \includegraphics[width=0.45\textwidth]{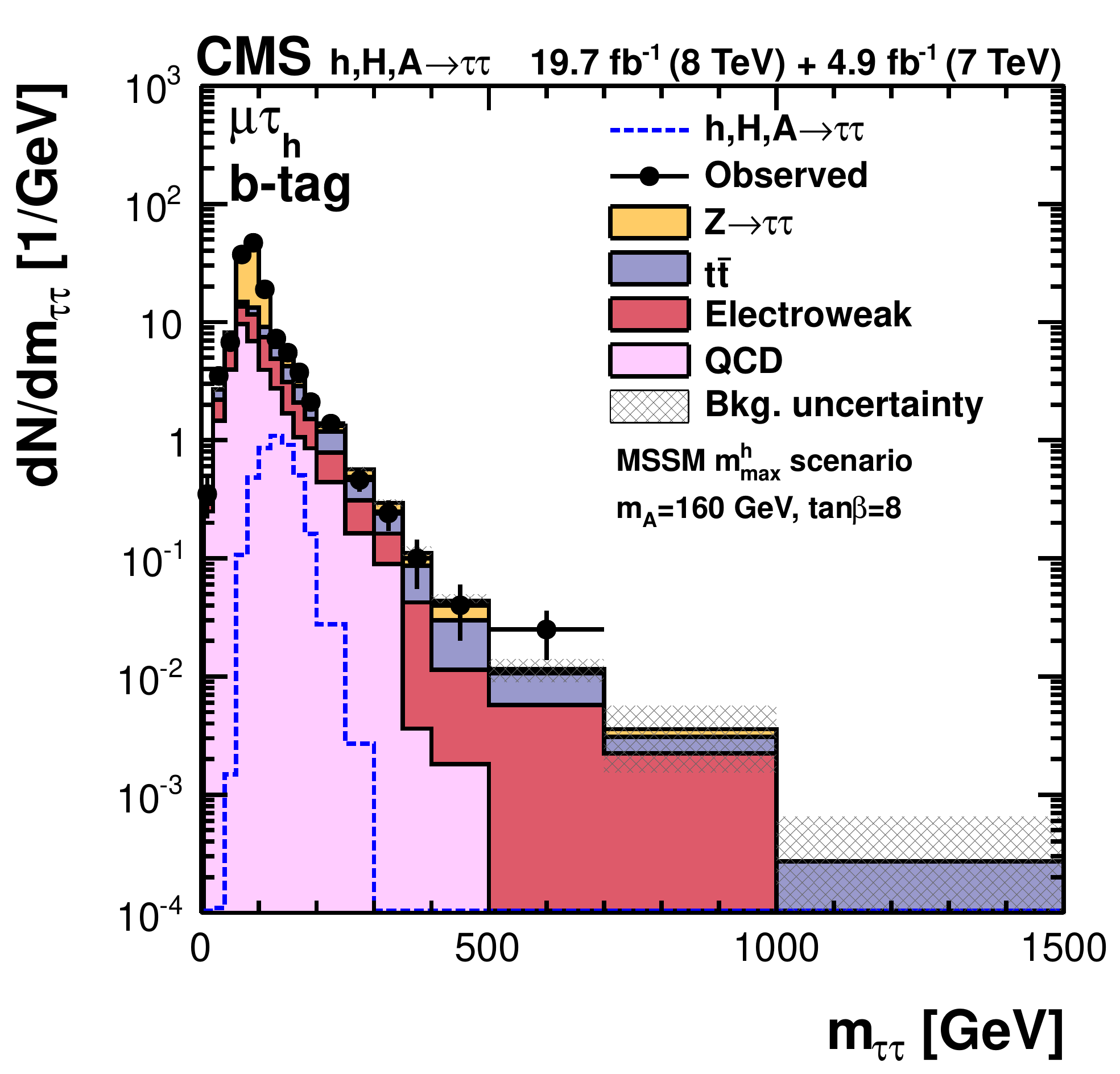}
 \includegraphics[width=0.45\textwidth]{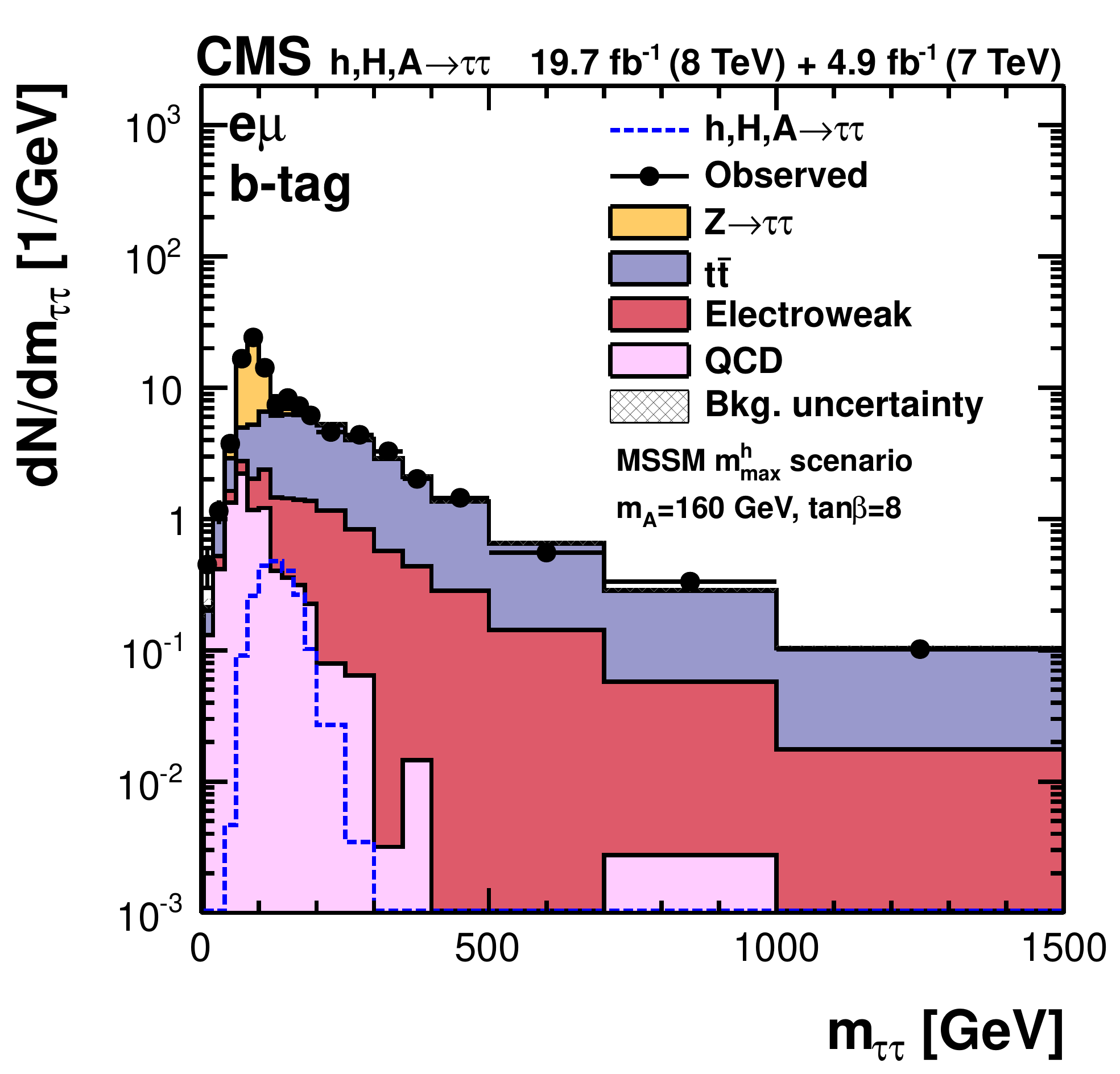}
 \includegraphics[width=0.45\textwidth]{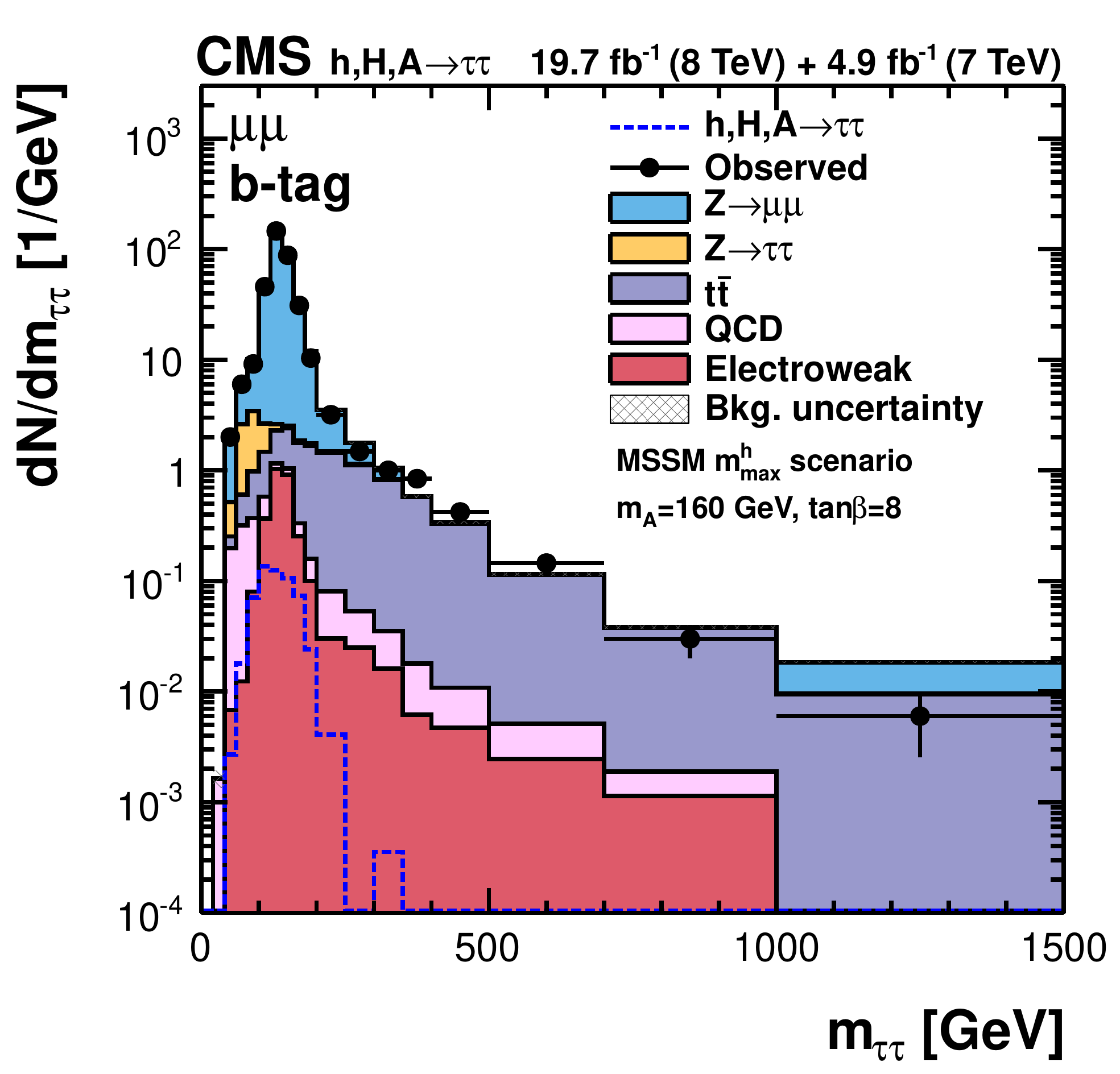}
 \includegraphics[width=0.45\textwidth]{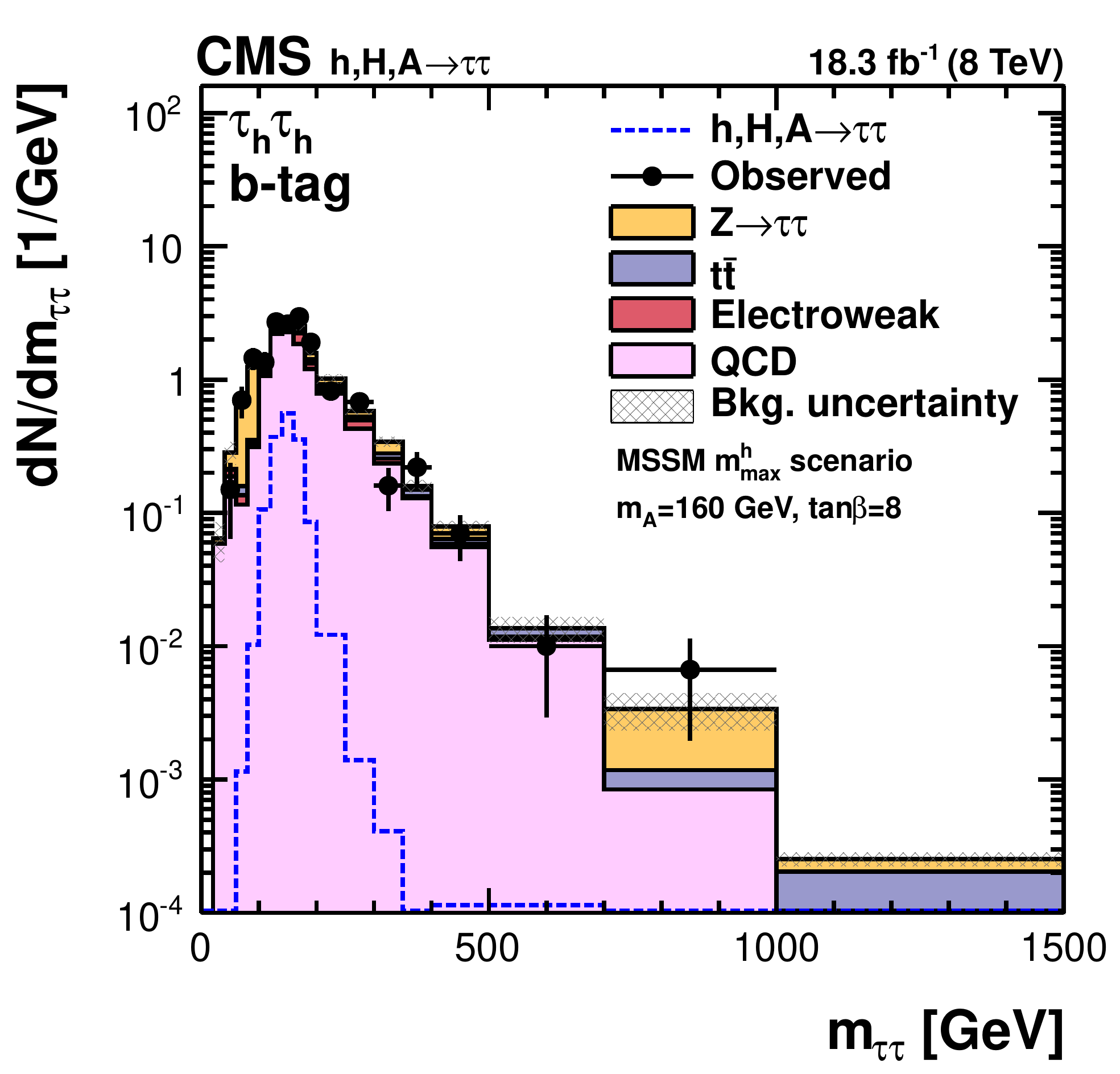}
 \caption{Reconstructed $\Pgt\Pgt$ mass in the b-tag category for the $\Pe\tauh$, $\Pgm\tauh$, $\Pe\Pgm$, $\Pgm\Pgm$ and $\tauh\tauh$ channels. Other details are as in Fig.~\ref{fig:mass_no_b_tag}.}
  \label{fig:mass_b_tag}\end{center}\end{figure*}

The distribution in $m_{\Pgt\Pgt}$ for the five final states studied,
$\Pe\tauh, \Pgm\tauh, \Pe\Pgm$, $\Pgm\Pgm$, and $\tauh\tauh$, compared with the background prediction in the no b-tag category is shown in Fig.~\ref{fig:mass_no_b_tag}. These events are more sensitive to the gluon fusion Higgs boson production mechanism.
Figure~\ref{fig:mass_b_tag} shows the $m_{\Pgt\Pgt}$ distribution in the b-tag category, which has an enhanced sensitivity to the b-quark associated Higgs boson production mechanism.

\clearpage

\section{Systematic uncertainties}
\label{sec:syst}
Various imperfectly known effects can alter the shape and the normalization of the invariant-mass spectrum. The main contributions to the normalization uncertainty, that affect the signal and the simulated backgrounds, include the uncertainty in the total integrated luminosity (2.2\% for 2011 and 2.6\% for 2012 data~\cite{CMS-PAS-SMP-12-008,CMS-PAS-LUM-13-001}), the jet energy scale (1--10\%), and the identification and trigger efficiencies of electrons (2\%) and muons (2-3\%).
The tau-lepton identification and trigger efficiency uncertainty is estimated to be 8\% from an independent study done using a ``tag-and-probe'' technique on $\cPZ\to \Pgt\Pgt$ events.
An extra uncertainty of $0.02\%\times\pt^{\tau}$\,[\GeVns{}], due to the extrapolation from the $\cPZ$-boson resonance region to larger tau lepton $\pt$ values, is also considered.
The b-tagging efficiency has an uncertainty between 2--7\%, and the mistag rate for light-flavor partons is accurate to 10--20\%~\cite{Chatrchyan:2012jua}.
The background normalization uncertainties from the estimation methods discussed in Section~\ref{sec:background} are also considered. Uncertainties that contribute to variations in the shape of the mass spectrum include the electron (1\%), muon (1\%), and tau lepton (3\%) energy scales.
The main experimental uncertainties and their effect on the yields in the two event categories are summarized in Table~\ref{tab:SystematicUncertainties}.

The theoretical uncertainties on the MSSM Higgs signal cross sections depend on $\tan\beta$, $m_{\PA}$, and the scenario considered, and can amount up to $\sim$25\%. In the cross section calculations the MSTW2008~\cite{Martin:2009iq} parton distribution functions are used, and the recommended prescription~\cite{Martin:2009iq,Martin:2009bu} to compute the uncertainties is followed. The renormalization and factorization scales used in the theoretical calculations and the variation considered are summarized in Ref.~\cite{Dittmaier:2011ti}.

\begin{table}[!h]
\begin{center}
\topcaption{
Systematic uncertainties that affect the estimated number of signal or background events. Several uncertainties are treated as correlated for the different decay channels and event categories. Some uncertainties vary with \pt, $\eta$, and final state, so ranges are given.}
\begin{tabular}{l|c|cc}
\multicolumn{2}{c}{} & \multicolumn{2}{|c}{Event yield uncertainty}  \\
\multicolumn{2}{c}{} & \multicolumn{2}{|c}{by event category} \\
\hline
  Experimental uncertainties                                 & Uncertainty      &  no b-tag    &  b-tag      \\
\hline
 Integrated luminosity 7 (8)\TeV                            & 2.2\;(2.6)\% & 2.2\;(2.6)\% & 2.2\;(2.6)\%        \\
 Jet energy scale                                            & 1--10\%  &   1--5\%     &   1--8\%          \\
 \MET                                                        & 1--5\%   &   1--2\%     &   1--2\%          \\
\hline
 Electron identification and trigger                          &     2\% &   2\%       &   2\%            \\
 Muon identification and trigger                              &     2--3\% &   2--6\%       &   2--6\%            \\
 Tau-lepton identification and trigger                        &     8\% &   8--19\%       &   8--19\%            \\
\hline
 b-tagging efficiency                                        & 2--7\%   &   2--4\%     &   2--9\%         \\
 b-mistag rate                                               & 10--20\% &   2\%       &   2--5\%            \\
\hline
 Normalization, $\cPZ$ production                            & 3.3\%   &   3.3\%     &   3.3\%          \\
 $Z\to\Pgt\Pgt$: category selection                  & 3\%     &   ---              &   1--3\%          \\
 Normalization, $\cPqt\cPaqt$                                & 10\%    &   10\%      &   10--17\%        \\
 Normalization, di-boson                                     & 15--30\% &   15--30\%   &   15--30\%        \\
 Normalization, QCD Multijet                                 & 10--35\%  &   10--35\%    &   20--35\%          \\
 Normalization, $\PW$+jets                                   & 10--30\%     & 10--30\% &  30\%             \\
 Normalization, $\cPZ\to\Pe\Pe$: $\Pe$ misidentified as $\tauh$    & 20\%    & 20\%     & 20\%         \\
 Normalization, $\cPZ\to\Pgm\Pgm$: $\Pgm$ misidentified as $\tauh$   & 30\%    & 30\%     & 30\%         \\
 Normalization, $\cPZ$+jets : jet misidentified as $\tauh$      & 20\%         & 20\%    & 20\%              \\
\hline
 Electron energy scale                                       &     1\% &   1\%       &   1\%            \\
 Muon energy scale                                           &     1\% &   1\%       &   1\%            \\
 Tau-lepton energy scale                                     &     3\% &   3\%       &   3\%            \\
\hline
\end{tabular}
\label{tab:SystematicUncertainties}
\end{center}
\end{table}

\section{Statistical analysis}\label{sec:stat}

To search for the presence of a MSSM Higgs boson signal in the selected events, a binned maximum likelihood fit is performed.
The invariant mass of the tau-lepton pairs is used as the input to the fit in the $\Pe\tauh, \Pgm\tauh, \Pe\Pgm$, and
$\tauh\tauh$ final states.
The sensitivity of the $\Pgm\Pgm$ channel is enhanced by fitting the two-dimensional distribution of $m_{\Pgt\Pgt}$
versus the mass of the visible decay products, $m_\text{vis}$,
utilizing the fact that most of the large $\cPZ\to \Pgm\Pgm$ background contributing to this channel
is concentrated in the $m_\text{vis}$ distribution within a narrow peak around the $\cPZ$-boson mass.
The fit is performed simultaneously for the five final states and the two event categories, b-tag and no b-tag.

The systematic uncertainties described in Section~\ref{sec:syst}  are incorporated in the fit via nuisance parameters and are treated according to the frequentist paradigm, as described in Ref.~\cite{LHC-HCG-Report}. The uncertainties that affect the shape of the mass spectrum, mainly those corresponding to the energy scales, are represented by the nuisance parameters whose variation results in a continuous perturbation of the spectrum.
Shape uncertainties due to limited statistics are incorporated via nuisance parameters that allow for uncorrelated single-bin fluctuations
of the background expectation, following the method described in  Ref.~\cite{Barlow:1993dm}. In the tail of the $m_{\Pgt\Pgt}$ distribution ($m_{\Pgt\Pgt} >150\GeV$), where the statistical uncertainties are large, a different method is used. A fit of the form $f=\exp\left[-m_{\Pgt\Pgt}/(c_{0}+c_1\cdot m_{\Pgt\Pgt})\right]$ is performed for each of the major backgrounds, the result of which replaces the nominal distribution.  The uncertainties in the fit parameters $c_{0}$ and $c_{1}$ are treated as nuisance parameters in the likelihood fit.

The invariant-mass spectra show no clear evidence for the presence of a MSSM Higgs boson signal so exclusion limits are obtained.
For the calculation of exclusion limits a modified frequentist criterion \CLs~\cite{Junk:1999kv,Read:2002hq} is used. The chosen test statistic $q$, used to determine how signal- or background-like the data are, is based on a profile likelihood ratio.

In this study, two searches are performed:
\begin{itemize}
\item a model independent search for a single narrow resonance $\phi$ for different mass hypotheses in the gluon fusion and b-quark associated Higgs boson production modes;
\item a search for the MSSM neutral Higgs bosons, $\Ph$, $\PA$, and $\PH$ in the $\Pgt\Pgt$ mass spectrum.
\end{itemize}

In the case of the model independent search for a single resonance $\phi$, the profile likelihood ratio is defined as
\begin{equation}
q_{\mu} = - 2\ln \frac{L\left( N_\text{obs} |\mu \cdot s + b, \hat{\theta}_{\mu} \right)}{L\left( N_\text{obs}|\hat{\mu}\cdot s + b, \hat{\theta} \right)}, \quad \text{with } \; 0 \leq \hat{\mu} \leq \mu,
\label{eq:teststatistic_LHC}
\end{equation}
where $N_\text{obs}$ is the number of observed events, $b$ and $s$ are the number of expected background and signal events, $\mu$ is the signal strength modifier, and $\theta$ are the nuisance parameters describing the systematic uncertainties. The value $\hat{\theta}_{\mu}$ maximizes the likelihood in the numerator for a given $\mu$, while $\hat{\mu}$ and $\hat{\theta}$ define the point at which the likelihood reaches its global maximum.
The ratio of probabilities to observe a value of the test statistic at least as large as the one observed in data, $q_{\mu}^\text{obs}$, under the signal-plus-background  ($\mu \cdot s + b$) and background-only~hypotheses,
\begin{equation}
\CLs(\mu)=\frac{P(q_\mu \ge q_{\mu}^\text{obs} | \mu \cdot s + b)}{P(q_\mu \ge q_{\mu}^\text{obs} | b)} \leq \alpha,
\label{eq:CLS}
\end{equation}
is used as the criterion for excluding the presence of a signal at the $1 - \alpha$ confidence level (CL).

Upper limits on $\sigma\cdot\mathcal{B}(\phi\to\Pgt\Pgt)$ at 95\% CL for gluon fusion  and b-quark associated neutral Higgs boson production for a single narrow resonance are obtained using Eqs.~\ref{eq:teststatistic_LHC}~and~\ref{eq:CLS}.
The expected limit is obtained by replacing the observed data by a representative dataset which not only contains the contribution from background processes but also a SM Higgs boson with a mass of 125\GeV. To extract the limit on the gluon fusion (b-quark associated) Higgs boson production, the rate of the b-quark associated (gluon fusion) Higgs boson production is treated as a nuisance parameter in the fit.

A search for MSSM Higgs bosons in the $\Pgt\Pgt$ final state is also performed, where the three neutral MSSM Higgs bosons are present in the
signal.
In light of the recent Higgs boson discovery at 125\GeV, a search for a MSSM signal versus a background-only hypothesis
has lost validity. A modified \CLs approach has been also adopted in this case, which tests the compatibility of the
data to a signal of the three neutral Higgs bosons $\Ph$, $\PH$, and $\PA$ compared to a SM Higgs boson hypothesis,
with inclusion of the backgrounds in both cases.
To achieve this a physics model is built according to
\begin{equation}
M(\mu) = \left[ \mu\cdot s(\mathrm{MSSM}) + (1-\mu)\cdot s(\mathrm{SM}) \right] + b.
\label{eq:physicsmodel_SM_MSSM}
\end{equation}
In the search, two well defined theories are tested so $\mu$ can only take the value of 0 or 1.
The test statistic used in the \CLs method is given by the ratio of likelihoods
\begin{equation}
q_{\mathrm{MSSM/SM}} = -2\ln  \frac{L\left( N_\text{obs} | M(1), \hat{\theta}_{1}\right) }{L\left( N_\text{obs}| M(0), \hat{\theta}_0 \right)},
\label{eq:teststatistic_MSSMvsSM}
\end{equation}
where the numerator and denominator are maximized by finding the corresponding nuisance parameters $\hat{\theta}_{1}$ for $\mu$=1 and $\hat{\theta}_{0}$ for $\mu$=0.

The MSSM Higgs boson signal expectation for each benchmark scenario studied is determined in each point of the parameter space as follows:
\begin{itemize}
\item At each point of $m_{\PA}$ and $\tan\beta$: the mass, the gluon fusion and associated-b production cross sections, and the branching fraction to $\Pgt\Pgt$ are determined for $\Ph$, $\PH$ and~$\PA$.
\item The contributions of all three neutral Higgs boson are added using the corresponding cross sections times branching fractions.
\end{itemize}
Limits on $\tan\beta$ versus $m_{\PA}$ at 95\% CL are obtained for different benchmark MSSM scenarios following the test statistic given in Eq.~\ref{eq:teststatistic_MSSMvsSM}.

\section{Results}
\label{sec:results}

The invariant-mass spectra show no clear evidence for the presence of a MSSM Higgs boson signal.
Therefore 95\% CL upper bounds on $\tan\beta$ as a function of the pseudoscalar Higgs boson mass $m_{\PA}$ are set for the traditional  MSSM benchmark scenario $m_{\Ph}^\text{max}$, and the recently
proposed benchmark scenarios, $m_{\Ph}^\text{mod$+$}$, $m_{\Ph}^\text{mod$-$}$,
light-stop, light-stau, and $\Pgt$-phobic. In the case of the low-$m_{\PH}$  scenario, limits on $\tan\beta$ as a function of the higgsino mass parameter $\mu$ are performed, where $m_{\PA}$ is set to a value of 110\GeV.

A test of the compatibility of the  data to a signal of the three neutral Higgs bosons $\Ph$, $\PH$, and~$\PA$ compared to a SM Higgs boson hypothesis is performed as described in Section~\ref{sec:stat}, using the test statistics given by Eq.~\ref{eq:teststatistic_MSSMvsSM}.
The simulation of the SM Higgs boson signal at 125\GeV used in the statistical analysis, is the same as in the dedicated SM Higgs boson search in the $\Pgt\Pgt$ decay mode~\cite{Chatrchyan:2014nva}, which includes the contributions from gluon fusion, vector boson fusion, and $\cPZ$ or $\PW$ boson and top-quark associated Higgs boson production. The contribution from SM b-quark associated Higgs boson production is expected to be small and is not included in this analysis.

\begin{figure*}[!hb]
\begin{center}
 \includegraphics[width=0.495\textwidth]{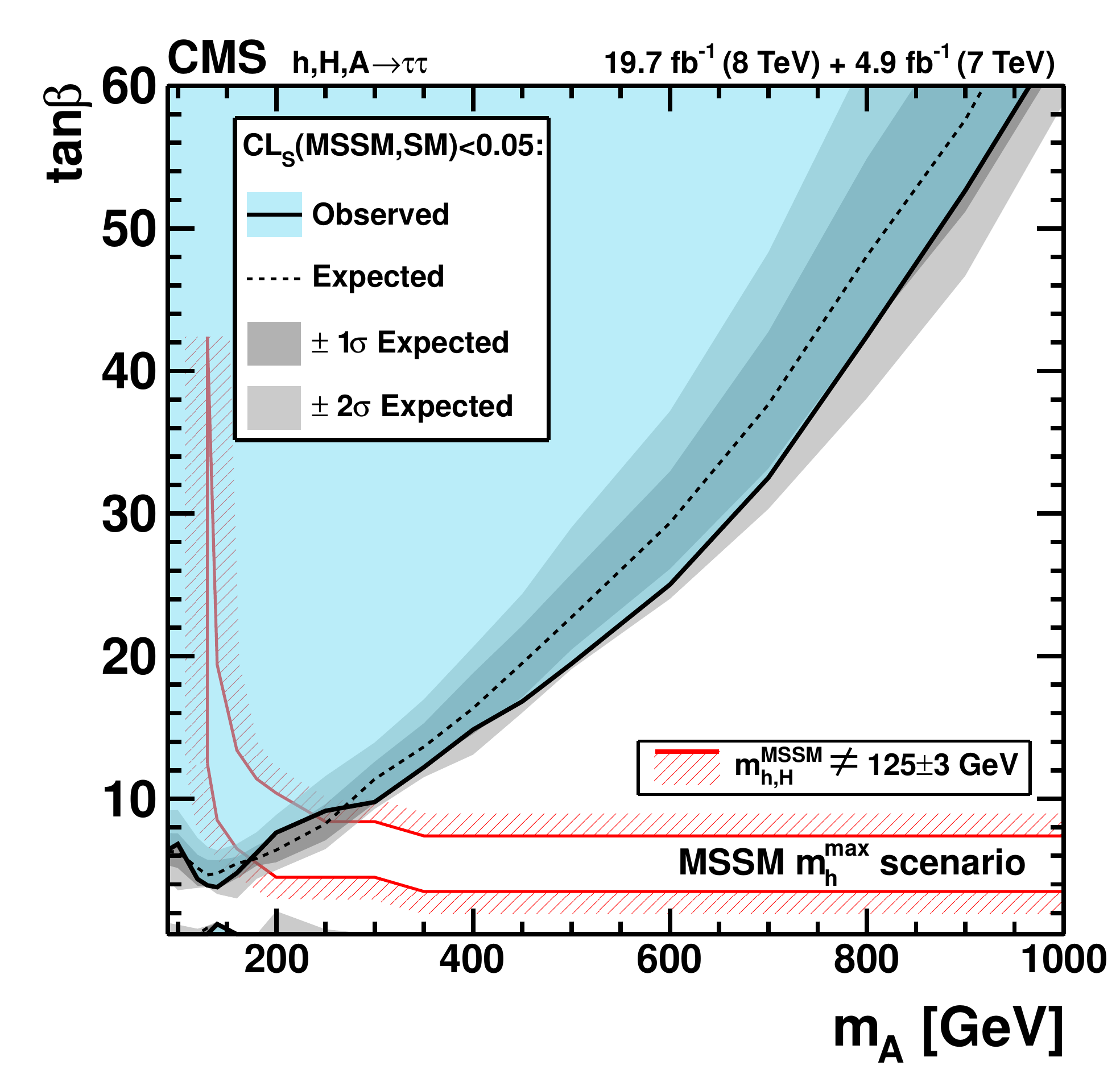} \\
 \includegraphics[width=0.495\textwidth]{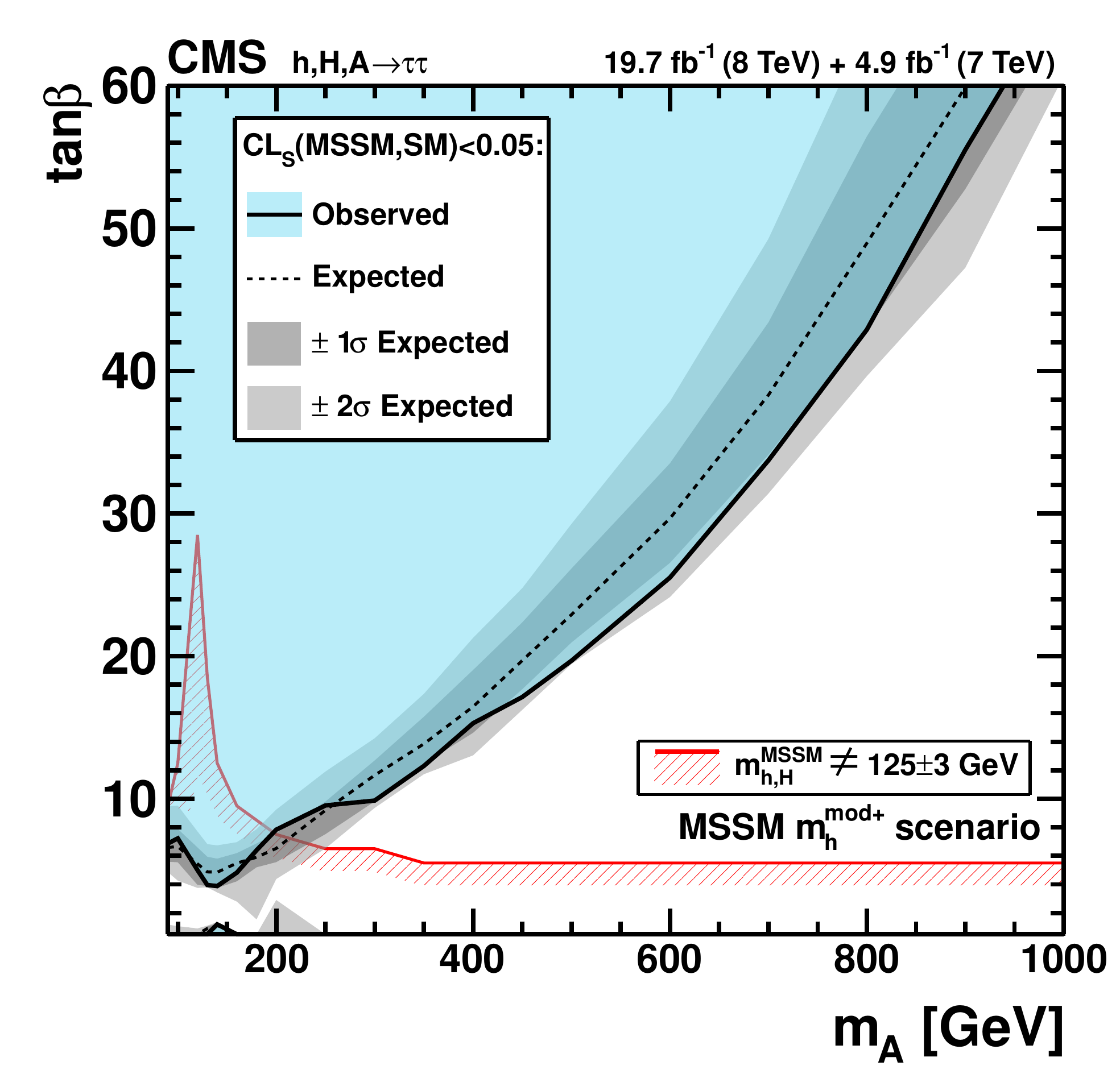}
 \includegraphics[width=0.495\textwidth]{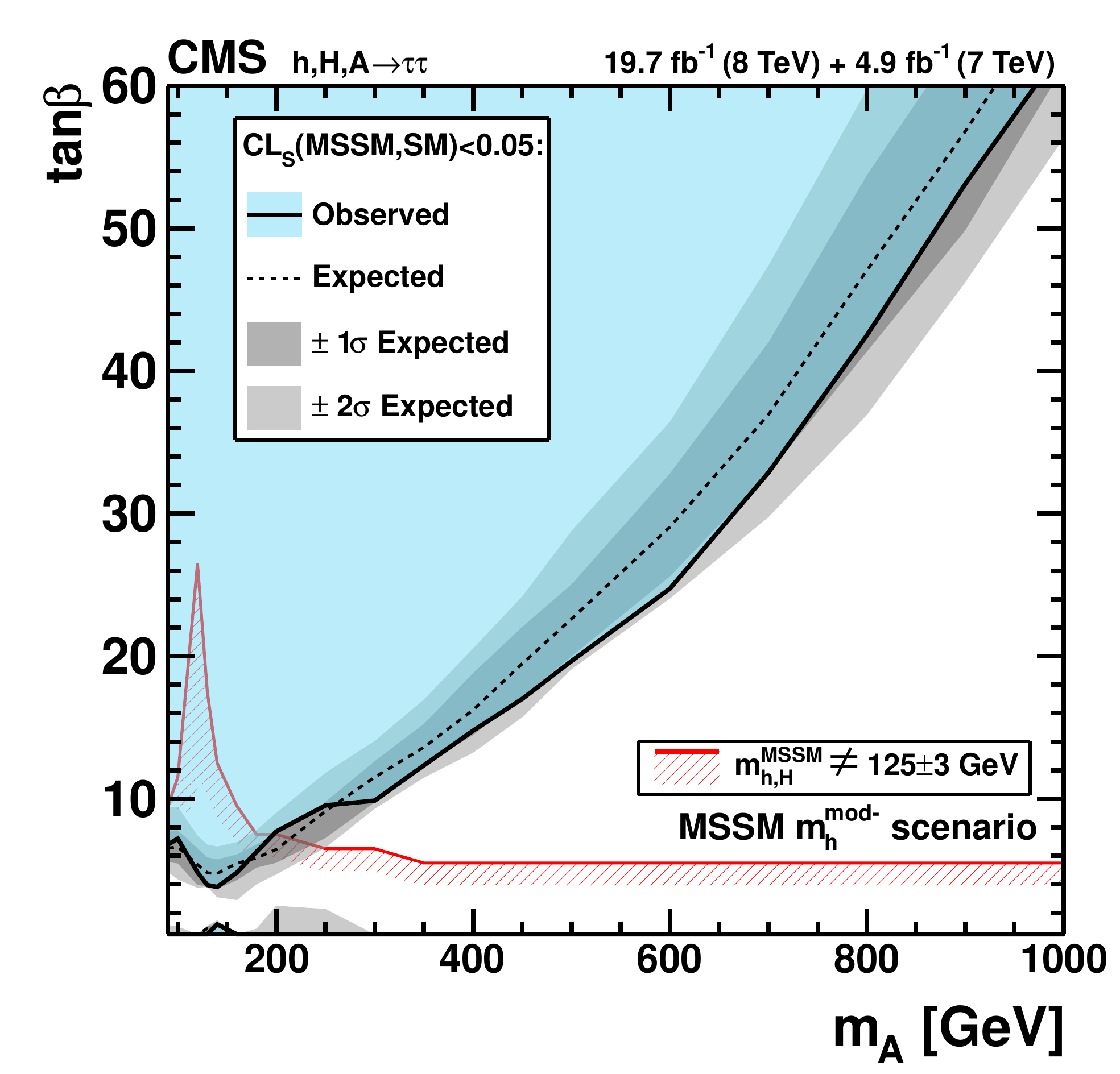}
 \caption{
Expected and observed exclusion limits at 95\% CL in the $m_{\PA}$-$\tan\beta$ parameter space for the MSSM $m_\mathrm{h}^\text{max}$, $m_{\Ph}^\text{mod$+$}$ and $m_{\Ph}^\text{mod$-$}$ benchmark scenarios, are shown as shaded areas.
The allowed regions where the mass of the MSSM scalar Higgs boson $\Ph$ or $\PH$ is compatible with the mass of the recently discovered boson of 125\GeV within a range of $\pm$3\GeV are delimited by the hatched areas. A test of the compatibility of the data to a signal of the three neutral Higgs bosons $\Ph$, $\PH$ and $\PA$ compared to a SM Higgs boson hypothesis is performed.
}
  \label{fig:tanbeta_ma_mhm}
\end{center}
\end{figure*}

Figure~\ref{fig:tanbeta_ma_mhm} shows the expected and observed exclusion limits at the 95\% CL in the $m_\mathrm{h}^\text{max}$ scenario and the modified scenarios  $m_{\Ph}^\text{mod$+$}$ and $m_{\Ph}^\text{mod$-$}$.
The allowed regions where the mass of the MSSM scalar Higgs boson $\Ph$ or $\PH$ is compatible with the mass of the recently discovered boson of 125\GeV within a range of $\pm$3\GeV are delimited by the hatched areas.
Most of the MSSM parameter space is excluded by the Higgs boson mass requirement in the $m_\mathrm{h}^\text{max}$ scenario, while in the modified scenarios the exclusion is mainly concentrated at low $\tan\beta$ values.

\newpage

\begin{figure*}[htbp]
\begin{center}
 \includegraphics[width=0.495\textwidth]{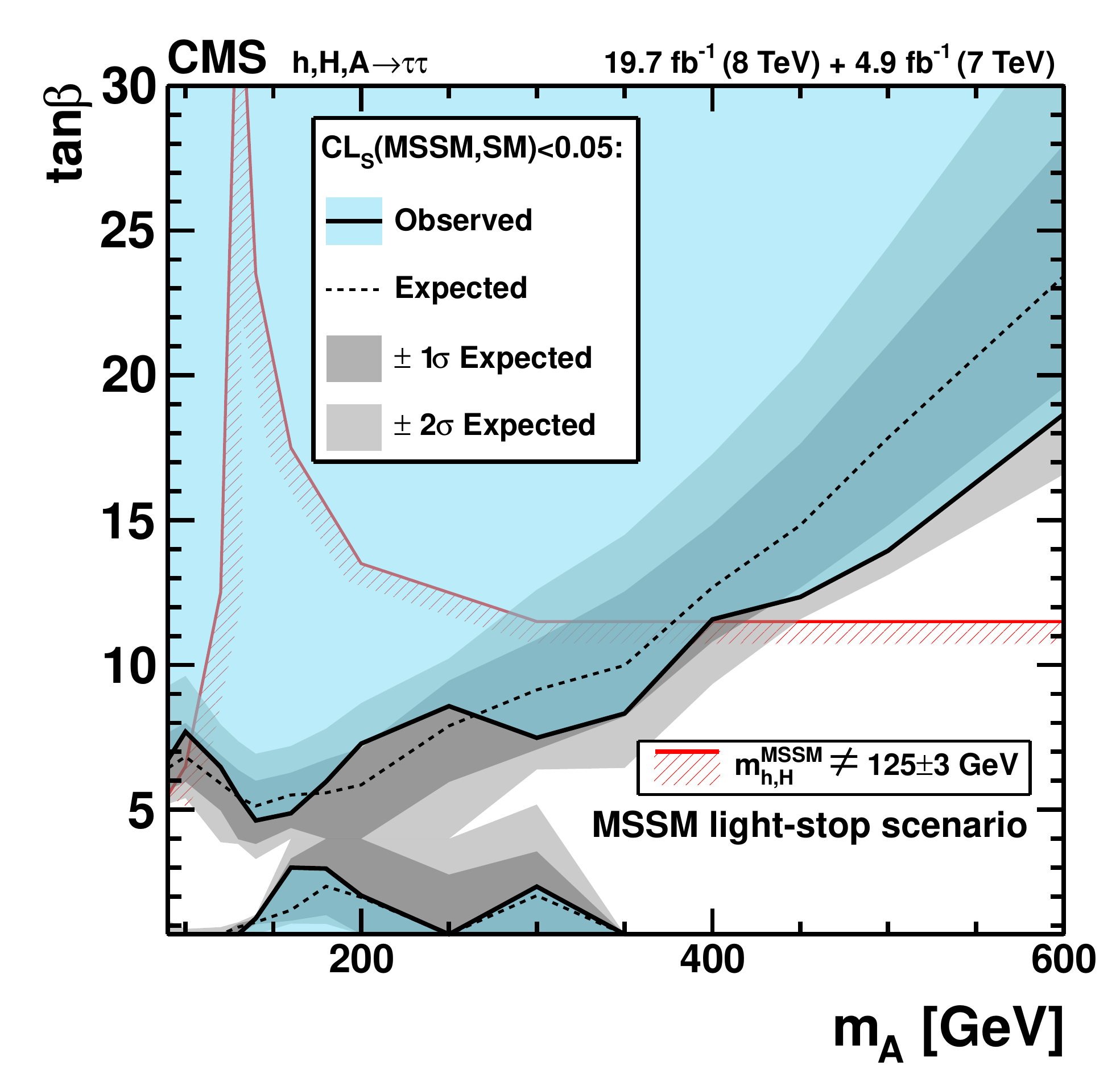}
 \includegraphics[width=0.495\textwidth]{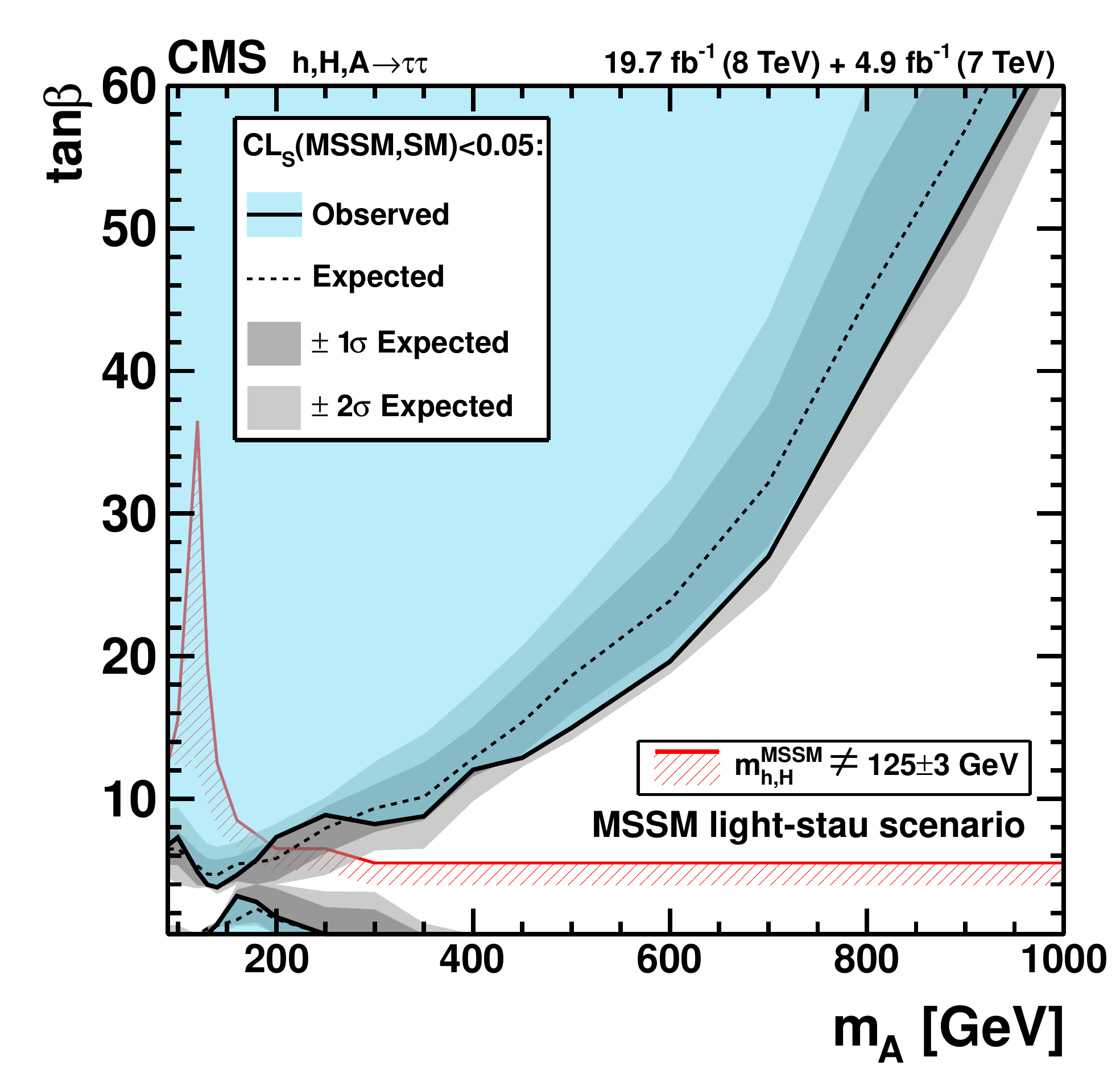}
 \includegraphics[width=0.495\textwidth]{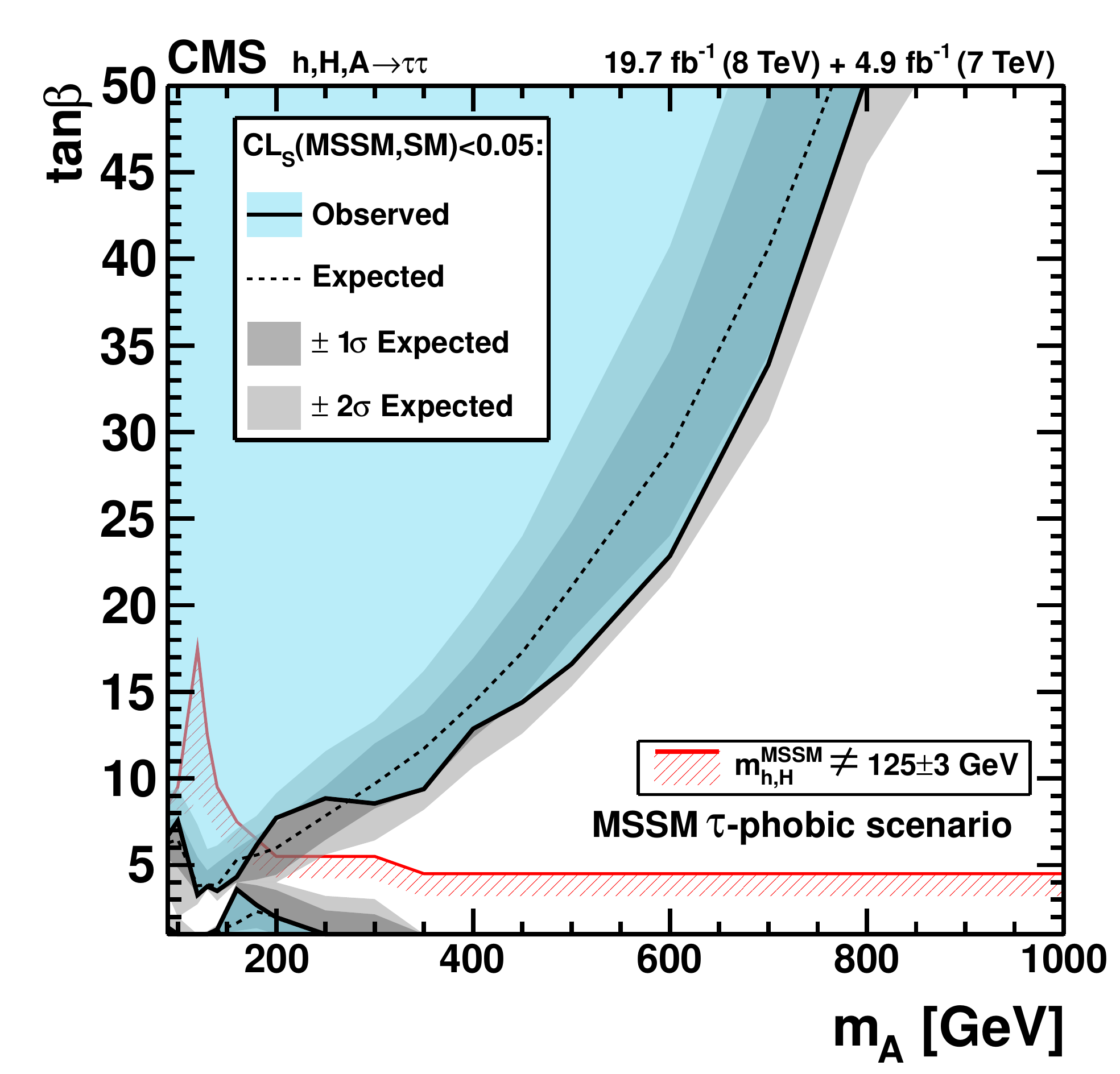}
 \includegraphics[width=0.495\textwidth]{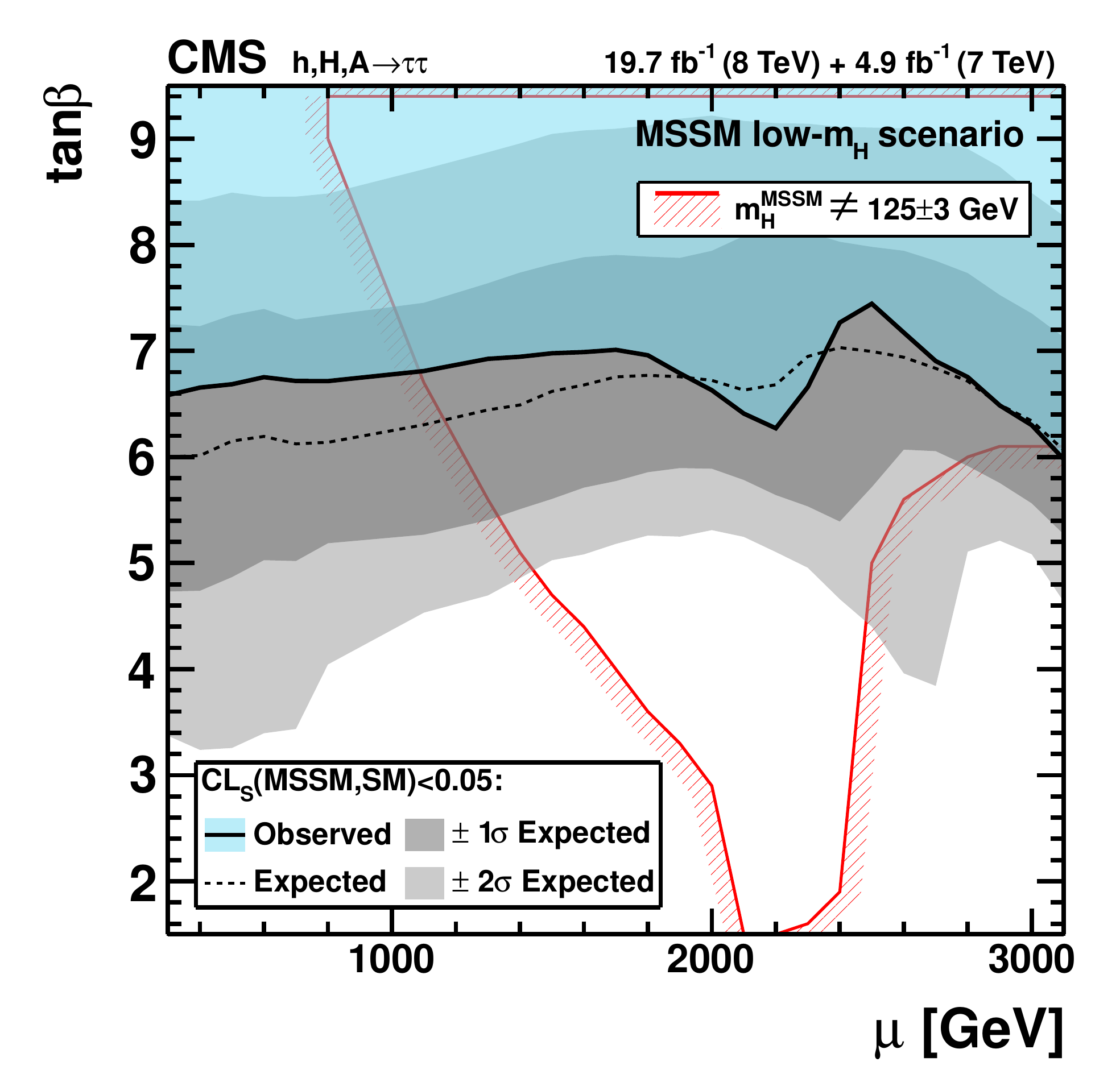}
 \caption{
Expected and observed exclusion limits at 95\% CL in the $m_{\PA}$-$\tan\beta$ parameter space for the MSSM light-stop, light-stau, and $\Pgt$-phobic benchmark scenarios. In the MSSM low-$m_{\PH}$ benchmark scenario the limits in the $\mu$-$\tan\beta$ parameter space are shown.
Other details are as in Fig.~\ref{fig:tanbeta_ma_mhm}.
}
  \label{fig:tanbeta_ma_other}
\end{center}
\end{figure*}

Figure~\ref{fig:tanbeta_ma_other} shows the exclusion limits at the 95\% CL in the light-stop, light-stau, $\Pgt$-phobic and  low-$m_{\PH}$ scenarios. In the light-stop scenario, most of the parameter space probed is excluded either by the direct exclusion of this search or by the Higgs boson mass compatibility requirement with the recent Higgs boson discovery at 125\GeV.
Numerical values for the expected and observed exclusion limits for all MSSM benchmark scenarios considered are given in Tables~\ref{tab:limits_mhmax}--\ref{tab:limits_lowmh} in Appendix~\ref{sec:tables}.

It should be noted that due to the interference effects of the bottom and top loops in the MSSM $\cPg\cPg\Ph$ cross section calculation, the direct search is also able to exclude some regions at low $\tan\beta$.
The excluded regions at low $\tan\beta$
can be seen better looking at the numerical values given in Tables~\ref{tab:limits_mhmax}--\ref{tab:limits_lowmh} in Appendix~\ref{sec:tables}.

\newpage

\begin{figure*}[htb]\begin{center}
\includegraphics[width=0.495\textwidth]{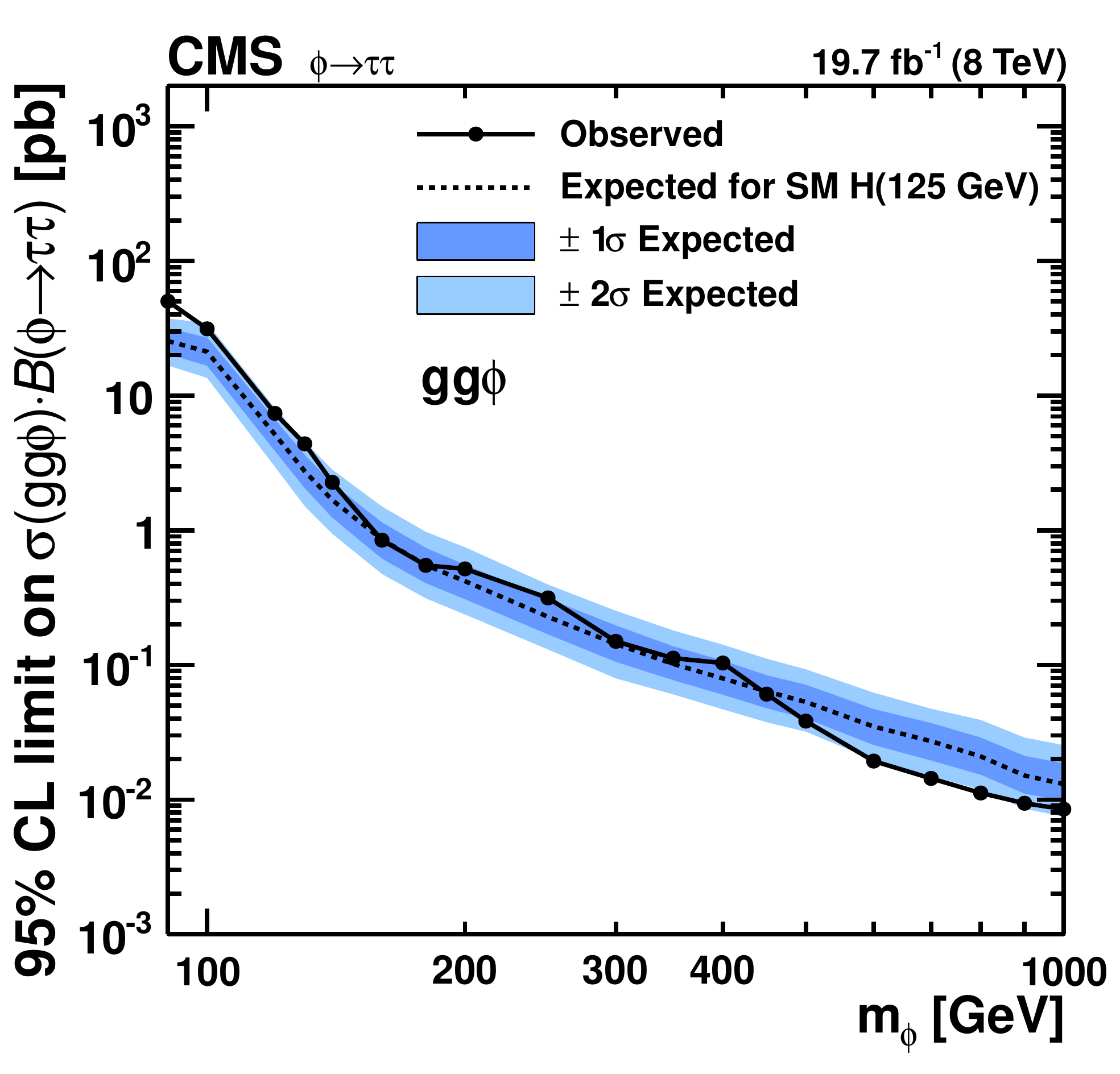}
\includegraphics[width=0.495\textwidth]{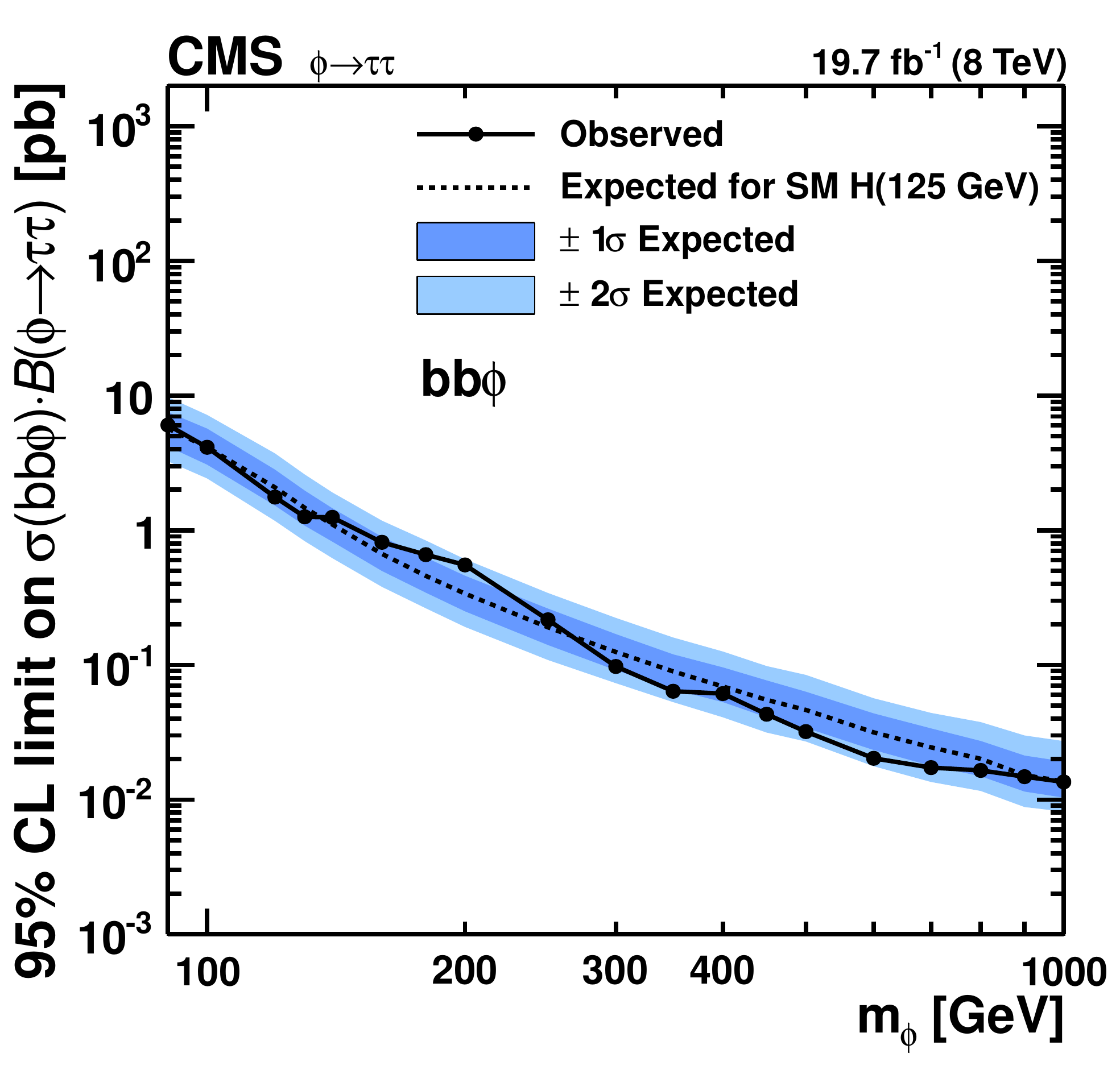}
\caption{
  Upper limit at 95\% CL on $\sigma(\cPg\cPg\phi)\cdot \mathcal{B}(\phi\to\Pgt\Pgt)$ (left) and  $\sigma(\cPqb\cPqb\phi)\cdot\mathcal{B}(\phi\to\Pgt\Pgt)$ (right) at 8\TeV center-of-mass energy as a function of $m_{\phi}$, where ${\phi}$ denotes a generic Higgs-like state. The expected and observed limits are computed using the test statistics given by Eq.~\ref{eq:teststatistic_LHC}. For the expected limits, the observed data have been replaced by a representative dataset which not only contains the contribution from background processes but also a SM Higgs boson with a mass of 125\GeV.}
  \label{fig:ggH_bbH_limit}\end{center}\end{figure*}

To allow to compare these results to other extensions of the SM apart from the MSSM, which have been proposed to solve the hierarchy problem, a search for a single resonance $\phi$ with a narrow width compared to the experimental resolution is also performed.  In this case, model independent limits on the product of the production cross section times branching fraction to $\Pgt\Pgt$,  $\sigma\cdot\mathcal{B}(\phi\to\Pgt\Pgt)$, for gluon fusion and b-quark associated Higgs boson production, as a function of the Higgs boson mass $m_{\phi}$ have been determined.   To model the hypothetical signal $\phi$, the same simulation samples as the neutral MSSM Higgs boson search have been used. These results have been obtained using the data with 8\TeV center-of-mass energy only and are shown in Fig.~\ref{fig:ggH_bbH_limit}.
The expected and observed limits are computed using the test statistics given by Eq.~\ref{eq:teststatistic_LHC}. To extract the limit on the gluon fusion (b-quark associated) Higgs boson production, the rate of the b-quark associated (gluon fusion) Higgs boson production is treated as a nuisance parameter in the fit. For the expected limits, the observed data have been replaced by a representative dataset which not only contains the contribution from background processes but also a SM Higgs boson with a mass of 125\GeV.
The observed limits are in agreement with the expectation.
The results are also summarized in Tables~\ref{ggH-limit} and~\ref{bbH-limit} in Appendix~\ref{sec:tables}.

\begin{figure*}[!Hp]\begin{center}
 \includegraphics[width=0.32\textwidth]{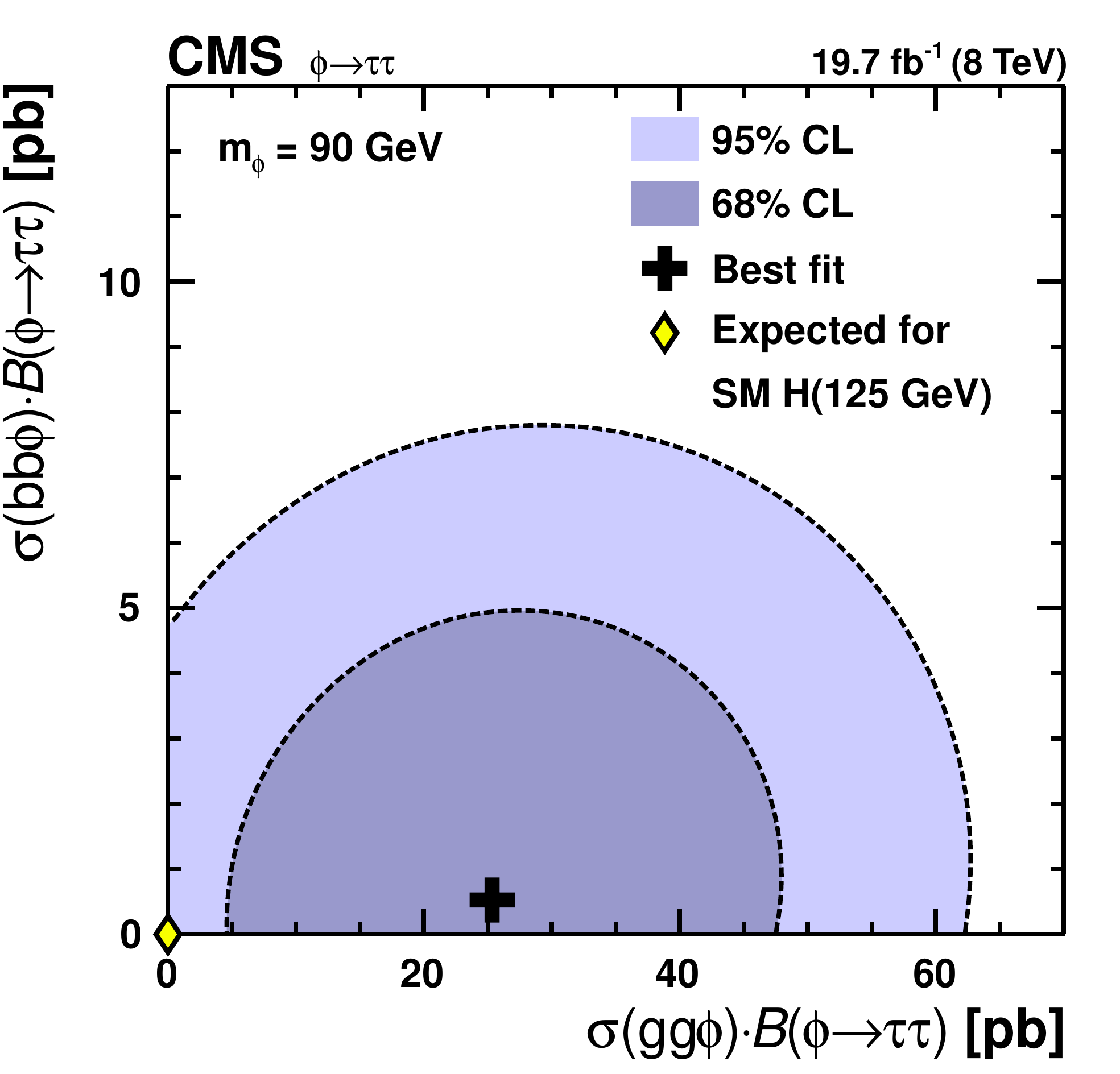}
 \includegraphics[width=0.32\textwidth]{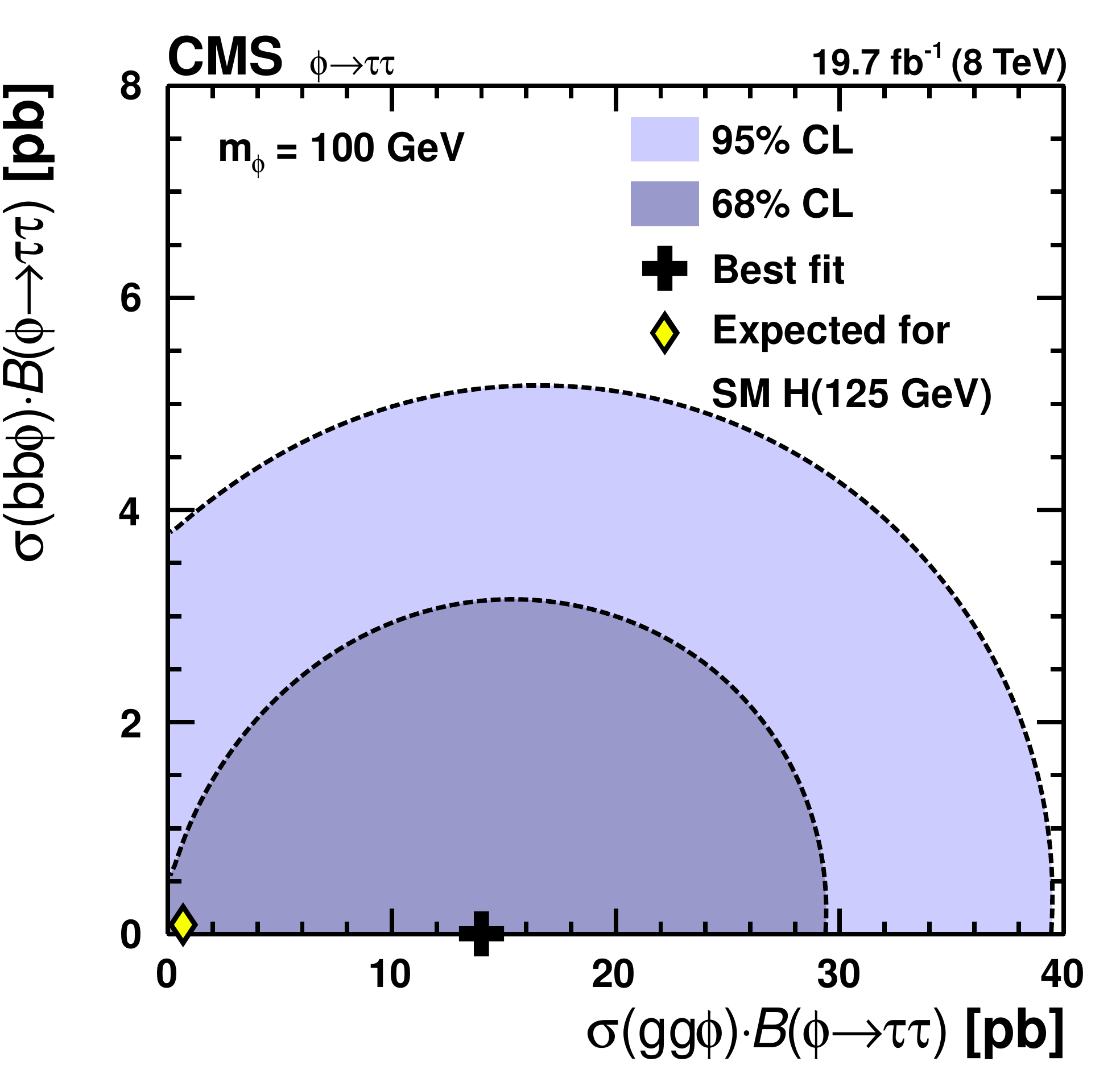}
 \includegraphics[width=0.32\textwidth]{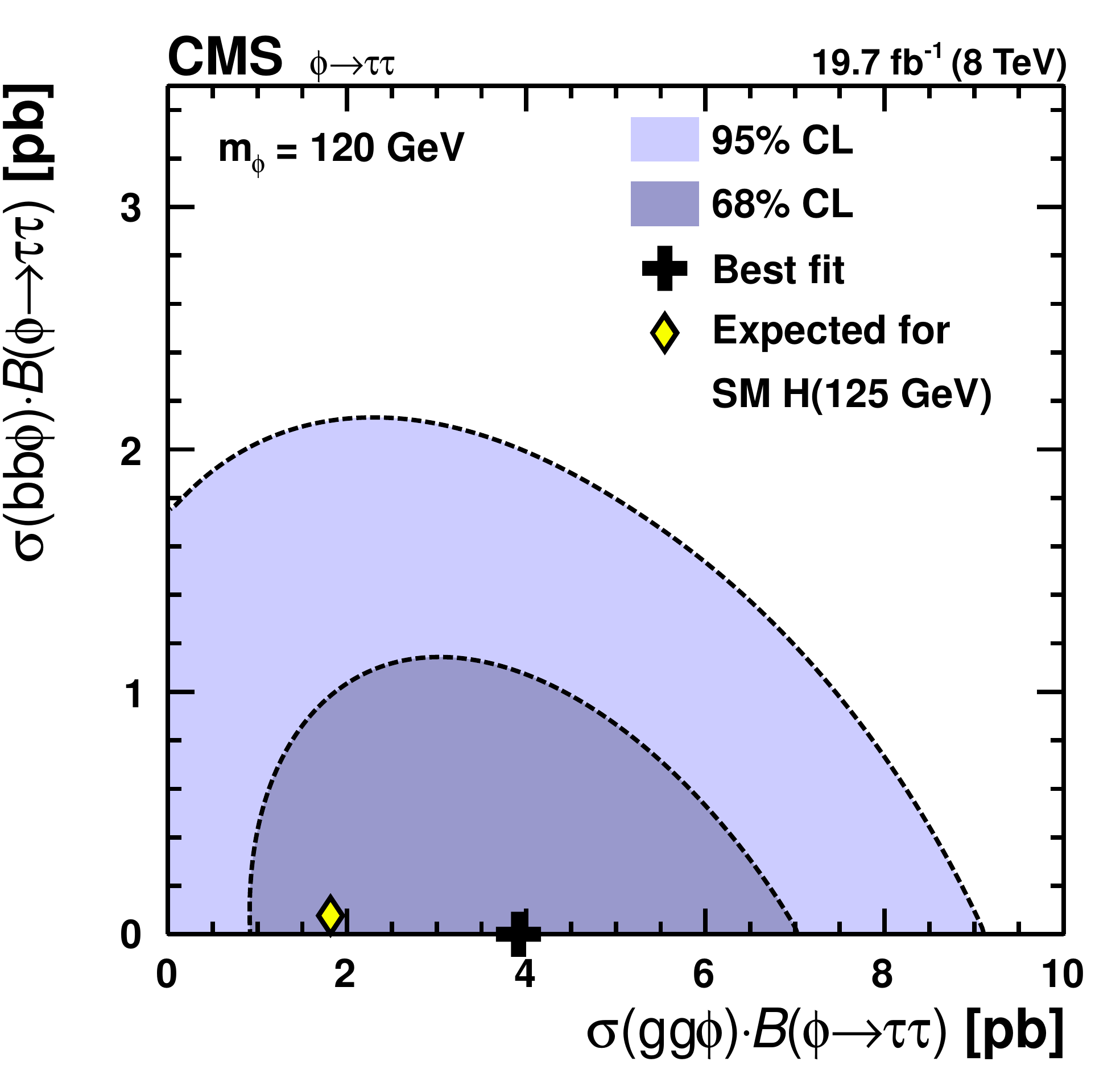}
 \includegraphics[width=0.32\textwidth]{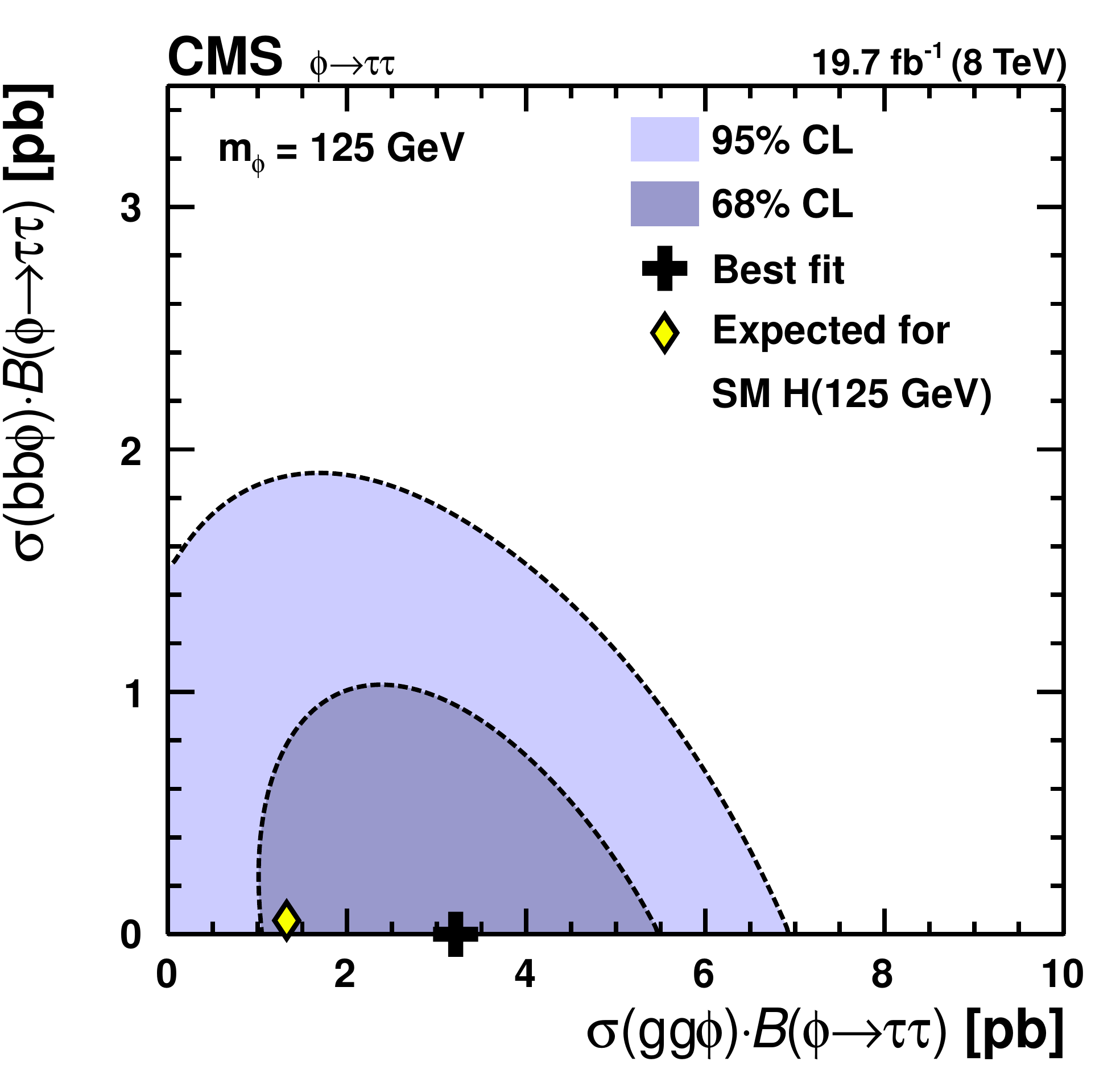}
 \includegraphics[width=0.32\textwidth]{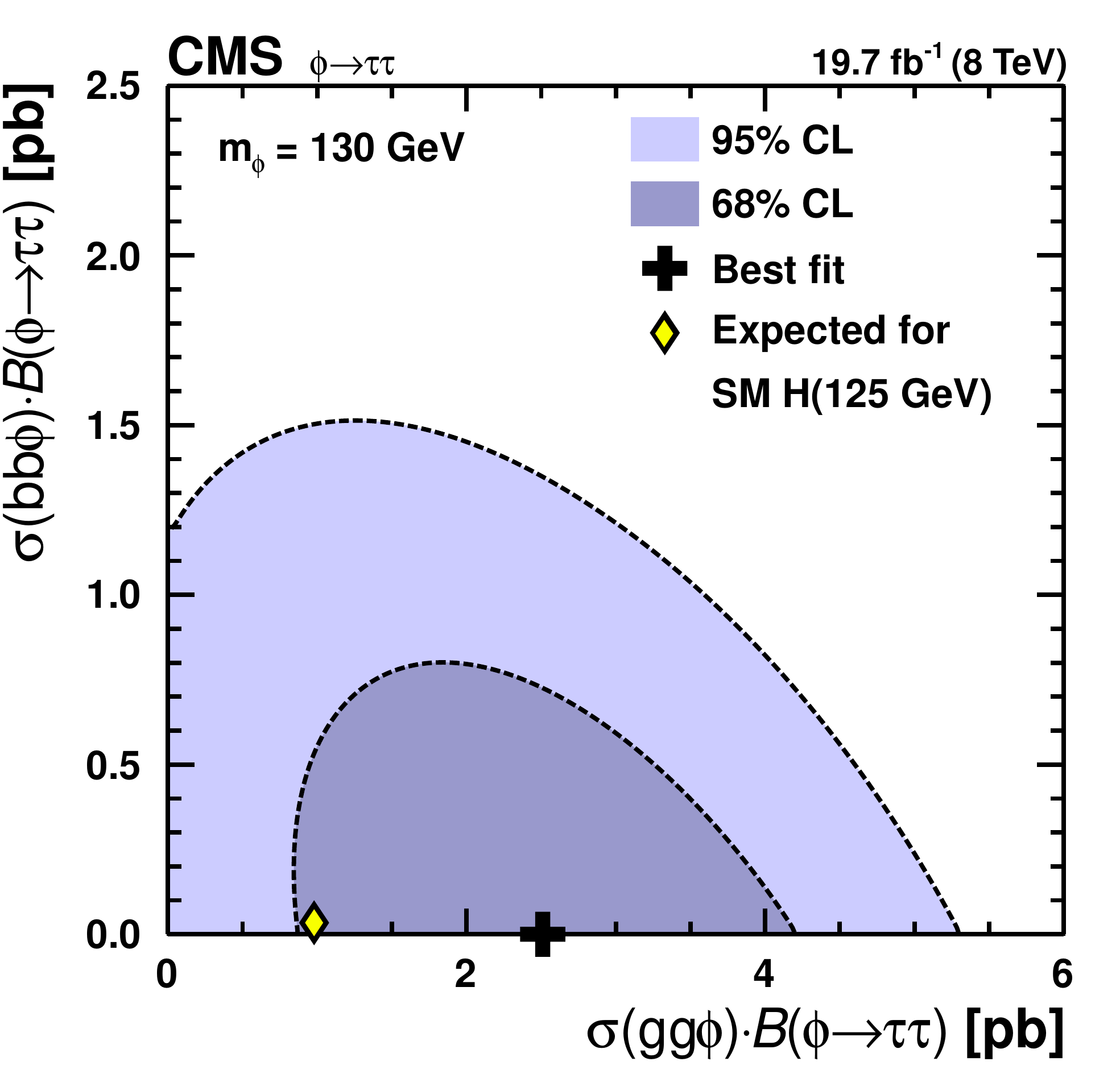}
 \includegraphics[width=0.32\textwidth]{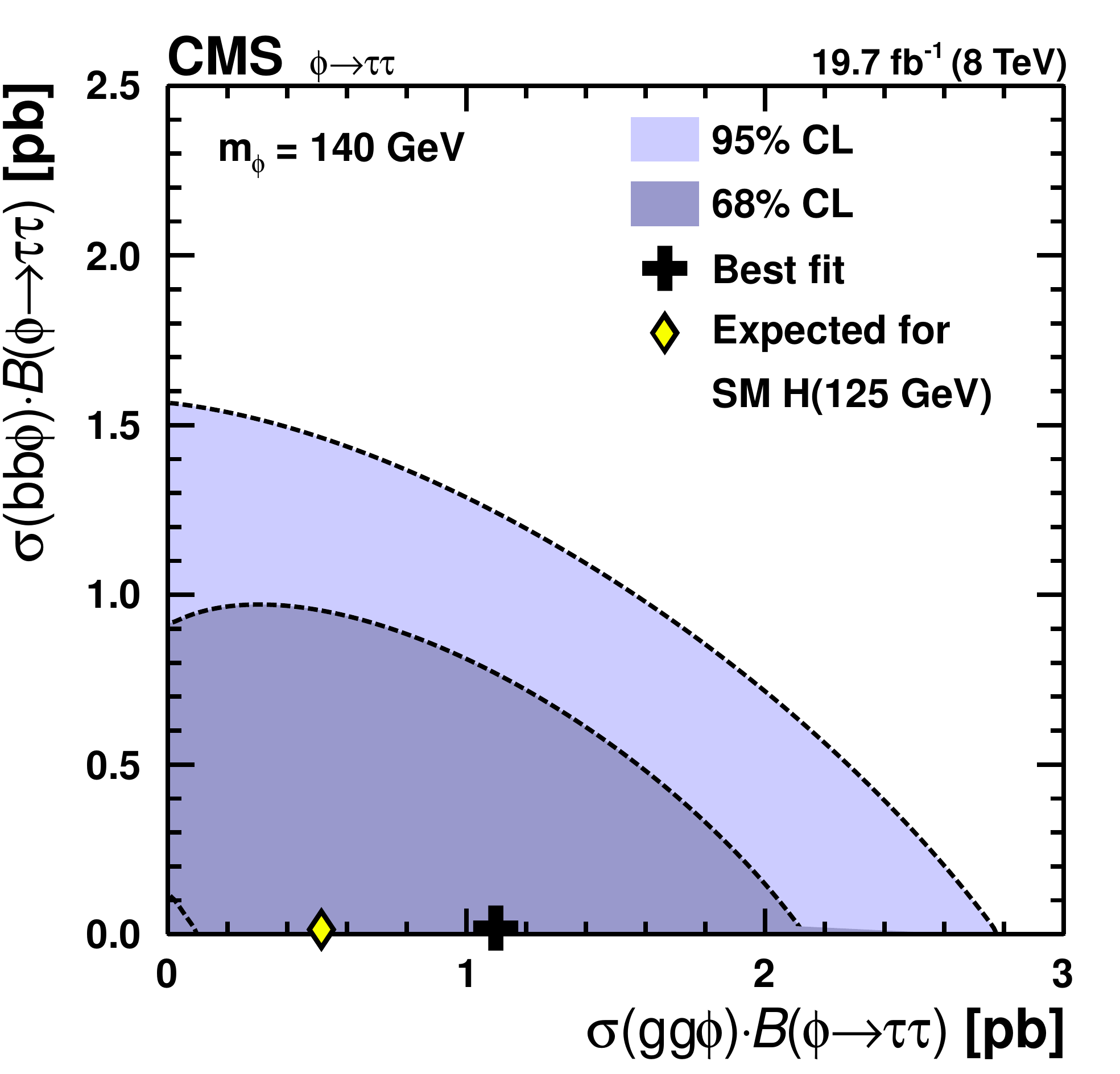}
 \includegraphics[width=0.32\textwidth]{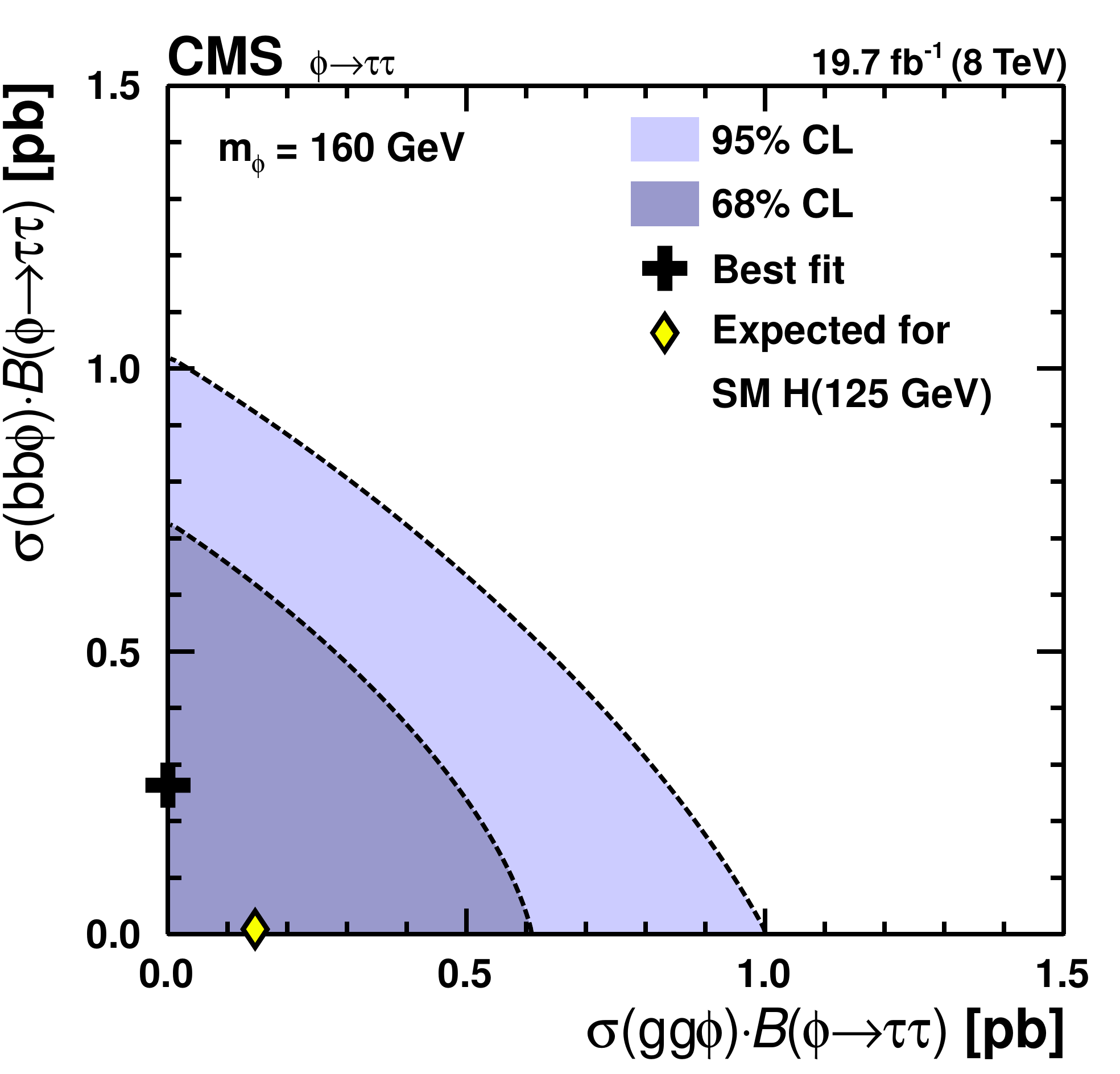}
 \includegraphics[width=0.32\textwidth]{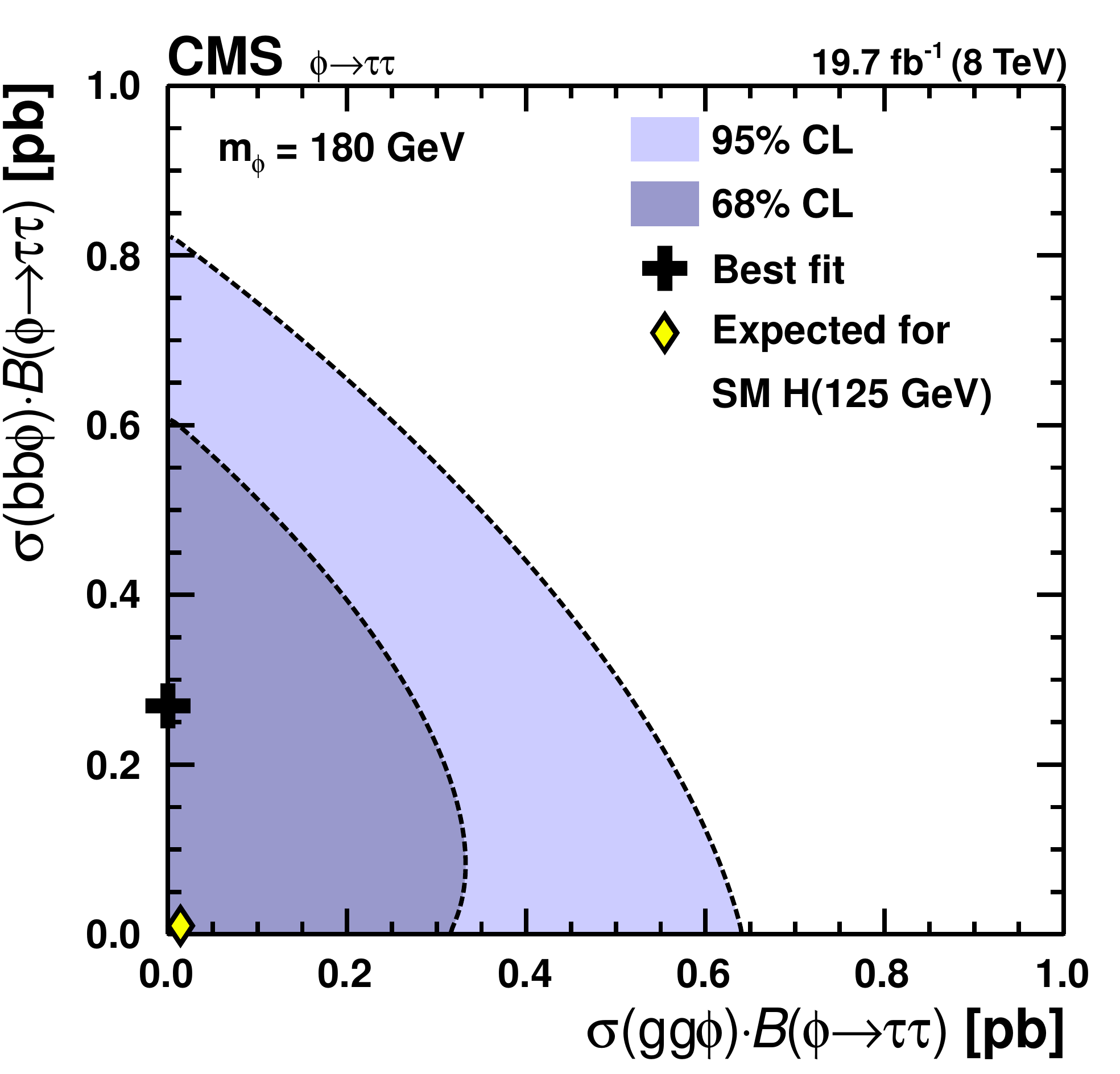}
 \includegraphics[width=0.32\textwidth]{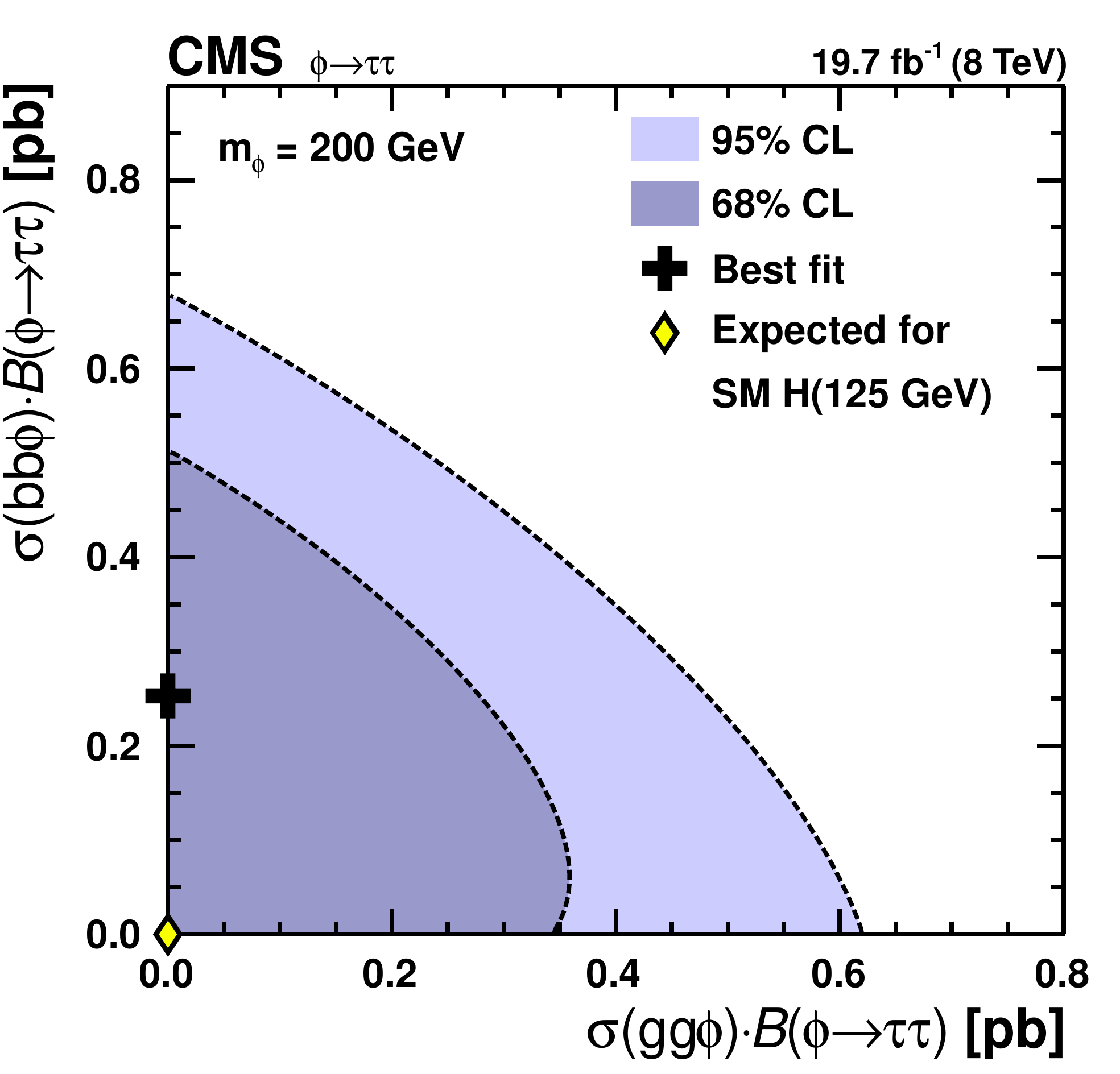}
 \includegraphics[width=0.32\textwidth]{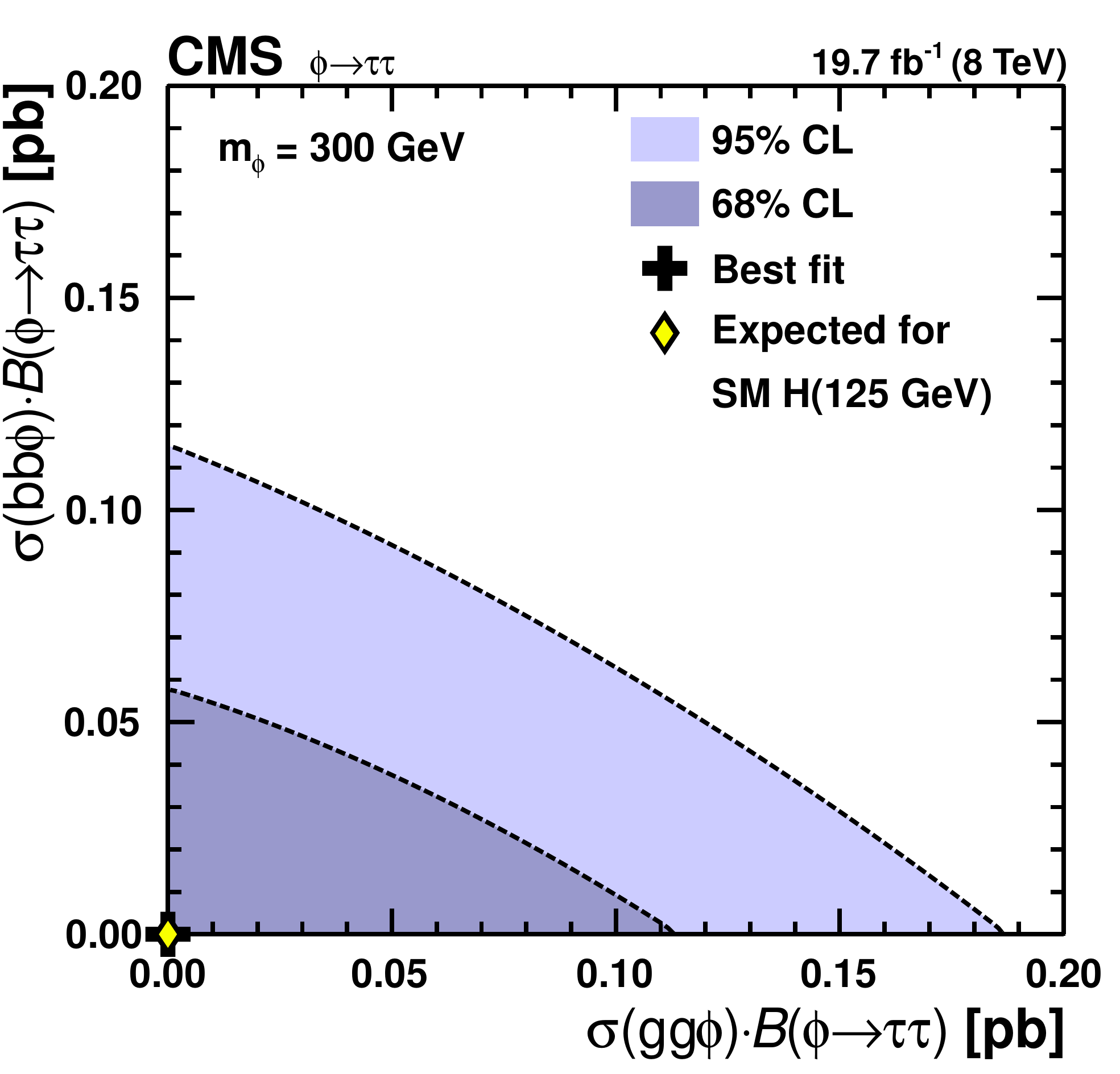}
 \includegraphics[width=0.32\textwidth]{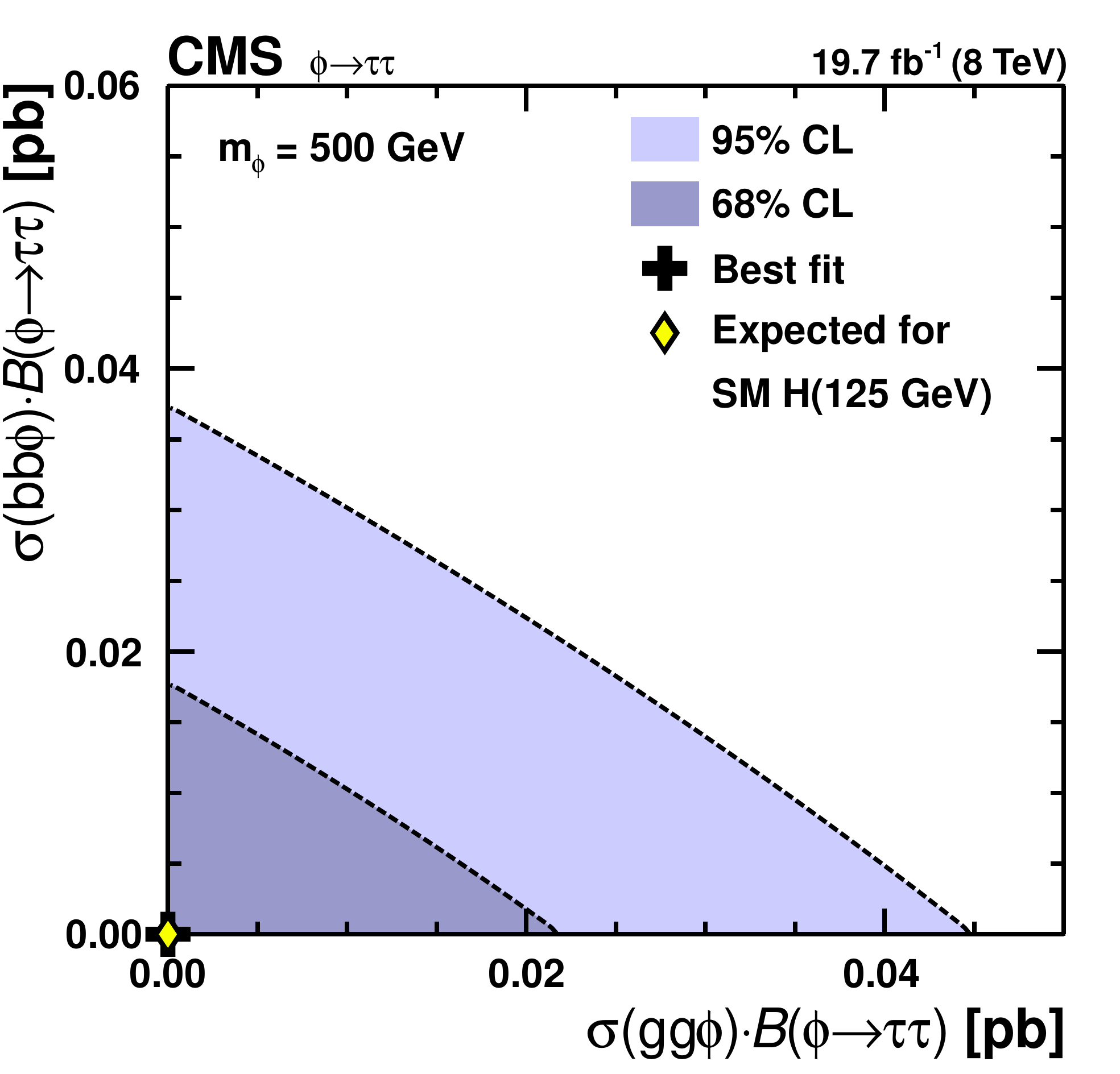}
 \includegraphics[width=0.32\textwidth]{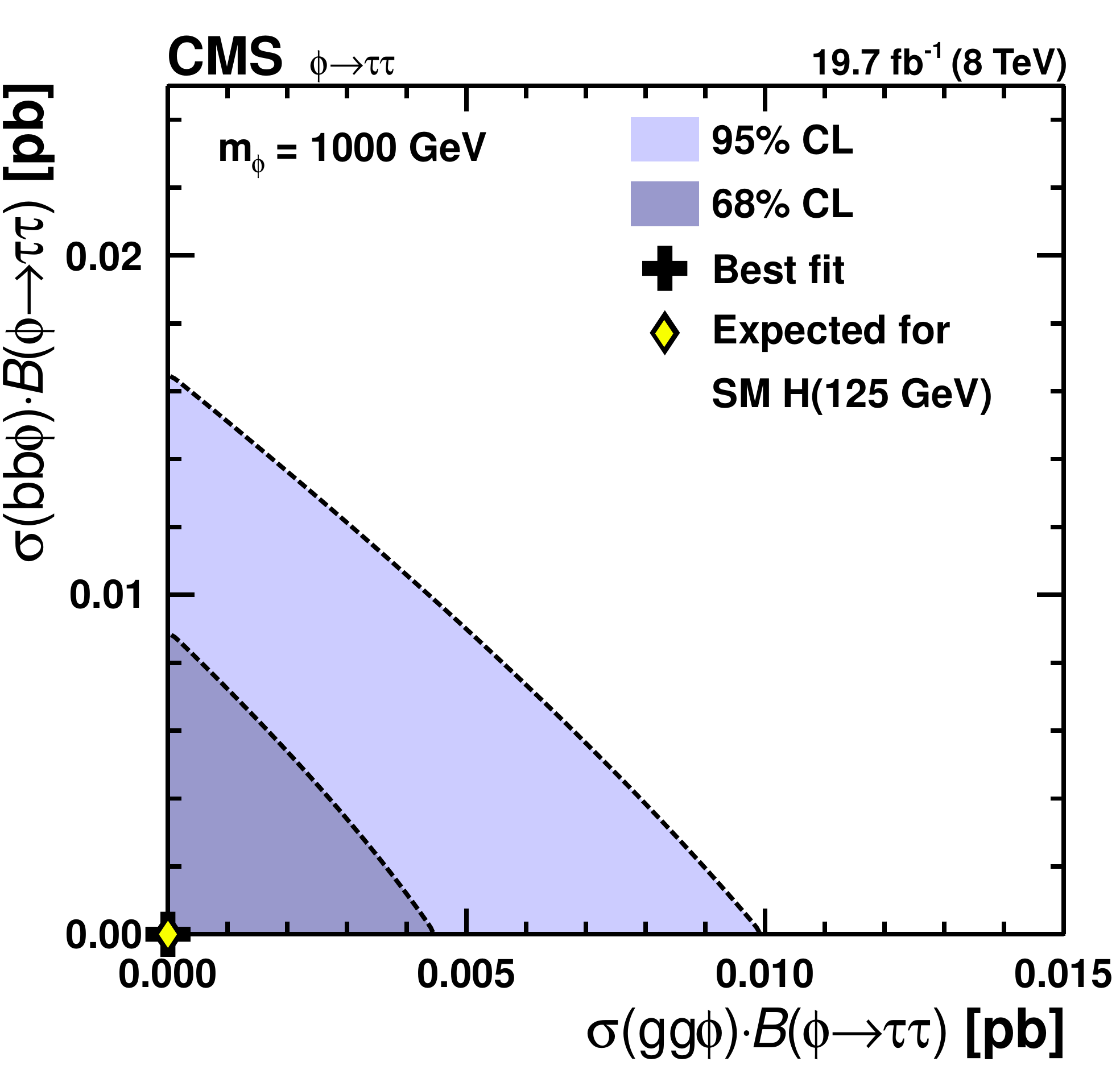}
 \caption{Likelihood contours of $\sigma(\cPqb\cPqb\phi)\cdot\mathcal{B}(\phi\to\Pgt\Pgt)$ versus $\sigma(\cPg\cPg\phi)\cdot\mathcal{B}(\phi\to\Pgt\Pgt)$ at 8\TeV center-of-mass energy for different values of the Higgs boson mass $m_{\phi}$. The best fit value (cross) and the expectation from a SM Higgs boson with a mass of 125\GeV (diamond) is also shown. In the case of $m_{\phi}$ = 300, 500 and 1000\GeV, the best fit value and the expectation from a SM Higgs boson at 125\GeV are both at (0,0).}
  \label{fig:contour1}\end{center}\end{figure*}

Finally, a 2-dimensional 68\% and 95\% CL likelihood scan of the cross section times branching fraction to $\Pgt\Pgt$ for gluon fusion and b-quark associated Higgs boson production,
$\sigma(\cPqb\cPqb\phi)\cdot\mathcal{B}(\phi\to\Pgt\Pgt)$ versus $\sigma(\cPg\cPg\phi)\cdot\mathcal{B}(\phi\to\Pgt\Pgt)$,
has also been performed.
 The results for different values of the Higgs boson mass $m_{\phi}$ are shown in Fig.~\ref{fig:contour1}.
The best fit value and the expectation from a SM Higgs boson with a mass of 125\GeV is also shown. The result from the likelihood scan for $m_{\phi}= 125\GeV$ is compatible with the expectation from a SM Higgs boson.

\section{Summary}
A search for neutral Higgs bosons in the minimal supersymmetric extension of the standard model (MSSM) decaying to tau-lepton pairs has been performed using events recorded by the CMS experiment at the LHC.
The dataset corresponds to an integrated luminosity of 24.6\fbinv, with 4.9\fbinv at 7\TeV and 19.7\fbinv at 8\TeV. Five different $\Pgt\Pgt$
signatures are studied,
$\Pe\tauh, \Pgm\tauh, \Pe\Pgm$, $\Pgm\Pgm$, and $\tauh\tauh$, where $\tauh$ denotes a hadronically decaying tau lepton.
 To enhance the sensitivity to neutral MSSM Higgs bosons, the search
 includes the case where the Higgs boson is produced in association with a b-quark jet.
No excess is observed in the tau-lepton-pair invariant mass spectrum. Exclusion limits are presented in the MSSM parameter space for different MSSM benchmark
scenarios,
$m_{\Ph}^\text{max}$, $m_{\Ph}^\text{mod$+$}$, $m_{\Ph}^\text{mod$-$}$, light-stop, light-stau, $\Pgt$-phobic and low-$m_{\PH}$.
Model independent upper limits on the Higgs boson production cross section times branching fraction for gluon fusion and b-quark associated production are also given.

\section*{Acknowledgements}

\hyphenation{Bundes-ministerium Forschungs-gemeinschaft Forschungs-zentren} We congratulate our colleagues in the CERN accelerator departments for the excellent performance of the LHC and thank the technical and administrative staffs at CERN and at other CMS institutes for their contributions to the success of the CMS effort. In addition, we gratefully acknowledge the computing centres and personnel of the Worldwide LHC Computing Grid for delivering so effectively the computing infrastructure essential to our analyses. Finally, we acknowledge the enduring support for the construction and operation of the LHC and the CMS detector provided by the following funding agencies: the Austrian Federal Ministry of Science, Research and Economy and the Austrian Science Fund; the Belgian Fonds de la Recherche Scientifique, and Fonds voor Wetenschappelijk Onderzoek; the Brazilian Funding Agencies (CNPq, CAPES, FAPERJ, and FAPESP); the Bulgarian Ministry of Education and Science; CERN; the Chinese Academy of Sciences, Ministry of Science and Technology, and National Natural Science Foundation of China; the Colombian Funding Agency (COLCIENCIAS); the Croatian Ministry of Science, Education and Sport, and the Croatian Science Foundation; the Research Promotion Foundation, Cyprus; the Ministry of Education and Research, Estonian Research Council via IUT23-4 and IUT23-6 and European Regional Development Fund, Estonia; the Academy of Finland, Finnish Ministry of Education and Culture, and Helsinki Institute of Physics; the Institut National de Physique Nucl\'eaire et de Physique des Particules~/~CNRS, and Commissariat \`a l'\'Energie Atomique et aux \'Energies Alternatives~/~CEA, France; the Bundesministerium f\"ur Bildung und Forschung, Deutsche Forschungsgemeinschaft, and Helmholtz-Gemeinschaft Deutscher Forschungszentren, Germany; the General Secretariat for Research and Technology, Greece; the National Scientific Research Foundation, and National Innovation Office, Hungary; the Department of Atomic Energy and the Department of Science and Technology, India; the Institute for Studies in Theoretical Physics and Mathematics, Iran; the Science Foundation, Ireland; the Istituto Nazionale di Fisica Nucleare, Italy; the Korean Ministry of Education, Science and Technology and the World Class University program of NRF, Republic of Korea; the Lithuanian Academy of Sciences; the Ministry of Education, and University of Malaya (Malaysia); the Mexican Funding Agencies (CINVESTAV, CONACYT, SEP, and UASLP-FAI); the Ministry of Business, Innovation and Employment, New Zealand; the Pakistan Atomic Energy Commission; the Ministry of Science and Higher Education and the National Science Centre, Poland; the Funda\c{c}\~ao para a Ci\^encia e a Tecnologia, Portugal; JINR, Dubna; the Ministry of Education and Science of the Russian Federation, the Federal Agency of Atomic Energy of the Russian Federation, Russian Academy of Sciences, and the Russian Foundation for Basic Research; the Ministry of Education, Science and Technological Development of Serbia; the Secretar\'{\i}a de Estado de Investigaci\'on, Desarrollo e Innovaci\'on and Programa Consolider-Ingenio 2010, Spain; the Swiss Funding Agencies (ETH Board, ETH Zurich, PSI, SNF, UniZH, Canton Zurich, and SER); the Ministry of Science and Technology, Taipei; the Thailand Center of Excellence in Physics, the Institute for the Promotion of Teaching Science and Technology of Thailand, Special Task Force for Activating Research and the National Science and Technology Development Agency of Thailand; the Scientific and Technical Research Council of Turkey, and Turkish Atomic Energy Authority; the National Academy of Sciences of Ukraine, and State Fund for Fundamental Researches, Ukraine; the Science and Technology Facilities Council, UK; the US Department of Energy, and the US National Science Foundation.

Individuals have received support from the Marie-Curie programme and the European Research Council and EPLANET (European Union); the Leventis Foundation; the A. P. Sloan Foundation; the Alexander von Humboldt Foundation; the Belgian Federal Science Policy Office; the Fonds pour la Formation \`a la Recherche dans l'Industrie et dans l'Agriculture (FRIA-Belgium); the Agentschap voor Innovatie door Wetenschap en Technologie (IWT-Belgium); the Ministry of Education, Youth and Sports (MEYS) of the Czech Republic; the Council of Science and Industrial Research, India; the HOMING PLUS programme of Foundation for Polish Science, cofinanced from European Union, Regional Development Fund; the Compagnia di San Paolo (Torino); the Consorzio per la Fisica (Trieste); MIUR project 20108T4XTM (Italy); the Thalis and Aristeia programmes cofinanced by EU-ESF and the Greek NSRF; and the National Priorities Research Program by Qatar National Research Fund.

\bibliography{auto_generated}   

\clearpage

\appendix
\section{Exclusion limits}
\label{sec:tables}

The 95\% CL exclusion limits in the $\tan\beta$-$m_{\PA}$ or $\tan\beta$-$\mu$ parameter space for different MSSM benchmark scenarios are given in Tables~\ref{tab:limits_mhmax}--\ref{tab:limits_lowmh}.

Model independent limits on $\sigma\cdot\mathcal{B}(\phi\to\Pgt\Pgt)$ for gluon fusion and b-quark associated Higgs boson production as a function of the Higgs boson mass $m_{\phi}$ are given in Tables~\ref{ggH-limit} and~\ref{bbH-limit}, for 8\TeV center-of-mass energy only.

\begin{table*}[!htb]
  \begin{center}
    \caption{Expected and observed 95\% CL limits for
             $\tan\beta$ as a function of $m_{\PA}$, for the MSSM search in the $m_{\Ph}^\text{max}$ benchmark scenario. For the statistical procedure a  test of the compatibility of the data to a signal of the three neutral Higgs bosons $\Ph$, $\PH$ and $\PA$ compared to a SM Higgs boson hypothesis is performed.
}
\begin{tabular}{c|ccccc|c}
\multicolumn{1}{c}{}  & \multicolumn{6}{c}{Neutral MSSM Higgs: $m_{\Ph}^\text{max}$ scenario}  \\
\cline{2-7}
  & \multicolumn{5}{c|}{Expected $\tan\beta$ limit}   &  Observed \\
\cline{2-6}
  $m_{\PA}$ [\GeVns{}] & $-2\sigma$  &   $-1\sigma$ &        Median &    $+1\sigma$ &  $+2\sigma$ & $\tan\beta$ limit \\
\hline
          \multirow{2}{*}{90} &  $>$4.3 &  $>$5.3 &  $>$6.3 &  $>$7.6 &  $>$8.3 &  $>$6.4  \\
                                   &  $<$1.3 &         &         &         &         &           \\
         \multirow{2}{*}{100} &  $>$3.6 &  $>$5.1 &  $>$6.3 &  $>$7.6 &  $>$9.0 &  $>$6.8  \\
                                   &  $<$1.2 &         &         &         &         &           \\

         \multirow{2}{*}{120} &  $>$3.8 &  $>$3.9 &  $>$5.2 &  $>$6.1 &  $>$6.6 &  $>$4.4  \\
                                   &  $<$0.9 &         &         &         &         &           \\

         \multirow{2}{*}{130} &  $>$3.8 &  $>$3.9 &  $>$4.7 &  $>$5.7 &  $>$5.5 &  $>$3.9  \\
                                   &  $<$1.1 &  $<$1.1 &  $<$0.9 &         &         &           \\
         \multirow{2}{*}{140} &  $>$3.3 &  $>$3.7 &  $>$4.8 &  $>$5.7 &  $>$6.2 &  $>$3.8  \\
                                   &  $<$1.4 &  $<$1.3 &  $<$1.1 &  $<$1.0 &         &  $<$1.2  \\
                         160 &  $>$3.0 &  $>$4.4 &  $>$5.5 &  $>$5.9 &  $>$7.9 &  $>$4.8  \\
                         180 &  $>$4.4 &  $>$5.3 &  $>$5.8 &  $>$6.7 &  $>$7.2 &  $>$6.2  \\
        \multirow{2}{*}{200} &  $>$5.0 &  $>$5.6 &  $>$6.4 &  $>$7.4 &  $>$7.8 &  $>$7.6  \\
                                  &  $<$2.1 &         &         &         &         &           \\
        \multirow{2}{*}{250} &  $>$6.5 &  $>$7.1 &  $>$8.3 &  $>$9.6 &  $>$10.0 &  $>$9.2  \\
                                  &  $<$0.8 &         &         &         &          &           \\
                         300 &  $>$9.3 &  $>$9.7 &  $>$11.4 &  $>$12.6 &  $>$13.5 &  $>$9.8  \\
                         350 &  $>$11.5 &  $>$12.4 &  $>$13.6 &  $>$15.3 &  $>$15.8 &  $>$12.2  \\
                         400 &  $>$13.1 &  $>$14.5 &  $>$16.3 &  $>$18.8 &  $>$19.6 &  $>$14.9  \\
                         450 &  $>$16.0 &  $>$17.1 &  $>$19.5 &  $>$22.1 &  $>$23.0 &  $>$16.8  \\
                         500 &  $>$19.1 &  $>$20.5 &  $>$22.7 &  $>$25.8 &  $>$26.4 &  $>$19.5  \\
                         600 &  $>$24.0 &  $>$26.1 &  $>$29.3 &  $>$33.0 &  $>$34.7 &  $>$25.0  \\
                         700 &  $>$30.3 &  $>$33.2 &  $>$37.7 &  $>$42.7 &  $>$45.0 &  $>$32.5  \\
                         800 &  $>$38.1 &  $>$42.6 &  $>$48.1 &  $>$54.9 &  --- &  $>$42.4  \\
                         900 &  $>$46.7 &  $>$51.2 &  $>$57.6 &  --- &  --- &  $>$52.7  \\
                        1000 &  $>$58.8 &  ---     &  ---     &  --- &  --- &  ---  \\
\hline
 \end{tabular}
    \label{tab:limits_mhmax}
  \end{center}
\end{table*}

\begin{table*}[!htb]
  \begin{center}
    \caption{Expected and observed 95\% CL limits for
             $\tan\beta$ as a function of $m_{\PA}$, for the MSSM search in the $m_{\Ph}^\text{mod$+$}$ benchmark scenario. For the statistical procedure a  test of the compatibility of the data to a signal of the three neutral Higgs bosons $\Ph$, $\PH$ and $\PA$ compared to a SM Higgs boson hypothesis is performed.
}
\begin{tabular}{c|ccccc|c}
\multicolumn{1}{c}{}  & \multicolumn{6}{c}{Neutral MSSM Higgs: $m_{\Ph}^\text{mod$+$}$ scenario}  \\
\cline{2-7}
  & \multicolumn{5}{c|}{Expected $\tan\beta$ limit}   &  Observed \\
\cline{2-6}
  $m_{\PA}$ [\GeVns{}] &$-2\sigma$  &   $-1\sigma$ &  Median &    $+1\sigma$ &  $+2\sigma$ & $\tan\beta$ limit \\
\hline
         \multirow{2}{*}{90} &   $>$4.9 &   $>$5.6 &   $>$6.6 &   $>$7.8 &   $>$8.3 &   $>$6.9  \\
                                  &   $<$1.2 &          &          &          &          &            \\
         \multirow{2}{*}{100} &   $>$4.3 &   $>$5.5 &   $>$6.7 &   $>$7.9 &   $>$9.0 &   $>$7.2  \\
                                   &   $<$1.1 &          &          &          &          &            \\
        \multirow{2}{*}{120} &   $>$3.8 &   $>$4.0 &   $>$5.4 &   $>$6.6 &   $>$7.1 &   $>$5.0  \\
                                  &   $<$0.9 &          &          &          &          &            \\
        \multirow{2}{*}{130} &   $>$3.8 &   $>$3.9 &   $>$4.9 &   $>$5.9 &   $>$6.0 &   $>$4.0  \\
                                  &   $<$1.1 &   $<$1.1 &   $<$0.9 &          &          &            \\
        \multirow{2}{*}{140} &   $>$3.4 &   $>$3.7 &   $>$4.8 &   $>$5.8 &   $>$6.3 &   $>$3.9  \\
                                  &   $<$1.4 &   $<$1.2 &   $<$1.1 &   $<$1.0 &          &   $<$1.2  \\
                         160 &   $>$2.8 &   $>$4.2 &   $>$5.5 &   $>$6.2 &   $>$8.2 &   $>$4.8  \\
                         180 &   $>$1.6 &   $>$5.2 &   $>$5.9 &   $>$6.9 &   $>$10.2 &   $>$6.4  \\
        \multirow{2}{*}{200} &   $>$4.4 &   $>$5.6 &   $>$6.5 &   $>$7.6 &   $>$8.7 &   $>$7.8  \\
                                  &   $<$2.9 &          &          &          &          &            \\
                         250 &   $>$6.5 &   $>$7.5 &   $>$9.2 &   $>$9.8 &   $>$11.9 &   $>$9.5  \\
                         300 &   $>$9.4 &   $>$9.8 &   $>$11.7 &   $>$12.7 &  $>$13.9 &   $>$9.9  \\
                         350 &   $>$11.7 &   $>$12.5 &   $>$13.9 &   $>$15.8 &   $>$16.0 &   $>$12.3  \\

                         400 &   $>$13.0 &   $>$14.6 &   $>$16.5 &   $>$19.0 &   $>$19.9 &   $>$15.3  \\
                         450 &   $>$16.2 &   $>$17.7 &   $>$19.7 &   $>$22.4 &   $>$23.2 &   $>$17.1  \\
                         500 &   $>$19.5 &   $>$21.0 &   $>$22.9 &   $>$26.1 &   $>$26.4 &   $>$19.7  \\
                         600 &   $>$24.1 &   $>$26.5 &   $>$29.7 &   $>$33.5 &   $>$35.2 &   $>$25.5  \\
                         700 &   $>$31.4 &   $>$34.1 &   $>$38.3 &   $>$43.4 &   $>$45.2 &   $>$33.7  \\
                         800 &   $>$39.6 &   $>$43.0 &   $>$49.0 &   $>$56.4 &   --- &   $>$42.9  \\
                         900 &   $>$47.2 &   $>$52.8 &   $>$59.9 &   --- &   --- &   $>$55.5  \\
                        1000 &   --- &   --- &   --- &   --- &   --- &   ---  \\
\hline
 \end{tabular}
    \label{tab:limits_mhmodp}
  \end{center}
\end{table*}

\begin{table*}[htbp]
  \begin{center}
    \caption{Expected and observed 95\% CL limits for
             $\tan\beta$ as a function of $m_{\PA}$, for the MSSM search in the $m_{\Ph}^\text{mod$-$}$ benchmark scenario. For the statistical procedure a  test of the compatibility of the data to a signal of the three neutral Higgs bosons $\Ph$, $\PH$ and $\PA$ compared to a SM Higgs boson hypothesis is performed.
}
\begin{tabular}{c|ccccc|c}
\multicolumn{1}{c}{}  & \multicolumn{6}{c}{Neutral MSSM Higgs: $m_{\Ph}^\text{mod$-$}$ scenario}  \\
\cline{2-7}
  & \multicolumn{5}{c|}{Expected $\tan\beta$ limit}   &  Observed \\
\cline{2-6}
  $m_{\PA}$ [\GeVns{}] &$-2\sigma$  &   $-1\sigma$ &  Median &    $+1\sigma$ &  $+2\sigma$ & $\tan\beta$ limit \\
\hline
         \multirow{2}{*}{90} &   $>$4.9 &   $>$5.6 &   $>$6.5 &   $>$7.7 &   $>$8.2 &   $>$6.8  \\
                                  &   $<$1.1 &          &          &          &          &            \\
        \multirow{2}{*}{100} &   $>$4.4 &   $>$5.5 &   $>$6.6 &   $>$7.8 &   $>$8.9 &   $>$7.2  \\
                                  &   $<$1.0 &          &          &          &          &            \\
                         120 &   $>$3.8 &   $>$4.0 &   $>$5.4 &   $>$6.5 &   $>$7.0 &   $>$4.8  \\
        \multirow{2}{*}{130} &   $>$3.8 &   $>$3.9 &   $>$4.8 &   $>$5.9 &   $>$5.8 &   $>$4.0  \\
                                  &   $<$1.1 &   $<$1.1 &   $<$0.9 &          &          &            \\
        \multirow{2}{*}{140} &   $>$3.1 &   $>$3.7 &   $>$4.8 &   $>$5.8 &   $>$6.4 &   $>$3.8  \\
                                  &   $<$1.5 &   $<$1.2 &   $<$1.1 &   $<$1.0 &          &   $<$1.2  \\
                         160 &   $>$2.9 &   $>$4.3 &   $>$5.5 &   $>$6.1 &   $>$8.0 &   $>$4.8  \\
        \multirow{2}{*}{180} &   $>$4.0 &   $>$5.2 &   $>$5.9 &   $>$6.8 &   $>$7.7 &   $>$6.3  \\
                                  &   $<$0.9 &          &          &          &          &            \\
        \multirow{2}{*}{200} &   $>$4.7 &   $>$5.5 &   $>$6.5 &   $>$7.5 &   $>$8.2 &   $>$7.7  \\
                                  &   $<$2.5 &          &          &          &          &            \\
        \multirow{2}{*}{250} &   $>$6.6 &   $>$7.3 &   $>$9.1 &   $>$9.8 &   $>$11.7 &   $>$9.5  \\
                                  &   $<$2.3 &          &          &          &           &            \\
                         300 &   $>$9.3 &   $>$9.8 &   $>$11.5 &   $>$12.6 &   $>$13.7 &   $>$9.9  \\
                         350 &   $>$11.5 &   $>$12.4 &   $>$13.6 &   $>$15.3 &   $>$15.7 &   $>$12.3  \\
                         400 &   $>$13.2 &   $>$14.5 &   $>$16.2 &   $>$18.8 &   $>$19.2 &   $>$14.8  \\
                         450 &   $>$15.7 &   $>$17.2 &   $>$19.5 &   $>$22.1 &   $>$23.2 &   $>$17.0  \\
                         500 &   $>$19.1 &   $>$20.0 &   $>$22.6 &   $>$25.0 &   $>$26.1 &   $>$19.7  \\
                         600 &   $>$24.0 &   $>$25.6 &   $>$29.1 &   $>$32.8 &   $>$34.1 &   $>$24.7  \\
                         700 &   $>$29.8 &   $>$32.8 &   $>$36.9 &   $>$42.0 &   $>$44.1 &   $>$32.9  \\
                         800 &   $>$36.9 &   $>$41.4 &   $>$47.1 &   $>$53.7 &   $>$57.2 &   $>$42.5  \\
                         900 &   $>$46.2 &   $>$49.9 &   $>$56.8 &   --- &   --- &   $>$53.1  \\
                        1000 &   $>$56.3 &   ---     &   ---     &   --- &   --- &   ---      \\
\hline

 \end{tabular}
    \label{tab:limits_mhmodm}
  \end{center}
\end{table*}

\begin{table*}[htbp]
  \begin{center}
    \caption{Expected and observed 95\% CL limits for
             $\tan\beta$ as a function of $m_{\PA}$, for the MSSM search in the light-stop benchmark scenario. For the statistical procedure a  test of the compatibility of the data to a signal of the three neutral Higgs bosons $\Ph$, $\PH$ and $\PA$ compared to a SM Higgs boson hypothesis is performed.
}
\begin{tabular}{c|ccccc|c}
\multicolumn{1}{c}{}  & \multicolumn{6}{c}{Neutral MSSM Higgs: light-stop scenario}  \\
\cline{2-7}
  & \multicolumn{5}{c|}{Expected $\tan\beta$ limit}   &  Observed \\
\cline{2-6}
  $m_{\PA}$ [\GeVns{}] &$-2\sigma$  &   $-1\sigma$ &  Median &    $+1\sigma$ &  $+2\sigma$ & $\tan\beta$ limit \\
\hline
         \multirow{2}{*}{90} &   $>$5.2 &   $>$5.7 &   $>$6.4 &   $>$7.7 &   $>$7.7 &   $>$6.7  \\
                                  &   $<$1.2 &          &          &          &          &            \\
        \multirow{2}{*}{100} &   $>$5.4 &   $>$6.0 &   $>$6.8 &   $>$8.0 &   $>$8.3 &   $>$7.7  \\
                                  &   $<$0.9 &          &          &          &          &            \\
        \multirow{2}{*}{120} &   $>$3.9 &   $>$5.0 &   $>$5.9 &   $>$6.9 &   $>$7.9 &   $>$6.5  \\
                                  &   $<$0.9 &          &          &          &          &            \\
        \multirow{2}{*}{130} &   $>$3.8 &   $>$4.0 &   $>$5.4 &   $>$6.4 &   $>$7.1 &   $>$5.5  \\
                                  &   $<$1.1 &   $<$1.1 &   $<$0.9 &          &          &            \\
        \multirow{2}{*}{140} &   $>$3.3 &   $>$3.8 &   $>$5.1 &   $>$6.0 &   $>$7.0 &   $>$4.6  \\
                                  &   $<$1.4 &   $<$1.2 &   $<$1.1 &   $<$1.0 &           &  $<$1.2  \\
        \multirow{2}{*}{160} &   $>$3.3 &   $>$4.4 &   $>$5.5 &   $>$6.3 &   $>$7.2 &   $>$4.9  \\
                                  &   $<$1.4 &   $<$3.3 &   $<$1.5 &   $<$1.2 &   $<$1.1 &   $<$3.0  \\
        \multirow{2}{*}{180} &   $>$3.3 &   $>$4.4 &   $>$5.6 &   $>$6.7 &   $>$7.8 &   $>$6.0  \\
                                  &   $<$1.4 &   $<$3.3 &   $<$2.4 &   $<$1.4 &   $<$1.1 &   $<$3.0  \\
        \multirow{2}{*}{200} &   $>$3.3 &   $>$4.4 &   $>$5.9 &   $>$7.1 &   $>$7.7 &   $>$7.3  \\
                                  &   $<$1.4 &   $<$3.3 &   $<$2.0 &          &          &   $<$2.0  \\
        \multirow{2}{*}{250} &   $>$3.3 &   $>$6.0 &   $>$7.9 &   $>$9.5 &   $>$11.8 &   $>$8.6  \\
                                  &   $<$1.4 &   $<$2.8 &          &          &           &            \\
        \multirow{2}{*}{300} &   $>$6.4 &   $>$7.1 &   $>$9.1 &   $>$10.9 &   $>$11.9 &   $>$7.5  \\
                                  &   $<$5.2 &   $<$3.6 &   $<$2.0 &           &           &   $<$2.3  \\
                         350 &   $>$6.4 &   $>$8.2 &   $>$10.0 &   $>$12.5 &   $>$13.5 &   $>$8.3  \\
                         400 &   $>$9.3 &   $>$10.8 &   $>$12.7 &   $>$14.9 &   $>$16.0 &   $>$11.6  \\
                         450 &   $>$11.6 &   $>$12.7 &   $>$14.8 &   $>$17.6 &   $>$18.1 &   $>$12.3  \\
                         500 &   $>$13.1 &   $>$14.8 &   $>$17.8 &   $>$21.1 &   $>$22.6 &   $>$13.9  \\
                         600 &   $>$16.6 &   $>$19.6 &   $>$23.4 &   $>$28.0 &   $>$30.3 &   $>$18.7  \\
\hline
 \end{tabular}
    \label{tab:limits_lightstop}
  \end{center}
\end{table*}

\begin{table*}[htbp]
  \begin{center}
    \caption{Expected and observed 95\% CL limits for
             $\tan\beta$ as a function of $m_{\PA}$, for the MSSM search in the light-stau benchmark scenario. For the statistical procedure a  test of the compatibility of the data to a signal of the three neutral Higgs bosons $\Ph$, $\PH$ and $\PA$ compared to a SM Higgs boson hypothesis is performed.
}
\begin{tabular}{c|ccccc|c}
\multicolumn{1}{c}{}  & \multicolumn{6}{c}{Neutral MSSM Higgs: light-stau scenario}  \\
\cline{2-7}
  & \multicolumn{5}{c|}{Expected $\tan\beta$ limit}   &  Observed \\
\cline{2-6}
  $m_{\PA}$ [\GeVns{}] &$-2\sigma$  &   $-1\sigma$ &  Median &    $+1\sigma$ &  $+2\sigma$ & $\tan\beta$ limit \\
\hline
         \multirow{2}{*}{90} &   $>$4.5 &   $>$5.4 &   $>$6.5 &   $>$7.7 &   $>$8.4 &   $>$6.7  \\
                                  &   $<$1.1 &          &          &          &          &           \\
        \multirow{2}{*}{100} &   $>$4.0 &   $>$5.4 &   $>$6.5 &   $>$7.8 &   $>$9.0 &   $>$7.3  \\
                                  &   $<$1.1 &          &          &          &          &            \\
                         120 &   $>$3.7 &   $>$3.9 &   $>$5.3 &   $>$6.4 &   $>$6.9 &   $>$4.9  \\
        \multirow{2}{*}{130} &   $>$3.8 &   $>$3.9 &   $>$4.7 &   $>$5.8 &   $>$5.7 &   $>$4.0  \\
                                  &   $<$1.1 &   $<$1.1 &   $<$0.9 &          &          &           \\
        \multirow{2}{*}{140} &   $>$3.3 &   $>$3.7 &   $>$4.7 &   $>$5.7 &   $>$6.0 &   $>$3.8  \\
                                  &   $<$1.4 &   $<$1.2 &   $<$1.1 &   $<$1.0 &          &   $<$1.2  \\
        \multirow{2}{*}{160} &   $>$3.3 &   $>$4.2 &   $>$5.4 &   $>$6.0 &   $>$7.0 &   $>$4.6  \\
                                  &   $<$1.4 &   $<$3.6 &   $<$1.5 &   $<$1.2 &   $<$1.1 &   $<$3.2  \\
        \multirow{2}{*}{180} &   $>$3.3 &   $>$4.2 &   $>$5.5 &   $>$6.6 &   $>$7.6 &   $>$5.7  \\
                                  &   $<$1.4 &   $<$3.6 &   $<$2.3 &   $<$1.3 &   $<$1.1 &   $<$2.8  \\
        \multirow{2}{*}{200} &   $>$3.3 &   $>$4.3 &   $>$5.8 &   $>$7.0 &   $>$7.6 &   $>$7.3  \\
                                  &   $<$1.4 &   $<$3.7 &   $<$1.6 &          &          &   $<$1.7  \\
        \multirow{2}{*}{250} &   $>$4.6 &   $>$6.2 &   $>$7.9 &   $>$9.5 &   $>$11.2 &   $>$8.9  \\
                                  &   $<$3.5 &   $<$2.4 &          &          &           &            \\
        \multirow{2}{*}{300} &   $>$6.4 &   $>$7.7 &   $>$9.3 &   $>$11.0 &   $>$12.3 &   $>$8.2  \\
                                  &   $<$3.5 &   $<$2.2 &          &           &           &           \\
        \multirow{2}{*}{350} &   $>$6.5 &   $>$8.5 &   $>$10.1 &   $>$12.5 &   $>$13.8 &   $>$8.8  \\
                                  &   $<$1.3 &          &          &           &           &           \\
                         400 &   $>$9.8 &   $>$11.6 &   $>$12.8 &   $>$15.0 &   $>$15.8 &   $>$12.0  \\
                         450 &   $>$12.2 &   $>$13.2 &   $>$15.4 &   $>$18.3 &   $>$18.5 &   $>$12.9  \\
                         500 &   $>$14.1 &   $>$16.0 &   $>$18.6 &   $>$21.6 &   $>$23.1 &   $>$15.0  \\
                         600 &   $>$18.8 &   $>$20.7 &   $>$23.9 &   $>$28.2 &   $>$29.0 &   $>$19.6  \\
                         700 &   $>$24.7 &   $>$27.7 &   $>$32.2 &   $>$37.6 &   $>$39.6 &   $>$27.0  \\
                         800 &   $>$34.8 &   $>$39.4 &   $>$45.1 &   $>$52.8 &   $-$ &   $>$39.5  \\
                         900 &   $>$45.2 &   $>$50.2 &   $>$56.9 &   --- &   --- &   $>$52.0  \\
                        1000 &   $>$59.6 &   ---     &   ---     &   --- &   --- &   ---      \\
\hline
 \end{tabular}
    \label{tab:limits_lightstau}
  \end{center}
\end{table*}

\begin{table*}[htbp]
  \begin{center}
    \caption{Expected and observed 95\% CL limits for
             $\tan\beta$ as a function of $m_{\PA}$, for the MSSM search in the $\Pgt$-phobic benchmark scenario. For the statistical procedure a  test of the compatibility of the data to a signal of the three neutral Higgs bosons $\Ph$, $\PH$ and $\PA$ compared to a SM Higgs boson hypothesis is performed.
}
\begin{tabular}{c|ccccc|c}
\multicolumn{1}{c}{}  & \multicolumn{6}{c}{Neutral MSSM Higgs: $\Pgt$-phobic scenario}  \\
\cline{2-7}
  & \multicolumn{5}{c|}{Expected $\tan\beta$ limit}   &  Observed \\
\cline{2-6}
  $m_{\PA}$ [\GeVns{}] &$-2\sigma$  &   $-1\sigma$ &  Median &    $+1\sigma$ &  $+2\sigma$ & $\tan\beta$ limit \\
\hline
         \multirow{2}{*}{90} &   $>$3.3 &   $>$5.4 &   $>$6.3 &   $>$7.5 &   $>$9.2 &   $>$6.7  \\
                                  &   $<$1.8 &   $<$1.4 &   $<$1.2 &   $<$1.1 &   $<$1.1 &   $<$1.4  \\
        \multirow{2}{*}{100} &   $>$3.3 &   $>$4.8 &   $>$6.5 &   $>$7.8 &   $>$10.9 &   $>$7.5  \\
                                  &   $<$1.8 &          &          &          &           &            \\
                         120 &   $>$2.7 &   $>$3.3 &   $>$3.8 &   $>$5.5 &   $>$4.9 &   $>$3.3  \\
        \multirow{2}{*}{130} &   $>$3.6 &   $>$3.7 &   $>$3.9 &   $>$4.7 &   $>$4.2 &   $>$3.8  \\
                                  &   $<$1.2 &   $<$1.1 &          &          &          &            \\
        \multirow{2}{*}{140} &   $>$2.9 &   $>$3.4 &   $>$3.8 &   $>$5.1 &   $>$4.7 &   $>$3.5  \\
                                  &   $<$1.5 &   $<$1.3 &   $<$1.1 &   $<$1.0 &          &   $<$1.3  \\
        \multirow{2}{*}{160} &   $>$2.9 &   $>$3.4 &   $>$5.3 &   $>$6.0 &   $>$7.1 &   $>$4.3  \\
                                  &   $<$1.5 &   $<$1.3 &   $<$1.7 &   $<$1.2 &   $<$1.1 &   $<$3.6  \\
        \multirow{2}{*}{180} &   $>$2.9 &   $>$4.2 &   $>$5.6 &   $>$6.7 &   $>$7.8 &   $>$6.1  \\
                                  &   $<$1.5 &   $<$3.8 &   $<$2.3 &   $<$1.4 &   $<$1.1 &   $<$2.7  \\
        \multirow{2}{*}{200} &   $>$2.9 &   $>$4.4 &   $>$6.0 &   $>$7.4 &   $>$8.0 &   $>$7.7  \\
                                  &   $<$1.5 &   $<$3.6 &   $<$2.0 &          &          &   $<$2.0  \\
        \multirow{2}{*}{250} &   $>$5.6 &   $>$6.4 &   $>$7.8 &   $>$9.6 &   $>$10.1 &   $>$8.8  \\
                                  &   $<$3.2 &   $<$2.4 &          &          &           &           \\
        \multirow{2}{*}{300} &   $>$6.4 &   $>$8.3 &   $>$9.7 &   $>$12.0 &   $>$13.0 &   $>$8.6  \\
                                  &   $<$3.0 &   $<$2.2 &          &           &            &          \\
                         350 &   $>$8.2 &   $>$9.4 &   $>$11.7 &   $>$13.7 &   $>$15.3 &   $>$9.4  \\
                         400 &   $>$10.7 &   $>$12.3 &   $>$14.4 &   $>$16.9 &   $>$18.1 &   $>$12.9  \\
                         450 &   $>$12.6 &   $>$14.6 &   $>$17.3 &   $>$20.6 &   $>$22.0 &   $>$14.4  \\
                         500 &   $>$15.3 &   $>$18.0 &   $>$21.1 &   $>$24.8 &   $>$26.8 &   $>$16.6  \\
                         600 &   $>$21.6 &   $>$24.0 &   $>$29.0 &   $>$34.7 &   $>$36.3 &   $>$22.9  \\
                         700 &   $>$30.6 &   $>$34.5 &   $>$40.6 &   $>$49.5 &   --- &   $>$33.9  \\
                         800 &   $>$45.5 &   --- &   --- &   --- &   --- &   ---  \\
                         900 &   --- &   --- &   --- &   --- &   --- &   ---  \\
                        1000 &   --- &   --- &   --- &   --- &   --- &   ---  \\
\hline
 \end{tabular}
    \label{tab:limits_tauphobic}
  \end{center}
\end{table*}

\begin{table*}[htbp]
  \begin{center}
    \caption{Expected and observed 95\% CL limits for
             $\tan\beta$ as a function of $\mu$, for the MSSM search in the low-$m_{\PH}$ benchmark scenario. For the statistical procedure a  test of the compatibility of the data to a signal of the three neutral Higgs bosons $\Ph$, $\PH$ and $\PA$ compared to a SM Higgs boson hypothesis is performed.
}
\begin{tabular}{c|ccccc|c}
\multicolumn{1}{c}{}  & \multicolumn{6}{c}{Neutral MSSM Higgs:  low-$m_{\PH}$ scenario}  \\
\cline{2-7}
  & \multicolumn{5}{c|}{Expected $\tan\beta$ limit}   &  Observed \\
\cline{2-6}
  $\mu$ [\GeVns{}] &$-2\sigma$  &   $-1\sigma$ &  Median &    $+1\sigma$ &  $+2\sigma$ & $\tan\beta$ limit \\
\hline
                         300 &   $>$3.4 &   $>$4.7 &   $>$6.0 &   $>$7.3 &   $>$8.6 &   $>$6.6  \\
                         400 &   $>$3.2 &   $>$4.7 &   $>$6.0 &   $>$7.2 &   $>$8.8 &   $>$6.7  \\
                         500 &   $>$3.3 &   $>$4.9 &   $>$6.1 &   $>$7.3 &  $>$9.0 &   $>$6.7  \\
                         600 &   $>$3.4 &   $>$5.0 &   $>$6.2 &   $>$7.4 &   $>$9.0 &   $>$6.8  \\
                         700 &   $>$3.4 &   $>$5.0 &   $>$6.1 &   $>$7.3 &   $>$8.8 &   $>$6.7  \\
                         800 &   $>$4.0 &   $>$5.2 &   $>$6.1 &   $>$7.3 &   $>$8.2 &   $>$6.7  \\
                        1100 &   $>$4.5 &   $>$5.3 &   $>$6.3 &   $>$7.5 &   $>$8.1 &   $>$6.8  \\
                        1300 &   $>$4.7 &   $>$5.4 &   $>$6.4 &   $>$7.6 &   $>$8.2 &   $>$6.9  \\
                        1400 &   $>$4.9 &   $>$5.5 &   $>$6.5 &   $>$7.7 &   $>$8.1 &   $>$6.9  \\
                        1500 &   $>$5.0 &   $>$5.6 &   $>$6.6 &   $>$7.8 &   $>$8.2 &   $>$7.0  \\
                        1600 &   $>$5.1 &   $>$5.7 &   $>$6.7 &   $>$7.9 &   $>$8.3 &   $>$7.0  \\
                        1700 &   $>$5.2 &   $>$5.8 &   $>$6.8 &   $>$7.9 &   $>$8.3 &   $>$7.0  \\
                        1800 &   $>$5.3 &   $>$5.9 &   $>$6.8 &   $>$7.9 &   $>$8.3 &   $>$7.0  \\
                        1900 &   $>$5.2 &   $>$5.9 &   $>$6.8 &   $>$7.9 &   $>$8.3 &   $>$6.8  \\
                        2000 &   $>$5.3 &   $>$5.9 &   $>$6.7 &   $>$7.9 &   $>$8.1 &   $>$6.6  \\
                        2100 &   $>$5.2 &   $>$5.8 &   $>$6.6 &   $>$8.1 &   $>$8.0 &   $>$6.4  \\
                        2200 &   $>$5.1 &   $>$5.6 &   $>$6.7 &   $>$8.1 &   $>$8.3 &   $>$6.3  \\
                        2300 &   $>$5.0 &   $>$5.5 &   $>$6.9 &   $>$8.1 &   $>$8.9 &   $>$6.7  \\
                        2400 &   $>$4.7 &   $>$5.4 &   $>$7.0 &   $>$8.0 &   $>$9.4 &   $>$7.3  \\
                        2500 &   $>$4.4 &   $>$5.7 &   $>$7.0 &   $>$8.0 &   $>$9.6 &   $>$7.4  \\
                        2600 &   $>$4.0 &   $>$6.1 &   $>$6.9 &   $>$7.9 &   $>$9.9 &   $>$7.2  \\
                        2700 &   $>$3.8 &   $>$6.1 &   $>$6.8 &   $>$7.8 &   $>$9.8 &   $>$6.9  \\
                        2800 &   $>$5.1 &   $>$5.9 &   $>$6.7 &   $>$7.7 &   $>$8.3 &   $>$6.8  \\
                        2900 &   $>$5.2 &   $>$5.8 &   $>$6.5 &   $>$7.5 &   $>$7.8 &   $>$6.5  \\
                        3000 &   $>$5.1 &   $>$5.6 &   $>$6.3 &   $>$7.4 &   $>$7.6 &   $>$6.3  \\
                        3100 &   $>$4.6 &   $>$5.3 &   $>$6.1 &   $>$7.1 &   $>$7.5 &   $>$6.0  \\
\hline
 \end{tabular}
    \label{tab:limits_lowmh}
  \end{center}
\end{table*}

\begin{table*}[htbp]
  \begin{center}
    \caption{Expected and observed 95\% CL upper limits for $\sigma(\cPg\cPg\phi)\cdot\mathcal{B}(\Pgt\Pgt)$ (pb) at 8\TeV center-of-mass energy as a function of the Higgs mass $m_{\phi}$, where ${\phi}$ denotes a generic Higgs-like state. The expected and observed limits are computed using the test statistics given by Eq.~\ref{eq:teststatistic_LHC}. For the expected limits, the observed data have been replaced by a representative dataset which not only contains the contribution from background processes but also a SM Higgs boson with a mass of 125\GeV.
}
\begin{tabular}{c|ccccc|c}
\multicolumn{1}{c}{}      & \multicolumn{6}{c}{$\sigma(\cPg\cPg\phi)\cdot\mathcal{B}(\Pgt\Pgt)$: $\sqrt{s} = 8\TeV$} \\
\cline{2-7}
     & \multicolumn{5}{c|}{Expected limit} & \multicolumn{1}{c}{Observed} \\
\cline{2-6}
  $m_{\phi}$ [\GeVns{}] &$-2\sigma$  &   $-1\sigma$ &  Median &    $+1\sigma$ &  $+2\sigma$ & limit \\
\hline

                90 &               $16.75$ &               $20.36$ &               $25.45$ &               $31.65$ &               $37.34$ &               $50.21$  \\
               100 &               $13.54$ &               $16.66$ &               $21.18$ &               $27.00$ &               $34.57$ &               $31.33$  \\
               120 &               $2.96$ &               $3.88$ &               $5.14$ &               $6.65$ &               $8.09$ &               $7.38$  \\
               130 &               $1.51$ &               $2.05$ &               $2.75$ &               $3.63$ &               $4.55$ &               $4.39$  \\
               140 &               $9.42 \times 10^{-1}$ &               $1.24$ &               $1.68$ &               $2.23$ &               $2.82$ &               $2.27$  \\
               160 &               $4.74 \times 10^{-1}$ &               $6.15 \times 10^{-1}$ &               $8.55 \times 10^{-1}$ &               $1.14$ &               $1.50$ &               $8.45 \times 10^{-1}$  \\
               180 &               $3.12 \times 10^{-1}$ &               $4.05 \times 10^{-1}$ &               $5.52 \times 10^{-1}$ &               $7.41 \times 10^{-1}$ &               $9.74 \times 10^{-1}$ &               $5.49 \times 10^{-1}$  \\
               200 &               $2.37 \times 10^{-1}$ &               $3.08 \times 10^{-1}$ &               $4.19 \times 10^{-1}$ &               $5.56 \times 10^{-1}$ &               $7.48 \times 10^{-1}$ &               $5.17 \times 10^{-1}$  \\
               250 &               $1.30 \times 10^{-1}$ &               $1.69 \times 10^{-1}$ &               $2.28 \times 10^{-1}$ &               $3.11 \times 10^{-1}$ &               $3.95 \times 10^{-1}$ &               $3.15 \times 10^{-1}$  \\
               300 &               $7.95 \times 10^{-2}$ &               $1.05 \times 10^{-1}$ &               $1.43 \times 10^{-1}$ &               $1.97 \times 10^{-1}$ &               $2.53 \times 10^{-1}$ &               $1.50 \times 10^{-1}$  \\
               350 &               $6.05 \times 10^{-2}$ &               $7.71 \times 10^{-2}$ &               $1.02 \times 10^{-1}$ &               $1.38 \times 10^{-1}$ &               $1.80 \times 10^{-1}$ &               $1.12 \times 10^{-1}$  \\
               400 &               $4.68 \times 10^{-2}$ &               $6.01 \times 10^{-2}$ &               $7.94 \times 10^{-2}$ &               $1.08 \times 10^{-1}$ &               $1.41 \times 10^{-1}$ &               $1.03 \times 10^{-1}$  \\
               450 &               $3.76 \times 10^{-2}$ &               $4.77 \times 10^{-2}$ &               $6.36 \times 10^{-2}$ &               $8.39 \times 10^{-2}$ &               $1.11 \times 10^{-1}$ &               $6.07 \times 10^{-2}$  \\
               500 &               $3.20 \times 10^{-2}$ &               $4.01 \times 10^{-2}$ &               $5.31 \times 10^{-2}$ &               $7.14 \times 10^{-2}$ &               $9.28 \times 10^{-2}$ &               $3.83 \times 10^{-2}$  \\
               600 &               $1.97 \times 10^{-2}$ &               $2.54 \times 10^{-2}$ &               $3.49 \times 10^{-2}$ &               $4.70 \times 10^{-2}$ &               $6.22 \times 10^{-2}$ &               $1.93 \times 10^{-2}$  \\
               700 &               $1.49 \times 10^{-2}$ &               $1.95 \times 10^{-2}$ &               $2.72 \times 10^{-2}$ &               $3.71 \times 10^{-2}$ &               $4.73 \times 10^{-2}$ &               $1.44 \times 10^{-2}$  \\
               800 &               $1.19 \times 10^{-2}$ &               $1.53 \times 10^{-2}$ &               $2.09 \times 10^{-2}$ &               $2.89 \times 10^{-2}$ &               $3.91 \times 10^{-2}$ &               $1.12 \times 10^{-2}$  \\
               900 &               $8.55 \times 10^{-3}$ &               $1.11 \times 10^{-2}$ &               $1.51 \times 10^{-2}$ &               $2.11 \times 10^{-2}$ &               $2.89 \times 10^{-2}$ &               $9.39 \times 10^{-3}$  \\
              1000 &               $7.47 \times 10^{-3}$ &               $9.80 \times 10^{-3}$ &               $1.30 \times 10^{-2}$ &               $1.86 \times 10^{-2}$ &               $2.54 \times 10^{-2}$ &               $8.50 \times 10^{-3}$  \\
\hline
 \end{tabular}
    \label{ggH-limit}
  \end{center}
\end{table*}

\begin{table*}[htbp]
  \begin{center}
    \caption{Expected and observed 95\% CL upper limits for $\sigma(\cPqb\cPqb\phi)\cdot\mathcal{B}(\Pgt\Pgt)$ (pb) at 8\TeV center-of-mass energy as a function of the Higgs mass $m_{\phi}$, where ${\phi}$ denotes a generic Higgs-like state. The expected and observed limits are computed using the test statistics given by Eq.~\ref{eq:teststatistic_LHC}. For the expected limits, the observed data have been replaced by a representative dataset which not only contains the contribution from background processes but also a SM Higgs boson with a mass of 125\GeV.}
\begin{tabular}{c|ccccc|c}
\multicolumn{1}{c}{}      & \multicolumn{6}{c}{$\sigma(\cPqb\cPqb\phi)\cdot\mathcal{B}(\Pgt\Pgt)$: $\sqrt{s} = 8\TeV$} \\
\cline{2-7}
                          & \multicolumn{5}{c|}{Expected limit} & \multicolumn{1}{c}{Observed} \\
\cline{2-6}
  $m_{\phi}$ [\GeVns{}] &$-2\sigma$  &   $-1\sigma$ &  Median &    $+1\sigma$ &  $+2\sigma$ & limit \\
\hline
                90 &               $3.25$ &               $4.19$ &               $5.69$ &               $7.54$ &               $9.74$ &               $6.04$  \\
               100 &               $2.43$ &               $3.07$ &               $4.14$ &               $5.69$ &               $7.19$ &               $4.13$  \\
               120 &               $1.18$ &               $1.53$ &               $2.08$ &               $2.82$ &               $3.72$ &               $1.76$  \\
               130 &               $8.31 \times 10^{-1}$ &               $1.08$ &               $1.47$ &               $1.97$ &               $2.60$ &               $1.26$  \\
               140 &               $6.21 \times 10^{-1}$ &               $8.24 \times 10^{-1}$ &               $1.10$ &               $1.46$ &               $1.91$ &               $1.25$  \\
               160 &               $3.80 \times 10^{-1}$ &               $4.99 \times 10^{-1}$ &               $6.68 \times 10^{-1}$ &               $8.94 \times 10^{-1}$ &               $1.18$ &               $8.14 \times 10^{-1}$  \\
               180 &               $2.64 \times 10^{-1}$ &               $3.45 \times 10^{-1}$ &               $4.59 \times 10^{-1}$ &               $6.27 \times 10^{-1}$ &               $8.40 \times 10^{-1}$ &               $6.59 \times 10^{-1}$  \\
               200 &               $1.92 \times 10^{-1}$ &               $2.50 \times 10^{-1}$ &               $3.39 \times 10^{-1}$ &               $4.60 \times 10^{-1}$ &               $6.10 \times 10^{-1}$ &               $5.53 \times 10^{-1}$  \\
               250 &               $1.09 \times 10^{-1}$ &               $1.41 \times 10^{-1}$ &               $1.90 \times 10^{-1}$ &               $2.62 \times 10^{-1}$ &               $3.42 \times 10^{-1}$ &               $2.16 \times 10^{-1}$  \\
               300 &               $7.33 \times 10^{-2}$ &               $9.15 \times 10^{-2}$ &               $1.25 \times 10^{-1}$ &               $1.70 \times 10^{-1}$ &               $2.24 \times 10^{-1}$ &               $9.75 \times 10^{-2}$  \\
               350 &               $5.30 \times 10^{-2}$ &               $6.58 \times 10^{-2}$ &               $8.95 \times 10^{-2}$ &               $1.20 \times 10^{-1}$ &               $1.60 \times 10^{-1}$ &               $6.38 \times 10^{-2}$  \\
               400 &               $4.09 \times 10^{-2}$ &               $5.30 \times 10^{-2}$ &               $6.92 \times 10^{-2}$ &               $9.59 \times 10^{-2}$ &               $1.26 \times 10^{-1}$ &               $6.13 \times 10^{-2}$  \\
               450 &               $3.15 \times 10^{-2}$ &               $4.09 \times 10^{-2}$ &               $5.51 \times 10^{-2}$ &               $7.68 \times 10^{-2}$ &               $9.81 \times 10^{-2}$ &               $4.31 \times 10^{-2}$  \\
               500 &               $2.70 \times 10^{-2}$ &               $3.41 \times 10^{-2}$ &               $4.63 \times 10^{-2}$ &               $6.33 \times 10^{-2}$ &               $8.44 \times 10^{-2}$ &               $3.20 \times 10^{-2}$  \\
               600 &               $1.77 \times 10^{-2}$ &               $2.34 \times 10^{-2}$ &               $3.16 \times 10^{-2}$ &               $4.37 \times 10^{-2}$ &               $5.66 \times 10^{-2}$ &               $2.03 \times 10^{-2}$  \\
               700 &               $1.35 \times 10^{-2}$ &               $1.78 \times 10^{-2}$ &               $2.44 \times 10^{-2}$ &               $3.38 \times 10^{-2}$ &               $4.40 \times 10^{-2}$ &               $1.73 \times 10^{-2}$  \\
               800 &               $1.16 \times 10^{-2}$ &               $1.49 \times 10^{-2}$ &               $2.01 \times 10^{-2}$ &               $2.73 \times 10^{-2}$ &               $3.77 \times 10^{-2}$ &               $1.65 \times 10^{-2}$  \\
               900 &               $8.81 \times 10^{-3}$ &               $1.15 \times 10^{-2}$ &               $1.52 \times 10^{-2}$ &               $2.12 \times 10^{-2}$ &               $3.00 \times 10^{-2}$ &               $1.48 \times 10^{-2}$  \\
              1000 &               $8.15 \times 10^{-3}$ &               $1.03 \times 10^{-2}$ &               $1.37 \times 10^{-2}$ &               $1.92 \times 10^{-2}$ &               $2.72 \times 10^{-2}$ &               $1.35 \times 10^{-2}$  \\
\hline
 \end{tabular}
    \label{bbH-limit}
  \end{center}
\end{table*}

\cleardoublepage \section{The CMS Collaboration \label{app:collab}}\begin{sloppypar}\hyphenpenalty=5000\widowpenalty=500\clubpenalty=5000\textbf{Yerevan Physics Institute,  Yerevan,  Armenia}\\*[0pt]
V.~Khachatryan, A.M.~Sirunyan, A.~Tumasyan
\vskip\cmsinstskip
\textbf{Institut f\"{u}r Hochenergiephysik der OeAW,  Wien,  Austria}\\*[0pt]
W.~Adam, T.~Bergauer, M.~Dragicevic, J.~Er\"{o}, C.~Fabjan\cmsAuthorMark{1}, M.~Friedl, R.~Fr\"{u}hwirth\cmsAuthorMark{1}, V.M.~Ghete, C.~Hartl, N.~H\"{o}rmann, J.~Hrubec, M.~Jeitler\cmsAuthorMark{1}, W.~Kiesenhofer, V.~Kn\"{u}nz, M.~Krammer\cmsAuthorMark{1}, I.~Kr\"{a}tschmer, D.~Liko, I.~Mikulec, D.~Rabady\cmsAuthorMark{2}, B.~Rahbaran, H.~Rohringer, R.~Sch\"{o}fbeck, J.~Strauss, A.~Taurok, W.~Treberer-Treberspurg, W.~Waltenberger, C.-E.~Wulz\cmsAuthorMark{1}
% ...
\vskip\cmsinstskip
% ...
1:~~Also at Vienna University of Technology, Vienna, Austria\\
2:~~Also at CERN, European Organization for Nuclear Research, Geneva, Switzerland\\
% ...

\end{sloppypar}
\end{document}